\documentclass[twocolumn]{aastex631}

\shorttitle{Evolutionary Period Changes for 25 X-ray Binaries}
\shortauthors{Schaefer}
\graphicspath{{./}{figures/}}

\begin{document}
\title{Evolutionary Period Changes for 25 X-ray Binaries and the Measurement of an Empirical Universal Law for Angular Momentum Loss in Accreting Binaries}

\author[0000-0002-2659-8763]{Bradley E. Schaefer}
\affiliation{Department of Physics and Astronomy,
Louisiana State University,
Baton Rouge, LA 70803, USA}



\begin{abstract}

I measure and collect timings of phase markers (like eclipse times) for the orbits of 25 X-ray binaries (XRBs) so as to calculate the steady evolutionary period change ($\dot{P}$).  I combine these with my observed $\dot{P}$ measures from 52 cataclysmic variables (CVs).  Further, I subtract out the contributions from gravitational radiation ($\dot{P}_{\rm GR}$) and mass transfer ($\dot{P}_{\rm mt}$), deriving the period change from the residual unknown angular momentum loss ($\dot{P}_{\rm AML}$=$\dot{P}$-$\dot{P}_{\rm GR}$-$\dot{P}_{\rm mt}$).  I have $\dot{P}_{\rm AML}$ measures for 77 XRBs and CVs, with these being direct measures of the driver of binary evolution.  The venerable Magnetic Braking Model (MBM) of binary evolution has its most fundamental predictions tested, with most systems having predictions wrong by over one order-of-magnitude.  Other proposed mechanisms to explain the AML also fail, so we are left with no known mechanism that dominates the AML.  An alternative path to the AML law is empirical, where my $\dot{P}_{\rm AML}$ measures are fitted to a power-law involving the fundamental binary properties.  With this, the dominant AML law for systems with orbital periods ($P$) from 0.13--1.0 days is $\dot{P}_{\rm AML} = -1500\times10^{-12} P^{1.29} M_{\rm prim}^{2.75}  M_{\rm comp}^{-1.00}\dot{M}^{0.43}_{-8}$, in appropriate units.  Similar AML laws for binaries below the Period Gap and for binaries with $P$$>$1.0 day are derived.  These three AML laws are of good accuracy and are the best representations of the actual evolution for all 77 XRBs and CVs of all classes, so the three taken together can be called `universal'.

\end{abstract}


\section{Introduction}

The most important issue involving accreting binaries is their evolution.  These include the X-ray binaries (XRBs) either with a black hole (BH) or neutron star (NS), and Cataclysmic Variables (CVs) with white dwarf (WD) accretors.  The evolution is always driven by the changes in the orbital period ($P$), for which the slow and steady changes in $P$ are measured with the time derivative $\dot{P}$.  The $\dot{P}$ drives the accretion, determines the lifetimes, and changes the periods, with all this being the primary determinant of the changes in individual systems and the overall population demographics.

The general evolution of XRBs and CVs (Patterson 1984, Knigge et al. 2011, henceforth K2011) is that when the natal binaries first come into contact, the $\dot{P}$ drives the evolution to shorter and shorter $P$.  The $\dot{P}$ is simply-related to the angular momentum loss (AML, $\dot{J}$) of the system, although I regard $\dot{P}$ as primary because it is observable and tangible.  Some of the period change is simply due to the mass transfer from the companion star (also called the mass `donor' star) to the primary star (here, always a compact star of either the BH, NS, or WD variety) adjusting the balance of angular momentum ($J$) in the binary.  In addition, there must be some source of angular momentum loss in the binary, as this is what relentlessly drives the binary to shorter and shorter $P$.  In this schematic evolution picture, the nature and magnitude and mechanisms for AML are unstated, and the nature of the AML is the whole question of evolution for XRBs and CVs.  As the $P$ decreases, at some time (typically when $P$$\sim$3 hours), the accretion of the binary will largely cut off, creating the Period Gap, where the number of systems with periods stereotypically from 2--3 hours (from $P_{\rm gap-}$ to $P_{\rm gap+}$) are relatively few compared to above the Gap and below the Gap.  After the system leaves the Period Gap, the primary mechanism for AML is the emission of gravitational radiation (GR for the mechanism).  Below the Gap, evolution is slow, with very low accretion rates ($\dot{M}$), and the ultimate contraction of the orbit stops at some minimum period ($P_{\rm min}$, i.e., with $\dot{P}$=0), after which the systems with increasing-$P$ are called `period bouncers'.  Everyone agrees that this general evolution picture must be schematically correct.  With the basic physical mechanisms being known (mass transfer, GR, and properties of the companion star), the primary question is the behavior of the AML.

Since the 1980s, the consensus model (K2011) for the specific evolution of XRBs and CVs is called the `magnetic braking model' (MBM).  In essence, the MBM is only a model of a specific power-law AML, producing the period change $\dot{P}_{\rm mb}$ that drives the binary evolution.  With this then applied to the properties of the stars plus well-known physical mechanisms, the time evolution of the binary properties can be calculated.  This model invokes a specific named AML mechanism as due to magnetic braking.  Magnetic braking is when the companion star emits a stellar wind, where the outgoing particles are frozen into co-rotation by the companion's surface magnetic field.  Much like in the oft-used analogy of a spinning skater extending their arms, the angular momentum given to the outgoing wind material slows down the rotation of the companion star.  The ejected wind material carries angular momentum away from the system, so the spin $J$ decreases and the total $J$ decreases.  The rotations of the companion stars in XRBs and CVs are always forced into synchronization with the orbit.  So the slowing of the spin $J$ of the companion is always transferred to the orbit, resulting in the orbital $\dot{J}$ and $\dot{P}_{\rm mb}$, both of which must be negative.  The magnetic braking mechanism does not work for fully convective stars (with corresponding $P$ of less than something like 3 hours), thus creating the Period Gap.  

The size and properties of this magnetic braking mechanism are largely unknown, with no empirical measures nor any detailed physical model.  As a path forward, all MBM calculations have postulated a conjectural law for all AML in close binaries as $\dot{J} \propto R_{\rm comp}^{\gamma}$, where $R_{\rm comp}$ is the radius of the companion star (K2011).  The considered values of $\gamma$ range from 0 to 4, and the negative constant of proportionality is taken from a range 10$^4$ in size.  Many MBM calculations adopt $\gamma$=3, using the specific formulation of Rappaport, Verbunt, \& Joss (1983).  With this choice, the MBM recovers the existence of the Period Gap, plus the values of $P_{\rm gap-}$, $P_{\rm gap+}$, and $P_{\rm min}$.  It is the successes of these indirect predictions that have made the MBM into the consensus model.

The magnetic braking mechanism has fundamental problems that make dubious its existence as a substantial source of AML.  A primary problem is that the existence of the mechanism has never been observed for any system.  Similar primary problems are that the existence of stellar winds and surface magnetic fields for XRB or CV companion stars have never been seen or measured, so the conformity to the MBM is mere hopeful conjecture, and where a likely chaotic reality has little resemblance to any assumed power law.   Further, I am not aware of any detailed physics calculation of the required wind strength and the required magnetic field strength so as to achieve any stated $\dot{J}$.  Another fundamental problem is that there is no real chance that any single power law can represent, even to within orders-of-magnitude, the diversity and complexity of stellar winds and surface magnetic fields of companion stars with 0.1--3 M$_{\odot}$ on the main sequence (MS) and out to evolved sub-giants.  Further, the strength of the AML power-law has uncertainties covering a range of 4 in the power-law index and a range of 10$^4$$\times$ in the constant (K2011 fig. 2), which is to say that no one has any idea as to the real strength of the magnetic braking power-law, or even whether it is a power-law at all. The basic mechanism of magnetic braking certainly exists, but there is no evidence that the magnitude of its effects are larger than completely-negligible.  It is dubious to base all of binary evolution on a physical mechanism that has never been observed, that requires stellar properties (the wind and magnetic field strengths) never before seen in the relevant situations, and with unknown detailed physics.

Nevertheless, modelers have forged onward with the MBM, largely because some formulation is desperately needed for evolution, demographics, and population synthesis studies of XRBs and CVs.  Detailed calculations depend on various assumptions and included corrections.  K2011 have recognized two primary variants, which they have labelled as the `standard model' (standard-MBM) and an `optimal' or `revised' model (revised-MBM).  Both the standard and revised models produce specific formulations for $\dot{P}_{\rm mb}$.  The differences are that standard-MBM uses the magnetic braking at the rate originally from Rappaport et al. (1983) and the GR rate at the level predicted by General Relativity, whilst the revised-MBM allows the strengths of magnetic braking and GR to vary arbitrarily\footnote{It sounds horrible to have the revised-MBM model postulate changes in Einstein's General Relativity, but what is really going on is that the revised-MBM model is simply postulating some additional mechanism added on top of GR, with this unknown added mechanism scaling on masses and periods identically as GR.  Still, it {\it is} horrible that the model is forced to postulate some mysterious and unknown new physical phenomenon invented solely to match the observed $P_{\rm min}$ and $P_{\rm gap-}$.} so as to match the observed $P_{\rm min}$, $P_{\rm gap-}$, and $P_{\rm gap+}$.  Both models require that all binary evolution quickly converges on to a single track, where the various system properties ($\dot{P}$, $\dot{M}$, ...) are functions of $P$.  In general, the revised-MBM model will be used, because the standard-MBM model fails badly in all the key predictions that made the MBM into the venerable model of old.

The MBM is largely untested.  The single most fundamental test would be to compare measured-$\dot{P}$ versus model-$\dot{P}$.  But such tests are hard, because a measure of the evolutionary period changes requires data of orbital phase markers (like eclipse times) well-measured over decades and a century, with data from long-ago being rare and precious.  For the case of recurrent novae (RNe) and classical novae (CNe), all $\dot{P}$ measures are by myself and fairly recent, published in a series of papers culminating in Schaefer (2023).  This preliminary study was extended to 52 CVs of all types (Schaefer 2024).  Stunningly, the MBM predictions fail comprehensively.  

This has motivated the current paper on XRBs, with this being a test of the generality of the MBM failures.  Prior good measures of the long-term evolutionary $\dot{P}$ for XRBs have been reported for a dozen systems.  These prior measures are all in disagreement with the MBM predictions, as noted by the authors in half the cases, with these disagreements explained away by invoking one or two, out of a dozen ideas, extra AML mechanism.  But the invoked AML mechanisms are different for each XRB and the mechanisms are of dubious application.  Falanga et al. (2015) have collected ten XRB $\dot{P}$ values, but they have not taken any global view, nor tested the MBM.

In this paper, I will measure, collect, improve, and update the measures of $\dot{P}$ for 25 XRBs.  These include 11 systems with high-mass companions (HMXBs), 10 systems with low-mass companions (LMXBs), 2 with intermediate-mass companions (IMXBs), and 2 with WD companions.  These will then be used to test the predictions of MBM.

\section{$\dot{P}$ MEASURES FOR 25 X-RAY BINARIES}

The immediate goal of this section is to measure and derive the $\dot{P}$ for each of 25 X-ray binaries with BH and NS compact objects.  The best tool for calculating the $\dot{P}$ is to construct and fit models to $O-C$ diagrams.  This is a means of representing a collection of observed times, say, times of minimum light, $T_{\rm min}$, as compared to some fiducial linear model.  The times for the calculated model are $T_{\rm O-C}=E_{\rm O-C}+N\times P_{\rm O-C}$, where $N$ is an integer that counts the orbits from the epoch.  The times are given as heliocentric Julian days (HJD) or as barycentric Julian days (BJD) when high accuracy is required.  The $O-C$ value for a particular observed time is calculated as $T_{\rm min}$-$T_{\rm O-C}$.  The $O-C$ diagram is a plot of the $O-C$ value as a function of time.  If the fiducial model is a good representation of the observed times, then the measured points will fall near to a flat line equal to zero.  If the true orbital period is somewhat different from that in the fiducial model, then the data points will follow a sloped line in the $O-C$ diagram.  The slope in the $O-C$ diagram is the orbital period $P$ at that time.  If the binary's period is changing steadily with some $\dot{P}$, then the $O-C$ data points will follow a parabola.  If the binary undergoes a sharp period change, say, as the result of some nova eruption, then the $O-C$ curve will show a sharp kink, with a sudden change of slope.  For details and full worked examples of my photometry from archival plates in the exact setting for this current paper, see Schaefer (2020a; 2020b) and Schaefer \& Patterson (1983).  

For each binary, I have either measured my own times or collected them from the literature.  My measured times come from a variety of data sources.  The oldest light curves uniquely come from the Harvard collection of archival plates, and I get data going back to 1889, with these magnitudes being valuable for providing the longest possible range for $O-C$ curves.  For two of my targets, I have made use of the large-scale program Digital Access to a Sky Century $@$ Harvard (DASCH)\footnote{\url{http://dasch.rc.fas.harvard.edu/lightcurve.php}}, with J. Grindlay (Harvard) as Principle Investigator, which has scanned and digitized the majority of Harvard plates, and used those scans to run a sophisticated program to calculate the $B$ magnitudes of all stars on the plates (Tang et al. 2013).  For seven targets, I have used the wonderful light curves (with 120--1800 s time resolution for all of $\sim$25 days) from the {\it Transiting Exoplanet Survey Satellite} ({\it TESS}), see Ricker et al. (2015)\footnote{\url{https://mast.stsci.edu/portal/Mashup/Clients/Mast/Portal.html}}.   Another set of light curves are those collected in the International Database\footnote{\url{https://www.aavso.org/data-download}} of the American Association of Variable Star Observers (AAVSO), with these having the advantage of many long runs covering many decades.  Another source of light curves is the Zwicky Transient Factory\footnote{\url{https://irsa.ipac.caltech.edu/cgi-bin/Gator/nph-scan?projshort=ZTF}} (ZTF), with the coverage being many hundred of magnitudes 2019--2022 for most stars brighter than 20.5 mag and north of $-$29$\degr$ declination (Bellm et al. 2019).  The All Sky Automated Survey\footnote{\url{http://www.astrouw.edu.pl/asas/?page=aasc}} (ASAS) covers the entire sky down to $\sim$14th mag, typically with hundreds of magnitudes 2000--2009 (Pojmanski 1997).  The Panoramic Survey Telescope and Rapid Response System (Pan-STARRS)\footnote{\url{https://catalogs.mast.stsci.edu/panstarrs/ }} is a sky survey covering 2010-2014, down to $g$=22.0 mag, north of $-$30$\degr$ declination, in five filters ({\it grizy}) from a 1.8-m telescope at Haleakala Observatory (Chambers et al. 2016).

With each light curve, I then calculate the phase marker (the times of mid-eclipse or the time of the photometric minimum light) with a chi-square fit to a template light curve.  These times are all collected to create standard $O-C$ curves.  Then the $O-C$ curves are fitted to a parabola with the form $E_0+NP+0.5P\dot{P}N^2$, with $\dot{P}$ dimensionless.  Details and methods are the same as in Schaefer (2023, 2024).  

In the following subsections, I will report the cases and measured period changes for 25 X-ray binaries with either a NS or BH as the compact object.  Several of these measures produce large uncertainties in $\dot{P}$, yet these will still be useful for constraining the AML.  

\begin{figure*}
	\includegraphics[width=2.0\columnwidth]{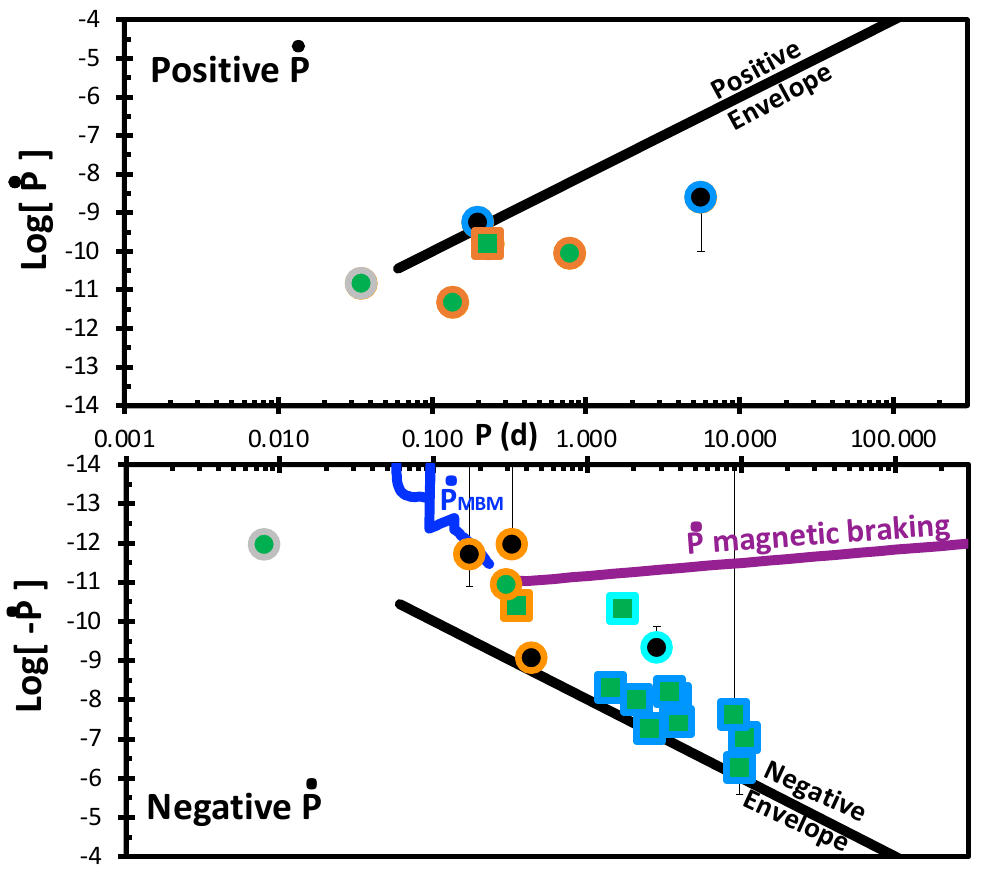}
    \caption{$\dot{P}$ versus $P$ for the XRBs.  The horizontal axis is the logarithm of $P$ in days, with this being the only way to cover the huge range for the binaries.  The vertical axis also has to cover a huge range of $\dot{P}$, so it must be logarithmic, but with negative and positive values, so I have divided the plot into the upper panel for positive-$\dot{P}$ and the lower panel for negative-$\dot{P}$.  This plot is just for the 23 XRBs that have a meaningful parabola in the $O-C$ curve, so it does not include V4580 Sgr or UY Vol.  The nature of the compact object in the binary is color coded by the core of the symbols, with black for BHs, green for NSs, and white for WDs.  The nature of the companion star is color coded by the `mantle' of the symbols, with red for the red giant star, orange for low-mass stars and sub-giants near the main sequence, blue for high-mass stars, cyan for intermediate-mass companions, and grey for white dwarfs.  The symbols are square-shaped for binaries where the primary star shows evidence of a dominating magnetic field, and circular for binaries with no visible evidence of a magnetic field.  The $\dot{P}$ values are mostly inside an envelope defined by $\dot{P} = \pm (P/10000)^2$, with $P$ measured in units of days, shown as the two black lines labeled `Positive Envelope' and `Negative Envelope'.  The required values from the revised version of the Magnetic Braking Model (Knigge et al. 2011) are displayed as the blue curve in the top-middle of the lower panel.  The revised-MBM does not extend its prediction to $P$$>$0.25 days, so I have used their equations to display the predicted effects of magnetic braking alone, as shown by the purple line.  The MBM was originally formulated to explain XRBs (Rappaport, Joss, \& Webbink 1982), so this purple line should be applicable for XRBs. }
\end{figure*}

The system properties and $\dot{P}$ measures are presented in Fig. 1 and Table 1, with ordering by increasing $P$.  The XRBs are named either by the official designation for optical variable stars from the {\it General Catalog of Variable Stars}\footnote{\url{http://www.sai.msu.su/gcvs/gcvs/}} (GCVS) or by one of the many designations from X-ray catalogs.  The third and fourth columns list the type of star for the primary and the secondary, with `MS' for low-mass main sequence stars, `Int-M' for intermediate mass main sequence stars, and `Hi-M' for high-mass main sequence stars.  The fifth column reports whether the primary star has any evidence for a significant magnetic field, such as pulses or cyclotron lines.  The sixth column points to the mode of the accretion flow, either by the companion's stellar wind or by Roche lobe overflow (RLOF).    The next column tells the type of evidence used to measure orbital phases, with `X' for X-ray data, `O' for optical data, as using eclipses (`ecl'), photometric minima (`min'), pulses (`pulse'), or radial velocity (`RV') data.  The next two columns give the year range for the $O-C$ data and the number of years that this spans ($\Delta Y$).  The tenth column quotes the orbital period $P$.  The last column collects my derived $\dot{P}$ measures in dimensionless units of $10^{-12}$.

\begin{table*}
	\centering
	\caption{Measured $\dot{P}$ for 25 time intervals in quiescence for XRBs}
	\begin{tabular}{lllllllrrrc}
		\hline
		GCVS   &   Name  &  Prim.   & Comp.  &    Mag?    &  Flow    &    $O-C$ data  &  $O-C$  &  $\Delta Y$  &  $P$  &  $\dot{P}$    \\
		 &  &   &   &    &   &    &  years  &  (yr)  &  (day)  &  ($10^{-12}$)   \\
		\hline	
...	&	4U 1820-30	&	NS	&	WD	&	No	&	RLOF	&	X-min	&	1976--2011	&	36	&	0.00793	&	$-$1.131$\pm$0.028	\\
V1405 Aql	&	4U 1916-053	&	NS	&	WD	&	No	&	RLOF	&	X-min	&	1978--2002	&	25	&	0.0347	&	14.6$\pm$0.3	\\
V4580 Sgr	&	SAX J1808.4-3658	&	NS	&	MS	&	Yes	&	RLOF	&	X-pulse	&	1998--2022	&	25	&	0.0839	&	`S' shaped	\\
...	&	XTE J1710-281	&	NS	&	MS	&	No	&	RLOF	&	X-ecl	&	2001--2017	&	17	&	0.137	&	4.7$\pm$0.3	\\
UY Vol	&	EXO 0748-676	&	NS	&	MS	&	No	&	RLOF	&	X-ecl	&	1985--2008	&	24	&	0.159	&	`W' shaped	\\
KV UMa	&	XTE J1118+480	&	BH	&	MS	&	No	&	RLOF	&	RV, O-min	&	2000--2024	&	25	&	0.170	&	$-$2$\pm$11	\\
V1521 Cyg	&	Cyg X-3	&	BH	&	Hi-M	&	No	&	Wind	&	X-min	&	1970--2018	&	49	&	0.199	&	562.9$\pm$0.2	\\
V691 CrA	&	4U 1822-371	&	NS	&	MS	&	Yes	&	RLOF	&	X-ecl, O-ecl	&	1977--2024	&	41	&	0.232	&	157$\pm$13	\\
V2134 Oph	&	MXB 1658-298	&	NS	&	MS	&	No	&	RLOF	&	X-ecl	&	1976--2019	&	44	&	0.296	&	$-$10.8$\pm$0.6	\\
V616 Mon	&	A0620-00	&	BH	&	MS	&	No	&	RLOF	&	RV, O-min	&	1983--2021	&	39	&	0.323	&	$-$1.1$\pm$1.3	\\
...	&	AX J1745.6-2901	&	NS	&	MS	&	Yes	&	RLOF	&	X-ecl	&	1994--2015	&	22	&	0.348	&	$-$40.3$\pm$3.2	\\
GU Mus	&	GS 1124-684	&	BH	&	MS	&	No	&	RLOF	&	RV	&	1992--2013	&	22	&	0.433	&	$-$890$\pm$170	\\
V818 Sco	&	Sco X-1 	&	NS	&	MS	&	No	&	RLOF	&	RV, O-min	&	1907--2020	&	114	&	0.787	&	87$\pm$51	\\
...	&	LMC X-4	&	NS	&	Hi-M	&	Yes	&	Wind	&	X-ecl	&	1976--2020	&	45	&	1.408	&	$-$4970$\pm$40	\\
HZ Her	&	Her X-1	&	NS	&	Int-M	&	Yes	&	RLOF	&	X-ecl, O-ecl	&	1894--2023	&	130	&	1.700	&	$-$48.5$\pm$1.3	\\
V779 Cen	&	Cen X-3	&	NS	&	Hi-M	&	Yes	&	RLOF	&	X-ecl	&	1971--2018	&	47	&	2.087	&	$-$10280$\pm$100	\\
...	&	M82 X-2	&	NS	&	Hi-M	&	Yes	&	RLOF	&	X-pulse	&	2014--2021	&	8	&	2.533	&	$-$56900$\pm$2400	\\
V4641 Sgr	&	XTE J1819-254	&	BH	&	Int-M	&	No	&	RLOF	&	O-min	&	1901--2022	&	122	&	2.817	&	$-$470$\pm$340	\\
V884 Sco	&	4U 1700-377	&	NS	&	Hi-M	&	Yes	&	Wind	&	X-ecl	&	1972--2013	&	42	&	3.412	&	$-$6400$\pm$2400	\\
QV Nor	&	4U 1538-522	&	NS	&	Hi-M	&	Yes	&	Wind	&	X-ecl	&	1972--2016	&	45	&	3.728	&	$-$9700$\pm$3800	\\
...	&	SMC X-1	&	NS	&	Hi-M	&	Yes	&	Wind	&	X-ecl	&	1971--2018	&	48	&	3.892	&	$-$37760$\pm$20	\\
V1357 Cyg	&	Cyg X-1	&	BH	&	Hi-M	&	No	&	Wind	&	RV, X-ecl	&	1939--2022	&	84	&	5.600	&	2500$\pm$2400	\\
GP Vel	&	Vela X-1 	&	NS	&	Hi-M	&	Yes	&	Wind	&	X-ecl	&	1972--2009	&	38	&	8.964	&	$-$24k$\pm$74k	\\
...	&	EXO 1722-363	&	NS	&	Hi-M	&	Yes	&	Wind	&	X-ecl	&	1998--2006	&	9	&	9.740	&	$-$560k$\pm$370k	\\
...	&	OAO 1657-415	&	NS	&	Hi-M	&	Yes	&	Wind	&	X-ecl	&	1991--2011	&	21	&	10.45	&	$-$97.3k$\pm$4.3k	\\

		\hline
	\end{tabular}
\end{table*}

\subsection{4U 1820-30}

4U 1820-30 is an X-ray burster that appears close to the centre of the globular cluster NGC 6624 in Sagittarius (Chou \& Jhang 2023 and references therein). As a burster, the primary must be a neutron star.  A roughly sinusoidal modulation with a period of 0.00793 days (685 seconds) is identified as the orbital period.  With such a short orbital period, the companion star can only be a low-mass helium WD.  Peuten et al. (2014) have measured and collected 34 times from eight X-ray satellites stretching from 1976--2011, and their derived $\dot{P}$ value is ($-$1.15$\pm$0.06)$\times$10$^{-12}$.  Chou \& Jhang (2023) added 6 timings from $NICER$ from 2017 to 2022, deriving a value of ($-$1.131$\pm$0.028)$\times$10$^{-12}$.  Figure 4 of Chou \& Jhang shows the nicely measured $O-C$ curve, displaying a nearly perfect concave-down parabola, so this plot is not repeated in this paper.  These $\dot{P}$ measures greatly disagree with the predictions of prior models, so workers have proposed seven wide-ranging speculations for unlikely explanations, all to attempt to save the old theory, and all with no positive evidence.

\subsection{V1405 Aql = 4U 1916-053}

V1405 Aql is a relatively steady source, discovered in the X-rays as far back as with the {\it Uhuru} satellite (Iaria et al. 2021 and references therein).  Its X-ray light curve shows X-ray bursts and also suffers periodic dips, which are tied to the orbital period.  The orbital period is short at 0.0347 days, which implies that the companion star is a low mass ($\sim$0.07 M$_{\odot}$) degenerate star.  The latest collection of dip times covers 41 years (from 1978 to 2018), with the $O-C$ curve showing a deep concave-up parabola, with $\dot{P}$=($+$1.46$\pm$0.03)$\times$10$^{-11}$ (Iaria et al. 2021).  The full $O-C$ curve is already displayed in Figure 2 of the Iaria paper, and I have nothing to add, so it is not duplicated here.  The shape of the $O-C$ curve is not a perfect parabola, as there are slow variations superposed on the parabola, with the amplitude at 10 per cent of the parabola's sagitta.  Such $O-C$ bumps are common amongst CVs (Schaefer 2024), always with small amplitudes, so apparently they are common for XRBs also.  The causes of these superposed variations are variously small zero-centered shifts associated with the ordinary movement of the accretion hot spot, timing jitter due to ordinary flickering shifting the apparent time of eclipse minima, and observational errors from inconsistent time conventions and poor definitions of eclipse times (Schaefer 2023).  For V1405 Aql, these bumps have been modeled as a sinewave with a period of 24.9 years, but such a model is dubious because the duration of the $O-C$ curve ($\Delta Y$) equals the claimed cycle so the poorly-measured sinewave cannot be of even small significance.  The presence or absence of the bumps make little difference in the $O-C$ curvature, so we are left with a highly-significant, large, and {\it positive} $\dot{P}$.


\subsection{V4580 Sgr = SAX J1808.4-3658}

V4580 Sgr is a recurring X-ray transient that is an accreting millisecond X-ray pulsar and an X-ray burster (Illiano et al. 2023 and references therein).  Timing analysis of the X-ray pulses gives an orbital period of 7249.156980 seconds (near 0.0839 days), which requires a very low mass companion star.  Illiano et al. (2023) collect 9 times of the ascending node of the orbit (as based on X-ray pulse timings) from each of 9 outbursts from 1998 to 2022.  Their full $O-C$ diagram is displayed nicely in their Figure 3, so it is not copied here.  They fitted their 9 data points to a parabola+sinewave model (with 6 fit parameters), then reported a $\dot{P}$ value of ($-$2.82$\pm$0.69)$\times$10$^{-13}$.  Even with their 6 fit parameters, their residuals are 3--6 sigma deviations for 5 out of their 9 measures.  Their best-fitting chi-square is 117.9 for 3 degrees of freedom, indicating that their model is horrible, so any derived quadratic term can have no confidence.  Their plotted $O-C$ curve shows that their sinewave period nearly equals their data interval, which means that most any $\dot{P}$ over a wide range can be derived for simple trade-offs from the parabola to the sinewave.  More to the point, the $O-C$ curve has no plausible parabola that can represent the general run for the evolution over the last 24 years.  In all, we have no useable $\dot{P}$ for V4580 Sgr.  

Still, we have a well-measured well-sampled $O-C$ curve that shows highly-significant structure that can be called an `S' shape.  Roughly, the XRB can be described as having nearly a constant $P$ that suffers two relatively-fast period changes, one around the year 2005 with $\Delta P$ of $+$0.0011 s, and the other around 2017 with $\Delta P$ of $-$0.0013 s.  This is a snapshot of the real evolution of V4580 Sgr.  Any parabolic term must be small in size, with $\vert$$\dot{P}$$\vert$$<$2$\times$10$^{-12}$ or so.  The S-shape means that the period changes are being controlled by some unknown mechanism that is dominating over any evolutionary parabola.  In the framework of standard theory and of the MBM, it means that modern models are missing the dominating mechanism for at least this one system.

\subsection{XTE J1710-281}

\begin{figure}
	\includegraphics[width=1.0\columnwidth]{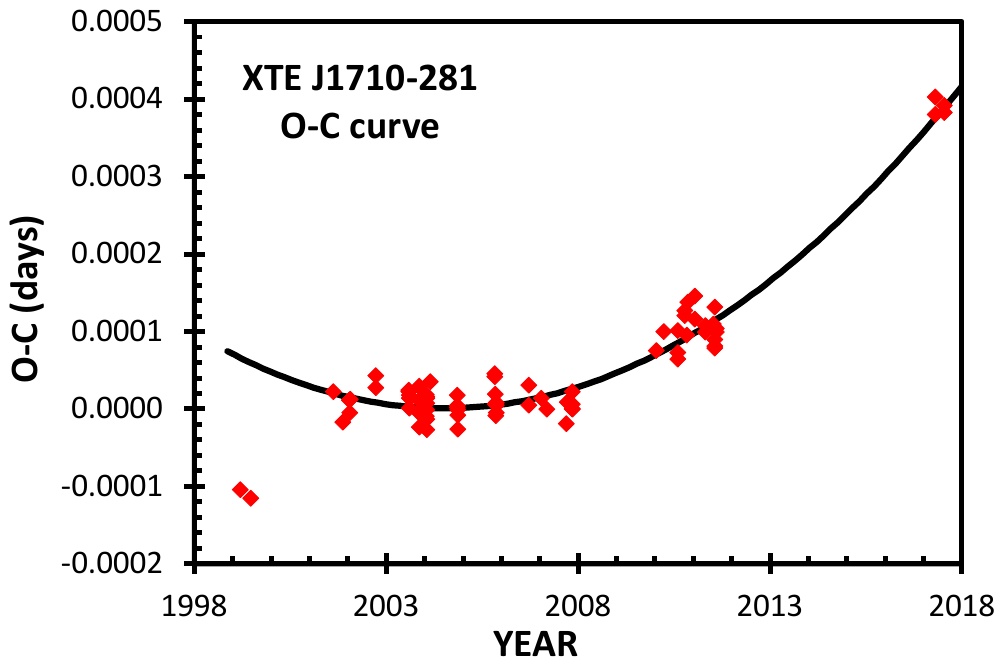}
    \caption{$O-C$ curve for XTE J1710-281.  This $O-C$ curve is constructed from the 78 times of mid-eclipse using data from five X-ray satellites over the years 1999 to 2017, as given by Jain, Sharma, \& Paul (2022).  The fiducial ephemeris for $O-C$ uses $P_{\rm O-C}$=0.1367109674 days and an epoch $E_{\rm O-C}$=2454410.541569.  The data after 2001.0 shows a nice concave-up parabola, with my best-fitting parabola shown as the black curve.  The two isolated times from 1999 are early (as compared to the long-term parabolic evolution) by 15 seconds, with this being a rather small example of bumps commonly seen in XRB and CV $O-C$ curves.}
\end{figure}

XTE J1710-281 is a highly variable X-ray source, discovered in 1998, that has dips, bursts, and eclipses (Jain, Sharma, \& Paul 2022 and references therein).  The orbital period is 0.137 days.  The eclipses are deep and sharp-edged, so their times are measured with high accuracy.  Jain, Sharma, \& Paul (2022) has collected 78 eclipse times from 1999--2017 from five X-ray satellites.  I have reproduced their $O-C$ plot in Figure 2.  They have idealized their resultant $O-C$ curve as consisting of four straight line segments, with three sudden period changes at the juncture of adjacent segments.  Each sudden period change is called a 'glitch', with no plausible physical mechanism, while the sharp period changes were not observed, so the existence of these idealized glitches is dubious.  Within this idealization, there is one well-observed segment, with 52 eclipse times from 2001.6 to 2007.7, that appears straight, and Jain et al. placed a limit  on $\dot{P}$ that is equivalent to ($-$9$\pm$10)$\times$10$^{-13}$.  

Any such limit on $\dot{P}$ for an isolated segment of time does not mean much for the evolution of the system, because it is the mysterious `glitches' that dominate the period change and determine the real long-term period evolution.  The effective $\dot{P}$ for evolution considerations is the overall shape.  This overall shape, from 2001 to 2017 is remarkably close to a simple parabola.  The alternative chosen by Jain et al. (breaking the $O-C$ curve into linear segments matching the segments of available data) does have a somewhat smaller chi-square than a simple parabola, but this is a necessary consequence of adding free fit parameters (6 fit parameters for the post-2001 $O-C$ broken into three straight lines, versus 3 parameters for the parabola), so the difference in chi-square is not significant.  In this case, the simple and expected description of the $O-C$ curve as a parabola is to be preferred over invoking a mysterious and unprecedented physical mechanism.

I have used the Jain et al. post-2001 eclipse times to fit to a simple parabola in the $O-C$ curve, as shown in Figure 2.  The best-fitting $\dot{P}$ is ($+$4.7$\pm$0.3)$\times$10$^{-12}$.  The residuals from this fit have the same RMS scatter as for the straight-segment fit, and there are no systematic trends or excursions as a function of the year.  This parabola is the best empirical representation of the real period change over long time-scales.  

In all cases, the two isolated eclipse times from 1999 are far from any extrapolation from later data.  I can think of three reasonable interpretations:  {\bf (1)} This might be from some real glitch, where the orbital period suddenly decreased at the level of $\Delta P$/$P$ equal to $-$0.13 parts per million around the year 2000.  For comparison, this sudden period change is over one order-of-magnitude smaller in size than has ever been measured in any nova or recurrent nova system across the nova event, and this size is roughly three orders-of-magnitude smaller than the typical nova $\Delta P$ (Schaefer 2023a).  {\bf (2)} As a possibility with much precedent, the 1999 $O-C$ measures might simply be part of a `bump' (common for XRBs and CVs) superposed on the evolutionary parabola.  Schaefer (2024) gives many examples of up-bumps and down-bumps, plus some discussion.  Such bumps are of unknown origin, and appear to have zero averages over time.  For the case of XTE J1710-281 in 1999, the downward bump would only be 15 seconds in amplitude.  {\bf (3)} Another possibility is that the first two isolated points have some sort of a normal measurement error.  The first two eclipse times were taken from different satellites than all the rest of the data, raising the possibility that these light curves were handled differently, resulting in some inconsistent offset.  In all three possibilities, the system's normal evolution is represented by my best-fitting $\dot{P}$.

\subsection{UY Vol = EXO 0748-676}

UY Vol is an X-ray transient that was in a high accretion state from $\sim$1982 to 2008, and it exhibits X-ray bursts, irregular dips, and eclipses (Mikles \& Hynes 2012 and references therein).  The eclipses show a periodicity of 0.159 days, and the companion star is an M2 main sequence star.  The deep eclipses have durations of 500 seconds, with ingress and egress durations of near 6 seconds, so the times of mid-eclipse have a typical uncertainty of 0.3 seconds.  Wolff et al. (2009) collected 443 X-ray eclipse timings from 1985 to 2008, with this being entirely during the high state.  Their full $O-C$ curve is in their Figure 3, and I have nothing to add, so their diagram is not copied here.  Their $O-C$ curve can be approximated with four straight-line segments, each 5--10 years in duration.  The period changes at the juncture of the segments are relatively sudden, lasting less than a year.  The three sharp period changes alternate positive (i.e., with a sudden period increase), negative, and positive, with all three having fractional period changes near 0.5 parts-per-million in size.  Superposed on these nearly straight line segments are many highly significant variations with time-scales ranging from a few days up to one year.  All of these variations have no plausible physical mechanism.  These variations are far from any idealized steady period change.  No parabola in the $O-C$ curve can provide a reasonable representation, and I have no useable $\dot{P}$ measure to add.  A limit on the parabolic contribution to the $O-C$ curve is $\vert$$\dot{P}$$\vert$$<$1$\times$10$^{-11}$ or so.

The sharp kinks in the $O-C$ curve are in 1990.6, 2000.2, and 2005.0.  These times do {\it not} correspond to times of changing $\dot{M}$, as can be seen from the X-ray luminosity history plot in figure 1 of Degenaar et al. (2011).  So the observed period changes cannot be ascribed to changes in the accretion rate.

The magnetic braking mechanism operates on the spin circularization timescale for the companion star, which is many centuries.  So the magnetic braking model cannot explain the three sharp kinks in the observed $O-C$ curve.  Similarly, for OY Car, Z Cha, V4580 Sgr, and SW Sex, the magnetic braking mechanism must be greatly smaller than some other unknown mechanism that dominates the evolution of the systems.

UY Vol provides proof that ordinary X-ray binaries are not all represented by a single evolutionary $\dot{P}$, albeit with occasional small perturbations superposed.  That is, any evolutionary parabola is greatly smaller than the large observed variations in the $O-C$ curve.  UY Vol provides proof that at least one system has its period changes dominated by some physical mechanism beyond the steady evolutionary effects.  The UY Vol $O-C$ curve provides a challenge to come up with a mechanism for which detailed physics calculations agree with the observed complex shape.

\subsection{KV UMa = XTE J1118+480}

\begin{figure}
	\includegraphics[width=1.0\columnwidth]{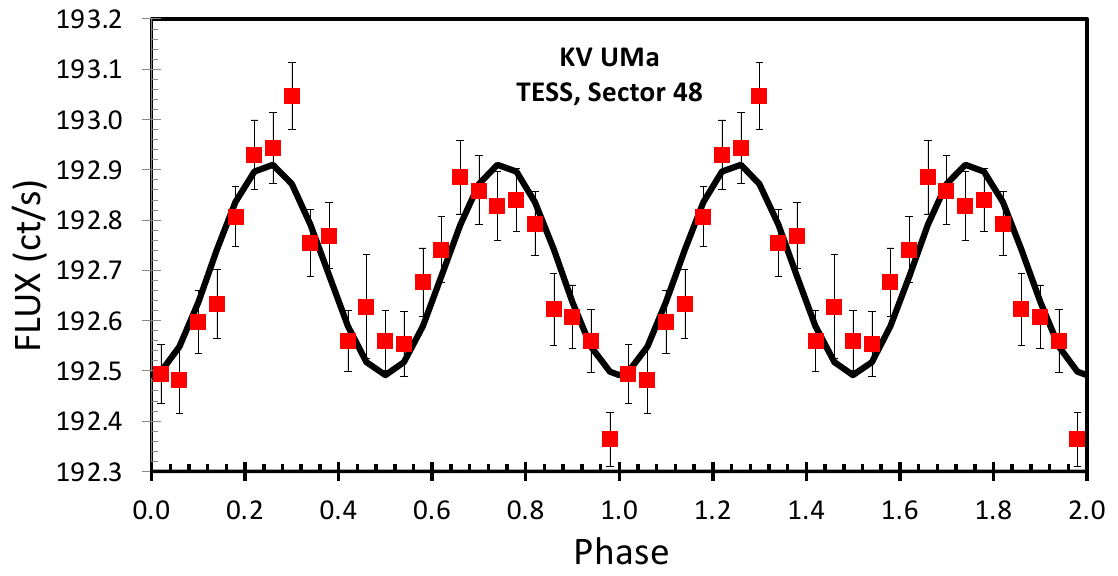}
    \caption{{\it TESS} folded light curve for Sector 48 for KV UMa (XTE J1118+480).  The phase folding is calculated with period of 0.16993394 days and zero phase at BJD 2459623.1474.  The 3069 individual fluxes have an RMS scatter (at a given phase) near 0.7 ct/s.  The individual points are phase averaged into bins 0.04 wide in phase, as shown by the red squares.  With 123 fluxes in each phase bin, the uncertainty in the averages is 0.063 ct/s.  This folded light curve shows that KV UMa has a small-amplitude ellipsoidal modulation.  The full unbinned light curve was modeled as a sinewave, as shown by the thick black curve.  The point of this light curve is to measure the time of the photometric minimum to be BJD 2459623.1474$\pm$0.0014.}
\end{figure}

\begin{figure}
	\includegraphics[width=1.0\columnwidth]{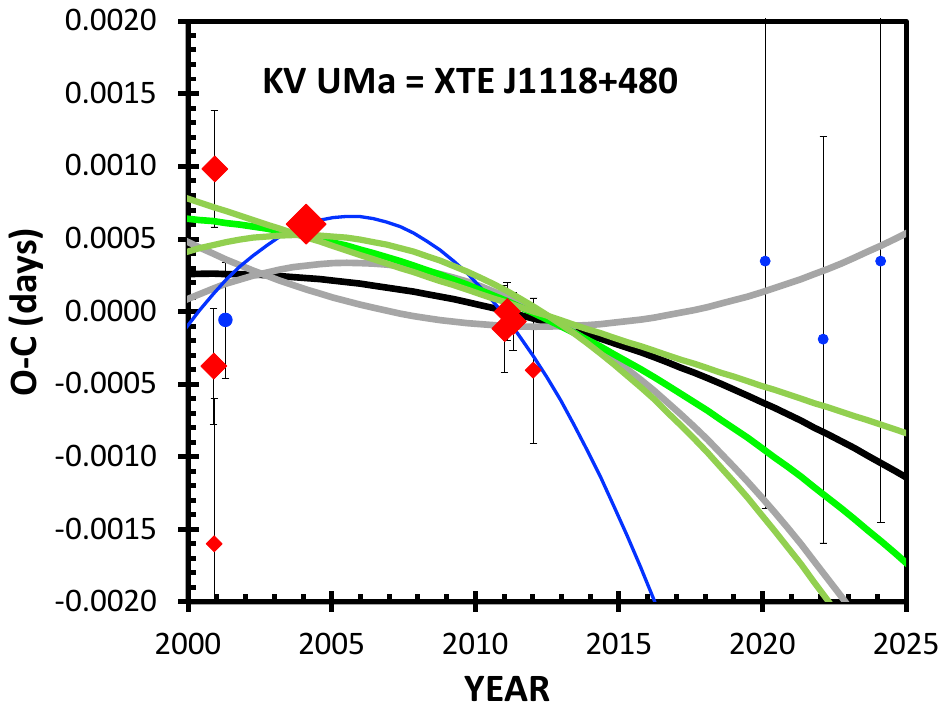}
    \caption{$O-C$ curve for KV UMa (XTE J1118+480).  This $O-C$ curve is constructed from the 12 times of inferior conjunction measured from radial velocity curves (red diamonds) and photometric minimum times (blue circles).  The $O-C$ is calculated with $P_{\rm O-C}$=0.16993394 days and $E_{\rm O-C}$=2455600.6413.  The size of the symbols is indicative of the weight of the datum, with the smallest points having the largest error bars.  The data from the three epochs around 2001, 2004, and 2011 were fitted by Gonz\'{a}lez Hern\'{a}ndez et al. (2014) to get the concave-down blue parabola.  But this parabola disagrees with the three {\it TESS} point from 2020--2024 by up to 11 minutes, so the real behavior apparently has a much flatter $O-C$ curve.  A straight chi-square fit to all 12 times returns the best fit as the green parabola, flanked by the grey-green curves for the 1-$\sigma$ range.  This fit has a large chi-square, indicating that some or all of the times have a systematic or intrinsic variance that needs to be added in.  With a 0.00027 day systematic error, the best fitting parabola is the thick black curve, flanked by the thick grey curves for the 1-$\sigma$ range.  We see that the $\dot{P}$ is consistent with zero, with both positive and negative curvature being allowed.  Taking the black and grey curves as the best representation, we have $\dot{P}$=($-$2$\pm$11))$\times$10$^{-12}$.  The uncertainty of nearly $\pm$10$^{-11}$ is actually good and useful for placing KV UMa in Figure 1. }
\end{figure}

KV UMa is a BH with a K5 main sequence star in a 0.1699 day orbit (Gonz\'{a}lez Hern\'{a}ndez et al. 2014 and references therein).  This X-ray transient source was discovered in outburst in the year 2000 (Hynes et al. 2000), and this was followed in the optical as a flat-topped light curve peaking at $V$=12.7 lasting for 110 days.  A second eruption was recorded in 2005 (Zurita et al. 2006), with the optical light peaking at $V$=13.5 for a fairly flat-topped light curve lasting near one month.  A superhump periodicity of 0.156 days was seen during the eruptions, and this is distinct from the coherent ellipsoidal modulations seen in quiescence.  Quiescence is near 19th magnitude, during which normal ellipsoidal modulations (nearly a sinewave at half the orbital period, where the secondary minimum is not quite as deep as the primary minimum) are seen with amplitude of roughly a tenth of a magnitude (McClintock et al. 2001).

Johannsen (2009) and Gonz\'{a}lez Hern\'{a}ndez et al. (2014) have collected 8 times of inferior conjunction of the companion star, mostly from published radial velocity (RV) curves.  However, one of their measures of inferior conjunction, in 2001.3, is derived from the time of photometric minimum in the light curve.  They have created $O-C$ curves and fitted parabolas so as to derive $\dot{P}$=($-$6.01$\pm$1.81)$\times$10$^{-11}$ for KV UMa.  This $O-C$ curve and their parabola is shown in Figure 4.  To this, I have added the time of inferior conjunction from McClintock et al. (2001).

I have added three times of inferior conjunction taken from {\it TESS} light curve from Sectors 21, 48, and 75 (in early 2020, 2022, and 2024 respectively).  The integration times are 120, 600, and 200 seconds, respectively, of nearly gap-free data over 27 days.  The folded light curve for Sector 48 is shown in Figure 3.  KV UMa displays a roughly sinusoidal modulation with half the orbital period, so we are seeing ordinary ellipsoidal modulation.  For the case of KVUMa, the amplitude of the ellipsoidal effects is 0.4 ct/s, while the RMS scatter of individual fluxes is 0.7 ct/s, so there is substantial scatter even in the phase-averaged light curve.  To measure the time of photometric minimum, I have used a chi-square formalism for fitting the full unbinned 27-days light curve to a sinewave.  The epoch of minimum was taken from near the center of the observing time interval.  The best fitting sinewave has its epoch of minimum for the value with the minimum $\chi^2$, and the 1-$\sigma$ range is that over which the $\chi^2$ is within 1.000 of the minimum.  These new times are BJD 2458884.2752$\pm$0.0017, 2459623.1474$\pm$0.0014, and 2460354.0333$\pm$0.0018.  

The three new photometric $O-C$ values from {\it TESS} lie far above the best-fitting parabola from the earlier works (see Figure 4) by up to 11 minutes.  These three 4-$\sigma$ deviations are far outside measurement errors.  A possible explanation is that the RV times and the photometric minimum times might have systematic offsets, perhaps because asymmetric brightness from the hot spot is slightly shifting the minimum from the ellipsoidal effect.  Correcting for such effects could shift the {\it TESS} times up or down by possibly large amounts in the $O-C$ diagram.  But any such offsets are apparently small, because the photometric minimum time from 2001.3 fits in nicely with the contemporaneous RV times.  So we have to take the {\it TESS} times at face value.  The three {\it TESS} times are consistent within their relatively large error bars, making for a good case that these are not from some long-lasting `bump' in the $O-C$ curve.  So the three {\it TESS} times make a good case that the $O-C$ curve is nearly flat.

So I have 12 timings of the inferior conjunction, as placed into the $O-C$ curve in Figure 4.  I have fitted these to a parabola, where the quoted measurement error bars are taken at face value.  The best-fitting parabola is shown as the green curve, with the 1-$\sigma$ range shown by the flanking grey-green curves.  The formal value is $\dot{P}$=($-$8$\pm$10)$\times$10$^{-12}$.  A substantial problem with this fit is that the $\chi$$^2$ equals 19.9 for 9 degree of freedom.  This means that the data has a larger variance than given by the measurement error bars.  The high $\chi$$^2$ can only come from unrecognized systematic errors applicable to any or all of the points, or from intrinsic variability in KV UMa that is applicable to all data points.  A second reason to see substantial systematic errors is that the 2000--2001 times show a scatter that is greatly larger than the error bars and cannot arise from the star.  For either reason, the individual $O-C$ values are likely to have a real uncertainty larger than given by the formal error bars.  In particular, the 2004.1 datum has a startlingly small reported error bar (at $\pm$6 seconds), with this claimed accuracy forcing all the fits to pass through this one point. 

The $\chi$$^2$ for the fit implies that the real error bars are substantially larger than quoted.  A reasonable case would be that there is some additional systematic or intrinsic uncertainty that is added to many or all of the 12 times.  The can be modeled by adding in some constant variance applied equally to all points.  The size of this constant variance is chosen so that the best fit has a reduced chi-square near unity.  In part, this standard procedure will make the weight of the 2004.1 datum smaller, so that it does not dominate the fit due to its unreasonably-small error bar.  With a systematic error of 0.00027 days added in quadrature, I get the best fit with a reduced chi-square near unity.  For this case, the best fit parabola is shown as the black curve in Figure 4.  The two gray flanking curves show the 1-$\sigma$ range.  For this, $\dot{P}$=($-$2$\pm$11))$\times$10$^{-12}$.  I judge this to be the best representation of the timing data in hand.

So we have a case for a near-zero $\dot{P}$, with the uncertainties putting a limit like $|\dot{P}|$$<$10$^{-11}$.  (Nevertheless, the $\dot{P}$ value from Gonz\'{a}lez Hern\'{a}ndez et al. (2014) is far outside the acceptable range, and their quoted error bars are greatly too small.)  To have an accuracy of $\pm$10$^{-11}$ is actually pretty good and useful (c.f. Figure 1).  At this point, the only way to improve the $\dot{P}$ is to get current X-ray eclipse timings in 2025.  

While searching for further light curves that might provide timings for an $O-C$ curve, I examined the DASCH light curve.  I found four prior eruptions.  These had been previously discovered by J. Grindlay (Harvard), and reported at a conference.  {\bf (1)} The first eruption was in 1927, with the best and brightest appearance on Harvard plate AY 1457 at $B$=12.64 at 1927.335.  The eruption was visible on five plates from 1927.315 to 1927.406.  {\bf (2)} The best observed eruption is in the year 1928, with good detections on 9 plates.  This event was first seen on 1928.064 at $B$=13.46, came to an observed peak at $B$=12.67 on 1928.121 (plate RH 62), and then smoothly faded until its last visibility on 1928.285 at $B$=13.63.  {\bf (3)} The third eruption was seen on only one plate (RH 6343) at $B$=13.54 in 1934.946, just at the end of the 1934 solar gap as the area was reappearing in the dawn sky, with all later plates showing no evidence of any source.  The image on plate RH 6343 is good and at the right position, so I am sure that this plate records the tail end of an eruption that peaked around 1934.9.  {\bf (4)} The Harvard plate DNB 366 shows a highly significant star image at the position of KV UMa at $B$=14.71 in the year 1971.228.  Unfortunately, no other plates where taken at a time and depth such that this eruption can be tested or confirmed, so only this one magnitude is available and there are no useful upper limits.

\subsection{V1521 Cyg = Cyg X-3}

Cyg X-3 (Antokhin \& Cherepashchuk 2019 and references therein) is one of the first discovered X-ray sources, because it is so bright.  However, in the optical band, the very high extinction makes the counterpart invisible at $V$$>$23.  In the infrared, the counterpart is identified, and the spectral type is of a rare Wolf-Rayet star, of class WN4/7, with a mass of something near 10 M$_{\odot}$.  In the X-ray band, a stable periodic modulation at $P$=0.1997 days is seen, with no eclipses.  The compact object shows no pulses, bursts, cyclotron features, or QPOs, so a NS is not indicated.  The X-ray spectra and spectral states show ``a general resemblance to those of the canonical spectral states of BH binaries'' (Zdziarski, Miko{\l}ajewska, \& Belczynski 2013), so the compact object is generally taken to be a BH.

The times of X-ray minima are a measure of the orbital phase, so their variations in an $O-C$ curve can be used to plot out the period changes.  The latest data set and analysis is by Antokhin \& Cherepashchuk (2019), and they have 270 times from 1970--2018.  Their Figure 2 shows the full O-C curve, which will not be duplicated here.  This $O-C$ curve is a well-measured parabola with concave-up.  As is often the case, the first few times show large scatter and large deviations from any parabolic fit, suggestive that the pioneering satellite has unrecognized systematic problems.  With or without these problematic data, the $\dot{P}$ value for Cyg X-3 is ($+$5.61$\pm$0.02)$\times$10$^{-10}$.

The residuals from this basic parabola are small, roughly 2 per cent of the sagitta of the parabola.  These residuals can be poorly described by a sinewave with a cycle of 15.79 years, where the RMS scatter is comparable to the claimed amplitude.  The sinewave speculation is only a response to two small bumps in the $O-C$ curve centered on the years 2000 and 2015, with such bumps being ubiquitous for CVs and XRBs.  And for the particular case of Cyg X-3, the origin of bumps in the $O-C$ curve were tracked down to the random phasing of the $RXTE$ observations.  This illustrates the too-common case where workers use insignificant bumps of dubious reality to propose periodicities that are large fractions of the observing interval, and then apply such to imaginative physical mechanisms.  The urge is strong to fit noise in the residuals to a sinewave.  The inclusion or exclusion of the sinewave in the $O-C$ makes only an insignificant difference in the derived $\dot{P}$.

\subsection{V691 CrA = 4U 1822-371}

\begin{figure}
	\includegraphics[width=1.0\columnwidth]{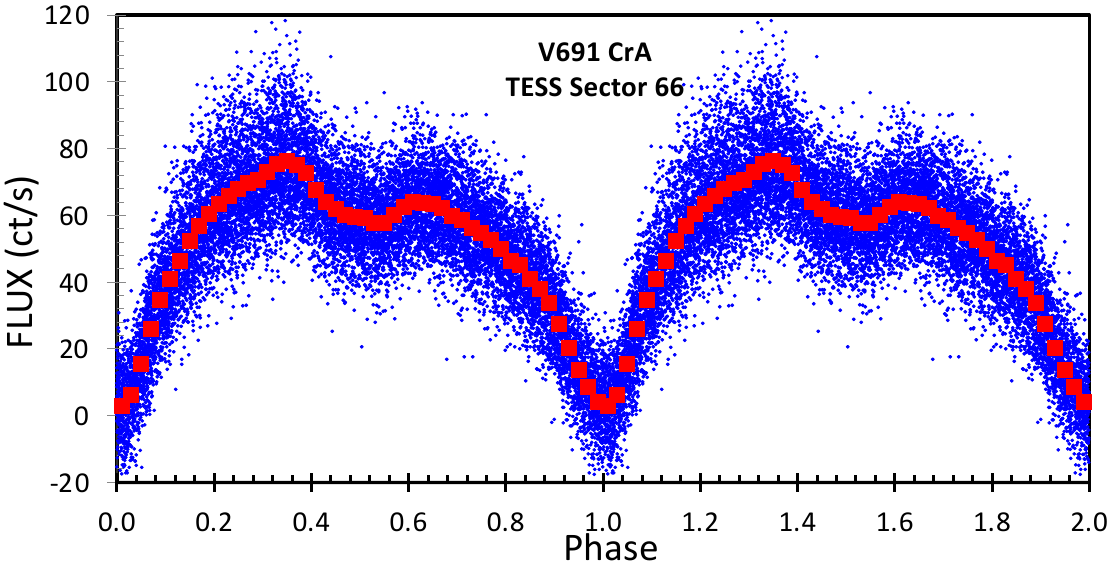}
    \caption{TESS curve for Sector 66 for V691 CrA (4U 1822-371). This light curve has been folded, with the phase calculated for a period of 0.2332109571 days and an epoch of 2460110.11180. The 15493 small blue dots are each for one flux measure for a 120 second integration. The red squares are phase-averages over bins 0.020 wide in phase.}
\end{figure}

\begin{figure}
	\includegraphics[width=1.0\columnwidth]{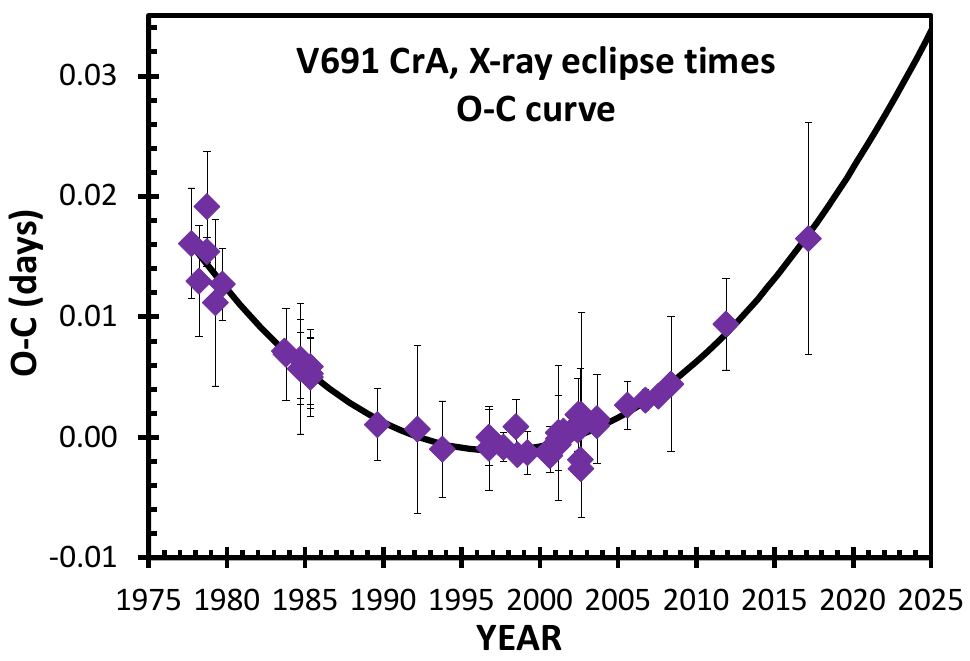}
	\includegraphics[width=1.0\columnwidth]{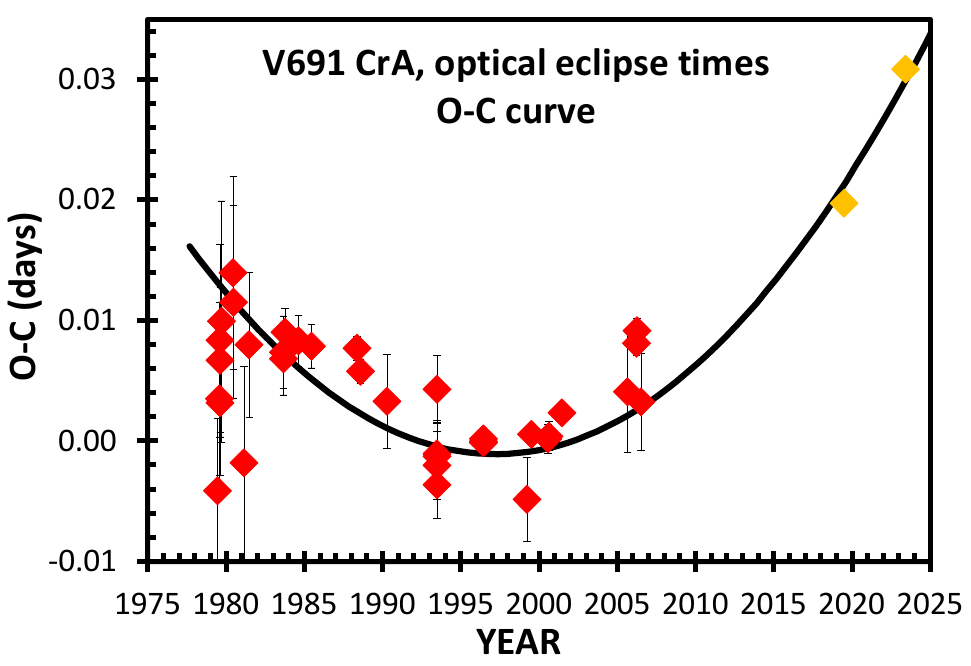}
    \caption{$O-C$ curve for V691 CrA (4U 1822-371).  The top panel is for the 48 X-ray eclipse times from Jain et al. (2010) and Mazzola et al. (2019).  The bottom panel shows the 35 optical eclipse times from Bayless et al. (2010) shown as red diamonds, plus the two times from {\it TESS} shown as orange diamonds.  I used $P_{\rm O-C}$=0.2332109571 days and $E_{\rm O-C}$=2450353.58728.  The {\it TESS} eclipse times are averages over 93 eclipses, resulting in an accuracy of $\pm$7 seconds.  The best fitting parabola (the thick black curves) is derived by a chi-square fit to both the optical and X-ray eclipse times.  With this, I have $\dot{P}$=($+$1.57$\pm$0.13)$\times$10$^{-10}$. }
\end{figure}

V691 CrA is an LMXB with an accretion powered X-ray pulsar and an accretion disc corona (Xing \& Li 2019 and references therein).  This system also shows narrow X-ray eclipses, which reveal an orbital period of 0.232 days.  Jain et al. (2010) collected 16 X-ray eclipse times from 1998--2007, finding $\dot{P}$ of ($+$1.3$\pm$0.3)$\times$10$^{-10}$.  Bayless et al. (2010) collected 35 optical eclipse times from 1979--2006, with $\dot{P}$ equal to ($+$2.12$\pm$0.18)$\times$10$^{-10}$.  Mazzola et al. (2019) collected 32 X-ray eclipse times from 1977--2017, and used these to derive ($+$1.51$\pm$0.07)$\times$10$^{-10}$ for $\dot{P}$.  These 83 eclipse times from these three sources are all for different eclipses.  Figure 6 shows the $O-C$ curve for the X-ray eclipse times (top panel) and the optical eclipse times (bottom panel).  The X-ray times have their quoted error bars substantially larger than the scatter about the best-fitting parabola, indicating that these uncertainties have been systematically over-estimated.

{\it TESS} has great coverage of many good eclipses for two time intervals; Sector 13 in mid-2019 and Sector 66 in mid-2024.  Both segments are 25 days of nearly gap-free photometry with 120 second time resolution.  The folded light curve for Sector 66 is shown in Figure 5.  With one gap in the middle, this light curve covers 93 eclipses.  We see an asymmetric eclipsing light curve, with broad primary and secondary eclipses.  I have fitted these light curves for the times of primary minima.  Averaging over all eclipses in each Sector, I have taken one eclipse time near the middle of the interval as the time for use in the $O-C$ curve.  Thus, I am adding minima times of BJD 2458670.09293$\pm$0.00008 for Sector 13, and BJD 2460110.11180$\pm$0.00011 for Sector 66.  The uncertainty is 7 seconds, with this being one order-of-magnitude better than all other measures.  This high accuracy is possible because each {\it TESS} Sector covers 93 eclipses.  These {\it TESS} times fall nicely on the parabola derived from all the other times.  

The best measure of the steady period change comes from using all 85 eclipse times with full coverage from 1977--2024.  For this, I find no significant systematic difference or offset between the X-ray and optical eclipse times, so all 85 eclipse times can be used in a single chi-square fit for a parabola in the $O-C$ curve.  With this, $\dot{P}$=($+$1.57$\pm$0.13)$\times$10$^{-10}$.

\subsection{V2134 Oph = MXB 1658-298}

\begin{figure}
	\includegraphics[width=1.0\columnwidth]{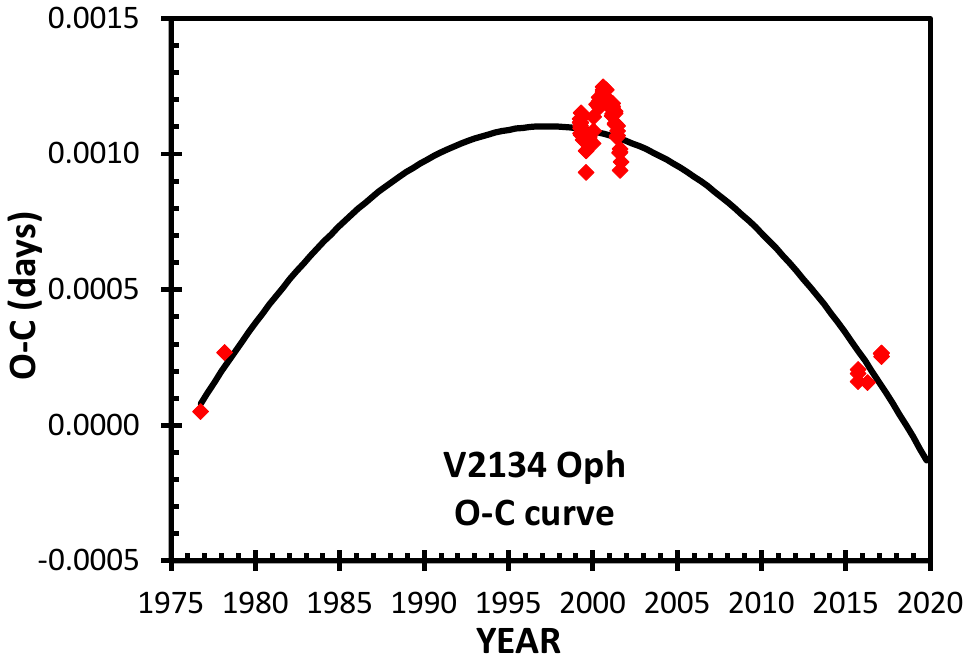}
    \caption{$O-C$ curve for V2134 Oph (MXB 1658-298).  This $O-C$ is calculated with the fiducial ephemeris of $P_{\rm O-C}$=0.29650453 days and $E_{\rm O-C}$=2443059.22595.  The X-ray eclipse times were only observable during outbursts starting in 1976, 1999, and 2015.  With only three widely separated outbursts, we can measure the parabolic term.  The best fitting parabola (the thick black curves) is concave-down, with a bump lasting one year around 2000.6.  With this, I have $\dot{P}$=($+$1.08$\pm$0.06)$\times$10$^{-11}$. }
\end{figure}

V2134 Oph is a low-mass X-ray binary, discovered in 1976, as an X-ray Burster (Iaria et al. 2018 and references therein).  The system is a transient, with outbursts lasting two years or so, starting in 1976, 1999, and 2015, and going invisible between outbursts.  Deep eclipses of duration 15-minutes prove the orbital period to be 0.2965 days, and that the companion star must be a low mass main sequence star.

Wachter, Smale, \& Bailyn (2000), Jain et al. (2017), and Iaria et al. (2018) have independently collected 6, 35, and 59 eclipse times from various X-ray satellites, with substantial duplication for individual data sets.  All three groups have recognized and measured the large period changes.  The residuals to the parabola have a small relative maximum around the year 2000.6.  This single maximum was used in an ill-advised attempt to fit a sinewave plus a cubic function.  One small bump does not make a sinewave, and a cubic term is meaningless when only three eruptions were observed.  Rather, a good description of the $O-C$ curve is just a parabola with a single small bump with amplitude of 10 seconds.  Such bumps are common for XRBs and CVs.

I have collected all 63 eclipse times from 1976--2019, with this largely being the list in Iaria et al. (2018).   These are displayed in an $O-C$ diagram in Figure 7.  These consist of 2 times during it discovery eruption (1976.8 and 1978.2), 54 times from 1999.3--2001.7, plus 7 eclipses from 2015.7 to 2017.1.  With data from these three short intervals, we have the minimum to measure a parabola.  I find $\dot{P}$ equal to ($-$1.08$\pm$0.06)$\times$10$^{-11}$.

\subsection{V616 Mon = A0620-00}

\begin{figure}
	\includegraphics[width=1.0\columnwidth]{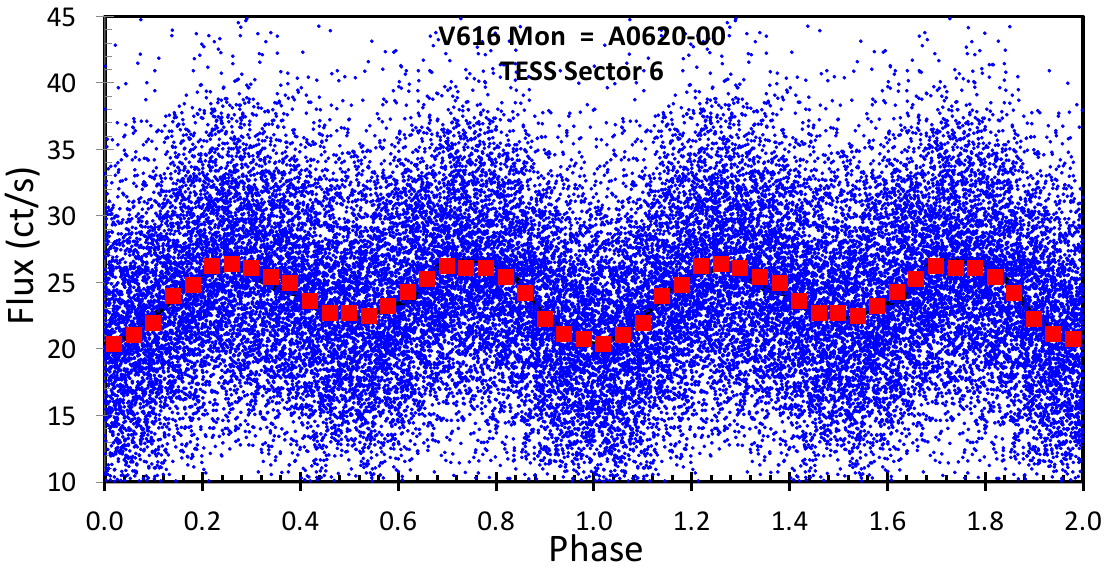}
    \caption{{\it TESS} folded light curve for Sector 6 for V616 Mon (A0620-00).  These data have 120 second time resolution, spread over a nearly-gap-free 25.8 days centered on 2019.0.  The phase folding is with zero phase at BJD 2459215.2884 and period of 0.32301415 days.  The 17458 individual fluxes (represented by the small blue diamonds) have an RMS scatter (at a given phase) of 6.1 ct/s, and this equals the average photometric error bar for an individual flux measure, so the observed scatter is entirely due to ordinary Poisson noise.  The individual points are averaged into 25 bins in phase, as shown by the red squares.  With 698 fluxes in each phase bin, the uncertainty in the averages is 0.23 ct/s, and is smaller than the red square symbols.  This folded light curve shows that V616 Mon has a typical small-amplitude ellipsoidal modulation, with a primary minimum and a shallower secondary minimum.  The full unbinned light curve was modeled by an appropriate template, as shown by the thick black curve barely visible under the red squares.  The point of this light curve is to measure the time of the photometric minimum to be BJD 2459215.2884$\pm$0.0002.}
\end{figure}

\begin{figure}
	\includegraphics[width=1.0\columnwidth]{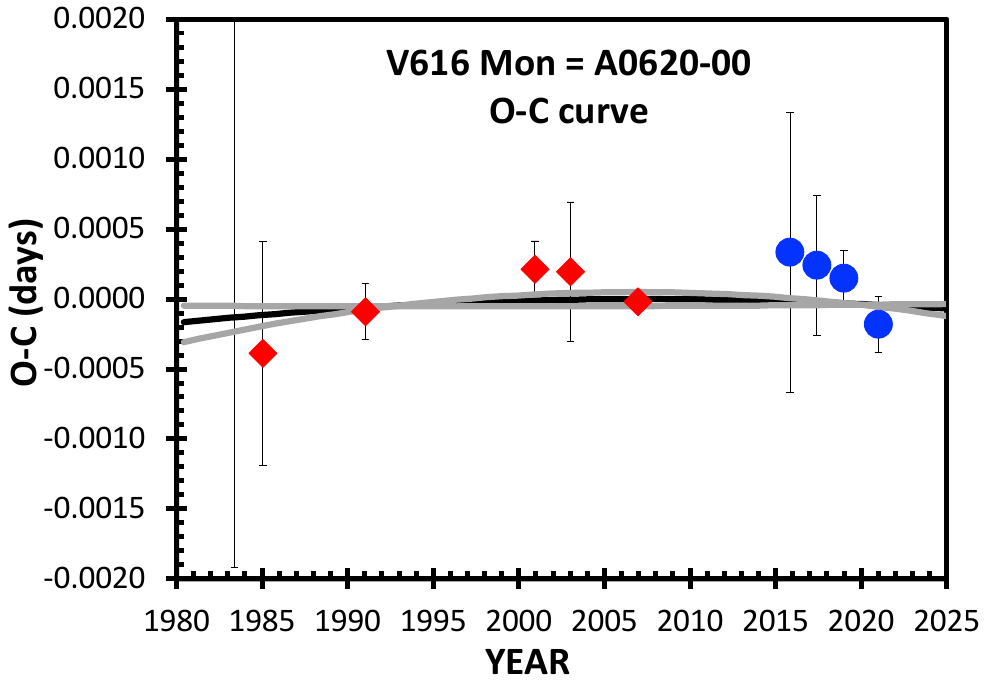}
    \caption{$O-C$ curve for V616 Mon (A0620-00) from 1983--2021.  This $O-C$ curve is constructed from the 10 times of inferior conjunction measured from radial velocity curves (red diamonds) and photometric minimum times (blue circles).  The earliest photometric minimum is only seen as an error bar extending from top to bottom.  The $O-C$ is calculated with a fiducial ephemeris of $P_{\rm O-C}$=0.323014053 days and $E_{\rm O-C}$=2454084.69487.  A straight chi-square fit to all 10 times returns the best fit as the thick black parabola, flanked by the thick grey curves for the 1-$\sigma$ range.  Taking the black and grey curves as the best representation, we have $\dot{P}$=($-$1.1$\pm$1.3)$\times$10$^{-12}$, which is consistent with zero.  The possible curvature in the $O-C$ diagram is actually over a very small range. }
\end{figure}

V616 Mon is the prototype X-ray nova, with a well observed eruption in 1975, at peak being the brightest X-ray source in the sky (Orosz et al. 1994, Gonz\'{a}lez Hern\'{a}ndez et al. 2014, and references therein).  Another eruption was discovered on the Harvard plates for the year 1917.  V616 Mon was the first of the BH binaries that had a widely accepted proof that the primary was indeed a black hole, as based on a RV curve yielding a mass function of 3.18$\pm$0.16 M$_{\odot}$ (McClintock \& Remillard 1986).  The companion is a K4 main sequence star in a 0.323 day orbit.  In quiescence, the system displays the usual ellipsoidal modulation, such that the secondary minimum is somewhat shallower than the primary minimum.

Gonz\'{a}lez Hern\'{a}ndez et al. (2014) have collected 5 times of inferior conjunction from radial velocities in the published literature over the years 1985--2006, and derived a $\dot{P}$ value.  I now add five further times, with these all being times of primary minimum light, which are at the times of inferior conjunction.  For the ZTF light curve from 2018--2022, I have used 229 {\it zg} magnitudes and 254 {\it zr} magnitudes, to derive two epochs for the primary minima, as based on a chi-square fit to a template for the ellipsoidal modulation.  I have combined the times for the two colors as a weighted average to get a minimum time in the middle of the interval as HJD 2457905.7899$\pm$0.0005.  For the two {\it TESS} light curves from Sectors 6 and 33, I have made the usual fits of the folded light curves to a template light curve shape for ellipsoidal variations, so as to derive a time for the primary minimum near the center of the observing interval (see Figure 8).  Each of the two Sectors records near 63 complete orbits with fairly noisy data, and this is what allows to beat down the accuracy for the time of minima to a useable level.  The resultant times of minima are HJD 2458479.1397$\pm$0.0002 and 2459215.2884$\pm$0.0002.  Further primary minimum times come from McClintock \& Remillard (1986) with HJD 2445477.827$\pm$0.005, and from Shugarov et al. (2016) with HJD 2457332.4400 and an estimated uncertainty of $\pm$0.0010.  With these five new times, my $O-C$ curve (see Figure 9) has substantially extended the coverage, to the years 1983--2021.  I derive a $\dot{P}$ of ($-$1.1$\pm$1.3)$\times$10$^{-12}$.

\subsection{AX J1745.6-2901}

AX J1745.6-2901 is a binary with a NS primary that is a magnetar and a Soft Gamma Repeater, while the companion is a main sequence star in a 0.378 day orbit (Ponti et al. 2018) and references therein).  The system was discovered in outburst in 2016, and it appears within 16 arc-seconds from the Galactic Center source Sgr A*.  With this proximity, the source can only be seen in the X-rays, for which it displays deep, sharp-edged, eclipses, with this feature allowing for highly accurate eclipse times.  The proximity to the Galactic Center also means that the source and its eclipses can be recovered in archival data going back to 1994.    Ponti et al. (2017) have collected 30 times of X-ray eclipses from {\it ASCA} and {\it XMM-Newton}.  They created an $O-C$ curve, as shown in their Figure 2.  They have eclipse times from three short time intervals (1994, 2007--2008, and 2013--2015), with this being minimal to measure the parabolic term.  They fit a parabola to derive $\dot{P}$ to be ($-$4.03$\pm$0.32)$\times$10$^{-11}$.

\subsection{GU Mus = Nova Muscae 1991 = GS 1124-684}

GU Mus is a classic X-ray nova, first seen in eruption in 1991 (Remillard, McClintock, \& Bailyn 1992 and references therein).  Spectra in quiescence show absorption lines for the companion star that have a temperature near 4400 K.  Three groups measured RV curves from these lines to derive the mass function to be near 3 M$_{\odot}$, so the primary is a BH.  These RV curves also determined the orbital period to be 0.433 days.  Gonz\'{a}lez Hern\'{a}ndez et al (2017) report times of inferior conjunction from RV curves for four separate times, although two of these times have multiple analyses for the same time.  I have not been able to find any additional times, for example, the quiescent counterpart is too faint for {\it TESS} to see any useable flux.  I have repeated their creation of an $O-C$ curve (see their Figure 3) and fitting of a parabola.  The $\dot{P}$ that I derive is ($-$8.9$\pm$1.7)$\times$10$^{-10}$.



\subsection{V818 Sco = Sco X-1}

\begin{figure}
	\includegraphics[width=1.0\columnwidth]{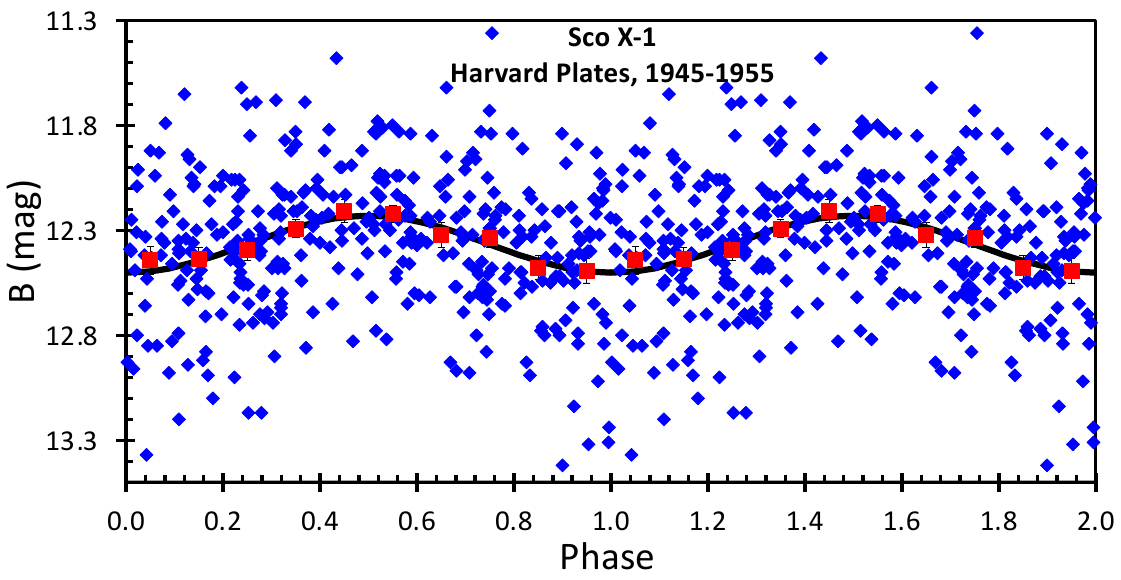}
    \caption{The Sco X-1 folded light curve for the Harvard plates magnitudes from 1945 to 1955.  The folding by phase adopts a period of 0.787311 days and has the time of the derived photometric minimum at HJD 2432471.742.  The interval contains 375 Johnson $B$-band magnitudes, as shown with the blue diamonds.  These have been folded into phase bins of width 0.10, then averaged together, to form the red squares.  The average RMS scatter of magnitudes within each phase bin is 0.32 mag.  With an average of 37 points in each bin, the error bars on the red squares is $\pm$0.016 mag, nearly the size of the red squares.  What we see is the expected sinusoidal modulation on the orbital period, with an amplitude of 0.27 mag.  The model for fitting the unbinned light curve is a simple sinewave, as shown by the thick black curve.  For this paper, the point of this figure is to show that Harvard data has a minimum on HJD 2432471.742$\pm$0.021 for an average year of 1947.8.}
\end{figure}

\begin{figure}
	\includegraphics[width=1.0\columnwidth]{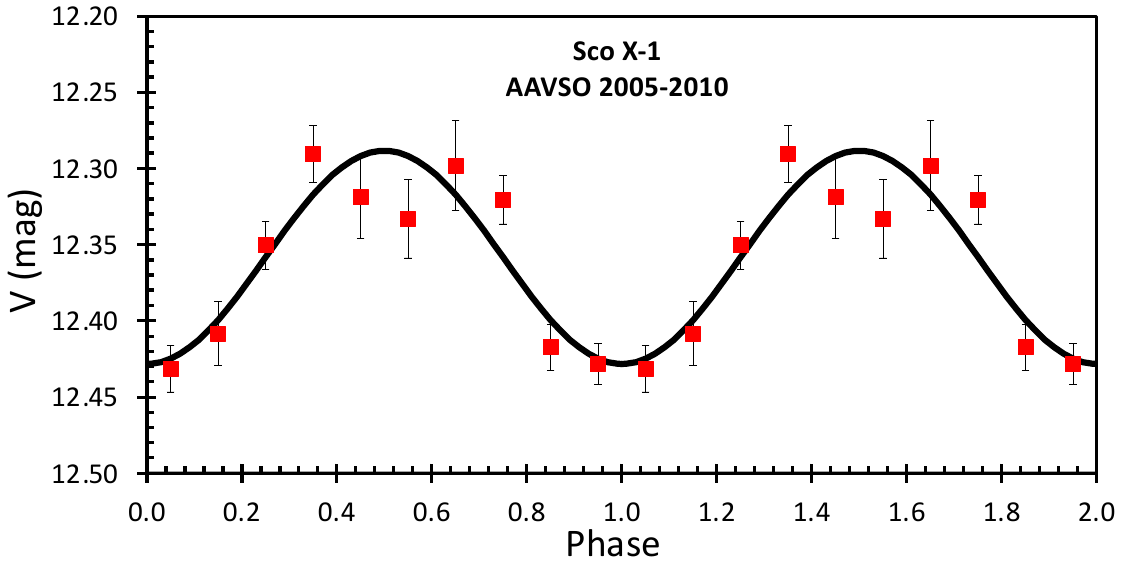}
    \caption{The Sco X-1 folded light curve for the AAVSO magnitudes from 2005 to 2010.  The folding by phase adopts a period of 0.787311 days and has the time of the derived photometric minimum at HJD 2454251.943.  The full folded light curve shows the two parallel sinewaves, corresponding to the Sco X-1 high-state and low-state (see Figure 2 of Hynes et al. 2016).  For this plot, I have screened out all points with $V$$>$12.6 mag, so as to select out the high state and hence to decrease the large dispersion in magnitudes.  The interval contains 228 Johnson $V$-band magnitudes.  The red squares show the phase averaged magnitudes in 0.10 bins.  Each bin has an average of 23 magnitudes, an RMS scatter of 0.093 mag, and a typical error bar of $\pm$0.020 mag.  The folded averaged light curve shows the expected sinusoidal modulation of Sco X-1 in its high state, with an amplitude of 0.14 mag.  The thick black curve  is the best-fitting sinewave, for which the middle time of minimum is HJD 2454251.943$\pm$0.016 for an average year of 2007.4.}
\end{figure}

Sco X-1 is the first-discovered, brightest, and prototype X-ray binary (Hynes et al. 2016 and references therein).  From sinusoidal modulations of its optical counterpart (11.1$<$$B$$<$14.1) and from its RV curve, $P$=0.787 days.  The RV curve plus allied information gives masses of 1.4 M$_{\odot}$ for the compact object and 0.7 M$_{\odot}$ for the companion star, both with fairly large uncertainties.  So Sco X-1 is an LMXB with a NS.  The accretion is driven by Roche lobe overflow (RLOF).  Sco X-1 has two distinct states (`high' and `low'), with the light curves for each state being good sinewaves with flares superposed.

 The use of conjunction times from RV curves is one method for measuring the $\dot{P}$ of Sco X-1.  The phasing of the RV curves for the hydrogen and helium lines are highly variable, greatly changing from line-to-line, and not tied with consistency to the orbit.  This is seen in the RV data of LaSala \& Thorstensen (1985, fig. 4 and Table {\rm II}) and in the trailed spectrograms of Killestein et al. (2023, fig. 1).  Indeed, when I make the best sinewave fits to the RV measures of these lines from many papers, I find a scatter of 0.3 in phase.  These emission lines are coming from the disc, and apparently are moving around in position, and so are not useable for purposes of an $O-C$ curve.
 
 In contrast, the narrow emission lines near 4640~\AA~ from Bowen fluorescence are emitted from the hotly irradiated inner hemisphere of the companion star, so their RV values must be stable and tied to the orbit.  Accurate measures of the Bowen lines' RVs started in 1999 and continued until 2019 as part of the Precision Ephemerides for Gravitational-wave Searches (PEGS) project (Killestein et al. 2023).  They do not report individual epochs for their observing years, nor do they consider any period change.  With all their 1999--2019 data, they report an epoch of BJD 2456723.3272$\pm$0.0004 and a period of 0.7873139$\pm$0.0000007 days.  Their duration of observations is too short to detect the $\dot{P}$.
 
A second method to measure $\dot{P}$ for Sco X-1 is to measure times of photometric minima for use in an $O-C$ curve.  I know of no prior attempts at this for Sco X-1.  For this, Sco X-1 is a good case, because it is bright ($V$$\sim$12.5) and the photometric modulation is readily detected (amplitude~0.15 mag in each state), so good minima times can be measured from now all the way back to 113 years ago.  For this task, the most important data source is the Harvard plates, with these plates extending back to 1890.  The most accurate time of minimum, by over a factor of 10$\times$ comes from the $K2$ mission of the {\it Kepler} spacecraft, where my analysis of the light curve (78.8 days nearly-continuous photometry with 54.2 second integrations) provides an epoch in 2014 that has a one-sigma accuracy of $\pm$0.0016 days (Hynes et al. 2016).  The AAVSO light curve is especially valuable for providing good accuracy minimum times with observing intervals of 2--10 years from 1980 to 2023.4.  Further times can be derived from ASAS, ZTF, and three papers in the literature, for which I have fitted sinewaves to the light curves.  All my times of photometric minima are tabulated in Table 2, and plotted into the $O-C$ curve in Fig. 12.

The Harvard archival plate collection contains half-a-million photographs with the developed emulsion on glass plates, covering 1888 to 1989.  All stars in the sky brighter than 12th mag are covered with 1--3 thousand plates, while all stars brighter than 16th mag are covered by at least $\sim$100 plates.  This is a unique resource in the world, and for most stars is the only way to get substantial amounts of photometry from 1888 to the 1950s.  For the purposes of this paper, the Harvard plates are the only means to get orbital phase information back before the 1960s, and the long lever arm of old times is required to pull out the $\dot{P}$.  For many stars, the Harvard light curves are more valuable scientifically than all the brief snapshots from {\it TESS} and {\it Kepler} combined.  For purposes of this section on Sco X-1, I have measured 1586 magnitudes that are on the modern Johnson $B$ magnitude system, covering 1890 to 1989, one full century.  These plates were measured both by the traditional by-eye methods and with the DASCH photometry.  In many extensive studies, the by-eye and the DASCH photometry have no detectable biases, and the measured photometric error bars are the same, and the same result is found for the U Sco measures.  Under typical conditions, the 1-$\sigma$ error bar is $\pm$0.10 or $\pm$0.15 mag.  For Sco X-1, such uncertainties are fine, because they are smaller than the usual flickering on all time scales.  (For stars with large intrinsic variability, one Harvard plate has nearly the same utility as one CCD measure with $\pm$0.01 mag accuracy, and nearly the same utility as some idealized magnitude measure with $\pm$0.000001 mag accuracy.)  The typical exposure time is 45-minutes, with a range of roughly 11--120 minutes, and the reported times are all converted into HJD for the middle of the exposure.  The Harvard light curve for Sco X-1 for 1945--1955 is shown in Figure 10.  In this case, I collected magnitudes from one decade so as to accumulate enough magnitudes to measure an accurate epoch of minimum light.  For the earliest HCO data, I had to use the interval 1890 to 1920 to collect 202 magnitudes.  For measuring $\dot{P}$, all the magnitudes inside the selected year range are then chi-square fitted to a sinewave.  The best fit is for the sinewave parameters that minimize the $\chi^2$.  The 1-$\sigma$ error bars are for the parameter range over which the $\chi^2$ is within unity of the minimum.  With the best fit sinewave, I then select a time of minimum near the average time of the magnitudes, and this is then reported in Table 2.

The AAVSO archives a huge database of fully-calibrated CCD magnitudes for most of the interesting variables in the sky.  These small telescope measures are of professional quality, and are often substantially better than the average professional photometrist.  A huge advantage of the small-telescope observers is that they provide, for many stars, orders-of-magnitude more data and often covering the last 2--3 decades.  For many stars of high interest, the AAVSO database provides a valuable resource, with wide coverage that cannot be obtained in any other way.  The photometric accuracy for Sco X-1 is typically $\pm$0.01 mag in $V$, $R$, and $I$, $\pm$0.02 mag in $B$, and $\pm$0.25 for visual observers.  For Sco X-1, the AAVSO database has 1545 $B$ magnitudes, 569 $V$ magnitudes (2001--2025), and 6244 $visual$ magnitudes (1967--2025).  Some of these time intervals do not have adequate numbers of magnitudes to provide a useful constraint on the epochs of minima.  The AAVSO $V$-band folded light curve for 2005--2010 is presented in Figure 11.  With these light curves, the chi-square calculations to find the times of minimum light are the same as for the HCO plates, and the other sources of light curves.
 
 The two times with the smallest error bars are both from 2014, my $K2$ epoch and the Bowen fluorescence line RV epoch of Killestein et al. (2023).  Both epochs are tied to the irradiated hemisphere of the companion star and are nominally the times of the inferior conjunction of the companion star.  Unfortunately, the $O-C$ for these two epoch differ by 0.0141 days (about 20 minutes), which is greatly larger than the quoted error bars.  This means that the difference must be intrinsic to the Sco X-1 system, perhaps by asymmetric illumination of the companion's inner hemisphere having different manifestation for optical light versus Bowen fluorescence.  With only one effective measure of the RV phase, this one point cannot be joined with the photometric data in an $O-C$ diagram.  
 
 So the $O-C$ diagram can only be constructed with the times of photometric minima.  However, the RV data do give a very accurate orbital period, and this is independent of the offset in phase.  So the chi-square fitting to a parabola is for the 19 times of photometric minima plus the one measure of $P$ at the time of the Killestein et al. epoch.  My $O-C$ curve (Fig. 12) is constructed with the fiducial linear ephemeris as given by the period at the epoch of Killestein et al.  This means that the slope of the best-fitting parabola should be closely flat in my diagram around the year 2014, as illustrated by the flat orange line segment in Fig. 12.  Further, as the one $K2$ time is $\ge$10$\times$ more accurate than all other times, the vertical position of the best-fitting parabola must pass close to this one point in 2014.  (To distinguish this point in Fig. 12, I have made the symbol with an orange core and made it somewhat larger than the other symbols.)  So we know the slope and position of the best-fitting parabola, it must pass nearly horizontally through the red/orange point in 2014.  This determines the $E_0$ and $P$ for the parabola, so roughly, the only remaining issue is to vary the $\dot{P}$ so as to minimize the chi-square.
 
 The best-fitting parabola seems to necessarily have a positive $\dot{P}$ (i.e., concave up with an {\it increasing} orbital period).  This is due to the $O-C$ measures being increasingly positive going back in time before the year 2000, and the only way to connect the negative slope from 1907 to 2000 with the near-zero slope in 2014 is to have a concave-up curvature.  That is, my $O-C$ diagram plus the Killestein et al. $P$ for 2014 forces the Sco X-1 $\dot{P}$ to be positive.
 
 The best-fitting parabola can be formalized with a chi-square fit.  This best-fit minimizes the chi-square from the 19 times in Table 2 plus the chi-square from the slope in the year 2014.  For this, I find $\dot{P}$=($+$8.7$\pm$5.1)$\times$10$^{-11}$.  This is shown as a thick black curve in Fig. 12.  If the $P$ in 2014 is constrained to exactly equal that of the Killestein et al. value, then the period change is $+$14.2$\times$10$^{-11}$.  The one-sigma range $\dot{P}$ is spanned by the two thick grey curves in Fig. 12.  The $\dot{P}$=0 case is rejected at the 1.6-sigma confidence level.

\begin{table}
	\centering
	\caption{Sco X-1 Times of Photometric Minima}
	\begin{tabular}{llrrl}
		\hline
		Minimum Time (HJD)   &   Year   &   $N$   &   $O-C$$^a$ & Ref$^b$ \\
		\hline
2417798.636	$\pm$	0.048	&	1907.6	&	-49440	&	0.108	&	[1]	\\
2425352.826	$\pm$	0.051	&	1928.3	&	-39845	&	0.021	&	[2]	\\
2429716.918	$\pm$	0.023	&	1940.2	&	-34302	&	0.032	&	[3]	\\
2432471.742	$\pm$	0.021	&	1947.8	&	-30803	&	0.045	&	[4]	\\
2438215.160	$\pm$	0.072	&	1963.5	&	-23508	&	0.008	&	[5]	\\
2439640.314	$\pm$	0.028	&	1967.4	&	-21698	&	0.123	&	[6]	\\
2441469.924	$\pm$	0.029	&	1972.4	&	-19374	&	0.016	&	[7]	\\
2444146.817	$\pm$	0.049	&	1979.7	&	-15974	&	0.042	&	[8]	\\
2444887.557	$\pm$	0.080	&	1981.8	&	-15033	&	-0.080	&	[9]	\\
2446401.664	$\pm$	0.037	&	1985.9	&	-13110	&	0.022	&	[10]	\\
2449831.182	$\pm$	0.022	&	1995.3	&	-8754	&	0.001	&	[11]	\\
2453439.412	$\pm$	0.031	&	2005.2	&	-4171	&	-0.029	&	[12]	\\
2453510.329	$\pm$	0.017	&	2005.4	&	-4081	&	0.030	&	[13]	\\
2454251.943	$\pm$	0.016	&	2007.4	&	-3139	&	-0.006	&	[14]	\\
2456287.170	$\pm$	0.019	&	2013.0	&	-554	&	0.015	&	[15]	\\
2456591.793	$\pm$	0.040	&	2013.8	&	-167	&	-0.053	&	[16]	\\
2456932.7386	$\pm$	0.0016	&	2014.8	&	266	&	-0.014	&	[17]	\\
2459021.445	$\pm$	0.037	&	2020.5	&	2919	&	-0.051	&	[18]	\\
2459075.841	$\pm$	0.016	&	2020.6	&	2988	&	0.020	&	[19]	\\
		\hline
	\end{tabular}	
	
\begin{flushleft}	
\
$^a$ $O-C$ is in units of days, with the fiducial linear ephemeris with a period of 0.7873139 days and an epoch of HJD 2456723.3272, from the ephemeris of Killestein et al. (2023).\\
$^b${\bf [1]}~HCO 1890.0--1920.0
{\bf [2]}~HCO 1920.0--1935.0
{\bf [3]}~HCO 1935.0--1945.0
{\bf [4]}~HCO 1945.0--1955.0
{\bf [5]}~HCO 1960--1970
{\bf [6]}~Wesselink \& Cesco (1972)
{\bf [7]}~van Genderen (1977)
{\bf [8]}~Bateson \& Venimore (1984)
{\bf [9]}~HCO 1970--1989
{\bf [10]}~AAVSO visual 1980.0--1990.0
{\bf [11]}~AAVSO visual 1990.0--2000.0
{\bf [12]}~AAVSO visual 2000.0--2010.0
{\bf [13]}~ASAS, Hynes \& Britt (2012)
{\bf [14]}~AAVSO V 2005.0--2010.0
{\bf [15]}~AAVSOnet 2012.0--2014.0
{\bf [16]}~AAVSO visual 2010.0--2020.0
{\bf [17]}~K2, Hynes et al. (2016)
{\bf [18]}~AAVSO V 2019.0--2023.5
{\bf [19]}~ZTF 2018.0--2023.0
\\

\end{flushleft}	

\end{table}

\begin{figure}
	\includegraphics[width=1.0\columnwidth]{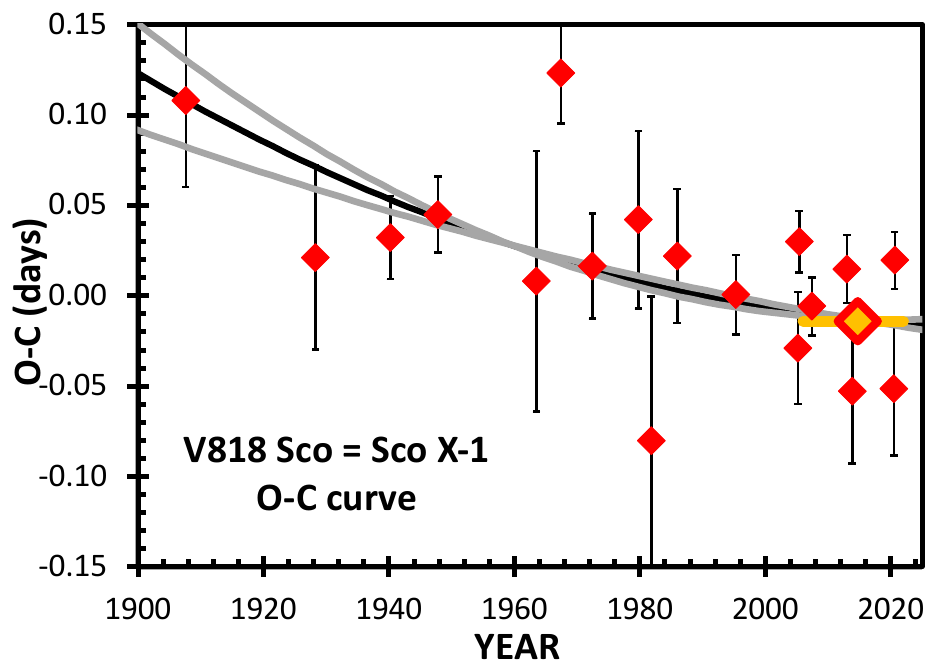}
    \caption{$O-C$ curve for Sco X-1.  This $O-C$ curve is constructed from the 19 times of primary minimum using data from 1890 to 2023.  The fiducial ephemeris for $O-C$ uses the $P$ of Killestein et al. (2023), so the best-fitting parabola must have a nearly flat slope around the year 2014, as illustrated with the flat orange line segment.  My $O-C$ value from $K2$ (depicted as a larger red diamond with an orange centre) has an uncertainly that is $\ge$10$\times$ better than all other times, so the best-fitting parabola must pass close to this one point in 2014.  This best-fitting parabola must also represent the distinctly negative slope of the $O-C$ from 1907 to 2000.  The only way to have a negative slope 1907--2000 and a near-zero slope around 2014 is for the best-fitting parabola to be curving upward.  That is, $\dot{P}$ is significantly positive, with its orbital $P$ increasing over the last century.}
\end{figure}

\subsection{LMC X-4}

LMC X-4 is an HMXB appearing in the Large Magellanic Cloud (Jain, Sharma, \& Paul 2024 and references therein).  The primary star is an X-ray pulsar with a spin period of 13.5 seconds.  The secondary star is an O8 {\rm III} star with high-mass and high-luminosity.  The orbital period is 1.408 days, as seen by deep eclipses in the X-rays.  The period change has been measured by Levine, Rappaport, \& Zojcheski (2000), Molkov, Lutovinov, \& Falanga (2015), and Falanga et al. (2015), with these measures for overlapping sets of eclipse timings.  Jain, Sharma, \& Paul (2024) added 10 new timings, bringing the total to 73 times from 1976 to 2020.  They constructed an $O-C$ curve (their Figure 4) that has a good parabolic shape with downward curvature.  They fit their $O-C$ curve to get $\dot{P}$=($-$4.97$\pm$0.04)$\times$10$^{-9}$.

\subsection{HZ Her = Her X-1}

\begin{figure}
	\includegraphics[width=1.0\columnwidth]{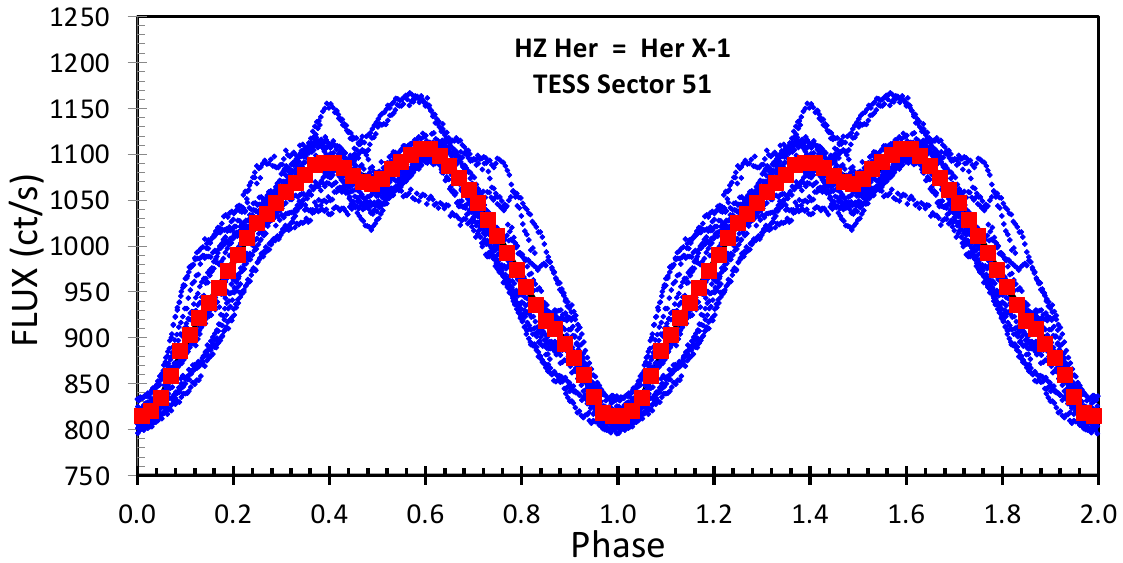}
    \caption{The Her X-1 folded light curve for {\it TESS} Sector 51 from 2022.3.  This unfiltered CCD light curve stretches 24.6 days, with few gaps, with 2821 integrations each 600 seconds in duration.  The folding by phase adopts a period of 1.70016759 days and the zero phase is at BJD 2459706.6871.  The individual fluxes are the small blue diamonds, while the folded light curve is averaged in bins of width 0.02 in phase, as represented by the red squares.  The photometric error bars on both types of magnitudes are much smaller than the plot symbols.  Strings of blue indicate the light curve for a given night.  The most striking thing about this folded light curve is that HZ Her greatly changes its light curve shape from orbit-to-orbit.  Most importantly is that the eclipse duration changes by up to a factor-of-two.  This is bad for a sparsely sampled light curve (like from HCO or the AAVSO), because then the deduced eclipse times will vary greatly depending on whether the isolated point is in the ingress/egress of a wide or narrow eclipse.}
\end{figure}

\begin{figure}
	\includegraphics[width=1.0\columnwidth]{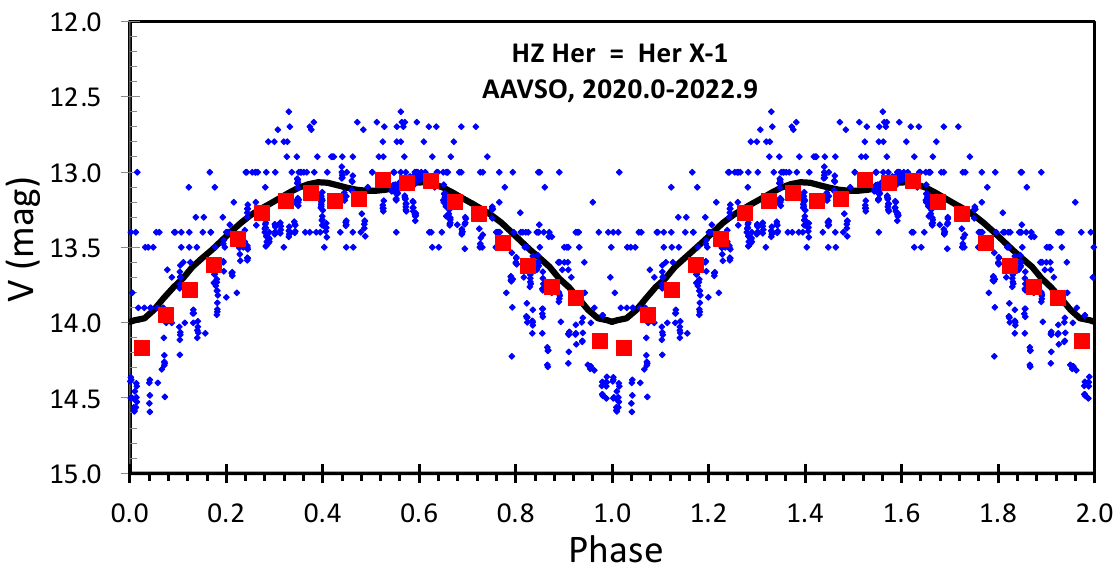}
    \caption{The Her X-1 folded light curve for the AAVSO magnitudes from 2020.0 to 2022.9.  This light curve is composed of 716 CCD Johnson $V$ magnitudes and $visual$ magnitudes.  The folding by phase adopts a period of 1.70016759 days and has the time of the derived photometric minimum at HJD 2459706.6871.  The small blue diamonds are for the individual measured magnitudes, while the red squares are averaged magnitudes in bins of 0.05 in phase.  The black curve is the template used for fitting to find the times of minima.  The large intrinsic variations in the light curve shape make for substantial uncertainty in recognizing the times of minima.  For this particular plot, the error on the time of minimum is $\pm$0.0057 days (8.2 minutes).  Unfortunately, this and other optical minimum times lead to an $O-C$ diagram with large scatter, substantially larger than the error bars, so only a poor limit is obtained for $\dot{P}$.  With this, the excellent X-ray $\dot{P}$ from Staubert, Klochkov, \& Wilms (2009) is the best answer, with $\dot{P}$=($-$4.85$\pm$0.13)$\times$10$^{-11}$.}
\end{figure}

Her X-1 is a famous X-ray binary, due to its brightness, long history, and the unusual phenomena it displays (Leahy \& Abdallah 2014 and references therein).  The basic orbital period is 1.700 days, as seen by its X-ray eclipses, with a neutron star in orbit around an A7 {\rm IV}e star.  The companion star is higher in mass than for LMXBs, and lower in mass than for HMXBs.  The accretion disc precesses with a period of 34.875 days, and this shows up as a prominent modulation in the X-rays.  The optical light curve varies wildly, both orbit-to-orbit and year-to-year, featuring a deep and sharp minimum at the time of the X-ray eclipses.  The overall modulation of the optical light covers a range from 12.8 to 15.2 mag (in $B$ light), with both reflection and ellipsoidal effects.  

Her X-1 is an X-ray pulsar with spin period of 1.238 s, and its X-ray pulses allow for the orbit to be closely tracked.   Staubert, Klochkov, \& Wilms (2009) made pulse timings from 1996--2007 with {\it RXTE} and {\it INTEGRAL}.  They combine their own timings with others from 1971--1992 to create an $O-C$ curve covering 36 years.  The shape of their curve is a good parabola, with concave-downward curvature.  Their derived $\dot{P}$ is ($-$4.85$\pm$0.13)$\times$10$^{-11}$.

Timings of the sharp minima in optical light curves can be used to track period changes in the orbit.  The best light curves for seeing the real behavior of the optical light curve are from {\it TESS} for Sectors 51 (see Figure 13) and 52 in the year 2022.  These light curves show that Her X-1 has factor-of-two changes in the duration of the primary minimum, substantial changes in the depth of the secondary minimum (ranging from one-quarter of the primary depth down to no secondary minimum at all), plus large changes in the peak fluxes on either side of the secondary minimum.  When dealing with sparse light curve sampling, this variability will lead to large random error in the derived minimum times.

The referee has requested that I address the issue that ``it is well known by exoplanet community that TESS shows some systematics'' that destroy real-and-large periodic modulations and create false-and-large coherent periodic modulations.  Indeed, researchers must be wary due to a variety of instrumental/software artifacts.  For the particular case of HZ Her, the creation of the light curve with the standard LightKurve package and the standard RC algorithm makes for all the large variability of HZ Her to be flattened out for all Sectors, while the alternative PLD algorithm retains the HZ Her variability but has added on large-amplitude baseline variations with timescales of days to weeks (for Sector 52, but not Sector 51 as in Figure 13).  Further, the standard SPOC light curves for HZ Her have the usual large-amplitude false-brightening and false-dimming superposed for the start and stop of each orbit (i.e., four times each Sector) last for up to a day or two each.  Partial solutions include clipping out the bad data, varying the input parameters, checking nearby stars, doing some independent intelligent de-trending, and avoiding the standard analysis packages.  I cannot regard any {\it TESS} periodicity or timing with any high reliability unless confirmed by independent evidence. 

The Harvard archival plates show that HZ Her was in a low-amplitude low-state from 1937--1945 and from 1949.5--1957.  In both intervals, the range is 14.3$<$$B$$<$14.8, and the primary and secondary minima have equal depths.  The transitions between the normal state time intervals (1894--1937, 1945--1949.5, and 1965--1990) are apparently sudden, with the changes appearing over less than a month.  

For constructing an $O-C$ curve from optical minimum timings, I have collected into Table 3 the minimum timings from many optical light curves covering 1894--2023.  For each light curve, I have used a template to the data in a chi-square fit, where the amplitude and offset of the template are allowed to vary to optimally represent the measured light curve.  The phasing is constructed with a period of 1.700167590  days, which is more than adequate to fold the data from the relatively short interval of observations.  The time of minimum light is set for a cycle near the centre of the observing interval, with this exact value varied to get a minimum chi-square, and the one-sigma uncertainty is the range of minimum times over which the chi-square value is within unity of the minimum.  A substantial problem is that the primary minimum often shows substantial asymmetry between the ingress and egress.  The end result is that any one eclipse has real uncertainties larger than the formal measurement errors, where the happenstance of the minimum shape and the light curve coverage make for a large jitter in the measured time of minimum.  My set of minimum timings is presented in Table 3.  The most valuable light curves are from the AAVSO (with its coverage with 12,032 magnitudes 1975--2023, see Figure 14) and from DASCH (with its unique coverage from 1894--1974).  The entire remainder of the available light curves is of much less utility because the temporal coverage is substantially lower than for AAVSO and DASCH.  To express the reality that the total uncertainties in the times of minimum are often much larger than the formal measurement errors in Table 3, I have added in quadrature a jitter error of 0.007 days, and this makes for reduced chi-square values close to unity.  With this, I have made a chi-square fit to the $O-C$ curve to a parabolic model, and I find $\dot{P}$=($+$7$\pm$6)$\times$10$^{-11}$.  

This result from the optical minimum times has a 46$\times$ larger error bar than does the $\dot{P}$ from the X-ray pulse timings.  The X-ray value is at the 2-sigma deviation level for the optical value, with this being adequate, even if uncommon.  My final value should be the weighted average of the two measures of $\dot{P}$, although this comes out close to the value from the X-ray pulse timings alone.  In the end, I conclude that $\dot{P}$=($-$4.85$\pm$0.13)$\times$10$^{-11}$ for Her X-1.

\begin{table}
	\centering
	\caption{Her X-1 Primary Minimum Times}
	\begin{tabular}{llrrl}
		\hline
		Minimum Time (HJD)   &   Year   &   $N$   &   $O-C$$^a$ & Ref$^b$ \\
		\hline
2416551.352	$\pm$	0.016	&	1904.2	&	-19967	&	0.0132	&	[1]	\\
2427410.295	$\pm$	0.046	&	1933.9	&	-13580	&	-0.0110	&	[2]	\\
2432216.69	$\pm$	0.02	&	1947.1	&	-10753	&	0.0116	&	[3]	\\
2442456.788	$\pm$	0.013	&	1975.1	&	-4730	&	0.0032	&	[4]	\\
2443793.1003	$\pm$	0.0071	&	1978.8	&	-3944	&	-0.0158	&	[5]	\\
2445426.9774	$\pm$	0.0052	&	1983.3	&	-2983	&	0.0007	&	[6]	\\
2446027.142	$\pm$	0.015	&	1984.9	&	-2630	&	0.0066	&	[7]	\\
2446904.4323	$\pm$	0.0054	&	1987.3	&	-2114	&	0.0104	&	[8]	\\
2447496.0766	$\pm$	0.0026	&	1988.9	&	-1766	&	-0.0034	&	[9]	\\
2448956.5238	$\pm$	0.0043	&	1992.9	&	-907	&	0.0003	&	[10]	\\
2450498.5785	$\pm$	0.0041	&	1997.1	&	0	&	0.0034	&	[11]	\\
2452537.06500	$\pm$	0.0048	&	2002.7	&	1199	&	-0.0104	&	[12]	\\
2454123.3232	$\pm$	0.0057	&	2007.1	&	2132	&	-0.0081	&	[13]	\\
2454347.7557	$\pm$	0.0020	&	2007.7	&	2264	&	0.0023	&	[14] 	\\
2454351.1664	$\pm$	0.0054	&	2007.7	&	2266	&	0.0127	&	[14] 	\\
2455311.730	$\pm$	0.002	&	2010.3	&	2831	&	-0.0181	&	[15]	\\
2456107.4243	$\pm$	0.0049	&	2012.5	&	3299	&	-0.0020	&	[16]	\\
2457863.7013	$\pm$	0.0082	&	2017.3	&	4332	&	0.0024	&	[17]	\\
2458739.2913	$\pm$	0.0025	&	2019.7	&	4847	&	0.0063	&	[18]	\\
2458739.2914	$\pm$	0.0021	&	2019.7	&	4847	&	0.0064	&	[19]	\\
2459274.8283	$\pm$	0.0057	&	2021.2	&	5162	&	-0.0093	&	[20]	\\
2459706.686	$\pm$	0.0014	&	2022.3	&	5416	&	0.0055	&	[21]	\\
2459732.1935	$\pm$	0.0085	&	2022.4	&	5431	&	0.0109	&	[22]	\\
		\hline
	\end{tabular}	
	
\begin{flushleft}	
\
$^a$ $O-C$ is in units of days, with the fiducial linear ephemeris with a period of 1.70016709 days and an epoch of HJD 2450498.5751.\\
$^b$ References:  
{\bf [1]}~DASCH and Kurochkin (1972) 1894--1916,~
{\bf [2]}~DASCH 1921.0--1937.0,~
{\bf [3]}~DASCH 1945.0--1949.5,~
{\bf [4]}~DASCH 1960.0--1980.0,~
{\bf [5]}~AAVSO 1975.0--1980.0,~
{\bf [6]}~AAVSO 1980.0--1985.0,~
{\bf [7]}~DASCH 1980.0--1990.0,~
{\bf [8]}~AAVSO 1985.0--1990.0,~
{\bf [9]}~Sazanov (2011),~
{\bf [10]}~AAVSO 1990.0--1995.0,~
{\bf [11]}~AAVSO 1995.0--2000.0,~
{\bf [12]}~AAVSO 2000.0--2005.0,~
{\bf [13]}~AAVSO 2005.0--2010.0,~
{\bf [14]}~Zasche (2010),~
{\bf [15]}~Diethelm (2010),~
{\bf [16]}~AAVSO 2010.0--2015.0,~
{\bf [17]}~AAVSO 2015.0--2020.0,~
{\bf [18]}~ZTF {\it zr},~
{\bf [19]}~ZTF {\it zg},~
{\bf [20]}~AAVSO 2020.0--2022.9,~
{\bf [21]}~{\it TESS} sector 51,~
{\bf [22]}~{\it TESS} sector 52.\\

\end{flushleft}	
	
\end{table}

\subsection{V779 Cen = Cen X-3}

Cen X-3 is a bright HMXB discovered back in 1967 by {\it Uhuru}, and was the first X-ray pulsar (Klawin et al. 2023 and references therein).  The primary star is a strongly-magnetic neutron star, with an X-ray spin period of 4.83 seconds.  The secondary star is an O6-8 {\rm III} supergiant.  The system shows deep X-ray eclipses with a period of 2.087 days and with a duration of 0.20 in phase.  In optical light, the system (called Krzeminski's Star) shows low-amplitude ellipsoidal modulations over a range of 13.25$<$$V$$<$13.46, with the secondary minima at the times of the X-ray eclipses.  

Raichur \& Paul (2010), Falanga et al. (2015), Klawin et al. (2023), and Liu \& Wang (2023) have all reported long lists (with substantial duplication) of X-ray eclipse times, for a total of 52 times from 12 spacecraft spanning the years 1971--2018.  Their $O-C$ curves (e.g., see Figure 3 of Klawin et al. 2023, or Figure 8 of Liu \& Wang 2023) show a perfect parabola, with concave-down curvature.  The derived $\dot{P}$ measures are (1.0280$\pm$0.0010)$\times$10$^{-8}$ from Raichur \& Paul (2010),  (1.0285$\pm$0.0010)$\times$10$^{-8}$ from Falanga et al. (2015), (1.03788$\pm$0.00027)$\times$10$^{-8}$ from Klawin et al. (2023), and ($-$1.01888$\pm$0.00006)$\times$10$^{-8}$ from Liu \& Wang (2023).  These measures are using the same methods and largely the same set of eclipse times, so I judge that the presented very-small error bars are not realistic.  I will adopt $\dot{P}$=($-$1.028$\pm$0.010)$\times$10$^{-8}$.

\subsection{M82 X-2}

M82 X-2 is an ultraluminous X-ray source in the famous `exploding' galaxy M82 (Bachetti et al. 2022).  The luminosity of this source is super-Eddington, and this is puzzling.  This source is an X-ray pulsar, with a spin period of 1.37 seconds, and must be a NS.  Analysis of the pulses shows that the system is in a binary, with the orbital parameters measured with high accuracy.  The system has an orbital period of 2.533 days, and the companion must be some high mass ($>$5 M$_{\odot}$) star.  Bachetti et al. (2022) collected 14 times from 2014--2021 for when the NS passed through the ascending node.  These were made into an $O-C$ curve (see their Figure 2), with this showing a closely-defined parabola.  The parabola curvature corresponds to $\dot{P}$=($-$5.69$\pm$0.24)$\times$10$^{-8}$.

\subsection{V4641 Sgr}

\begin{figure}
	\includegraphics[width=1.0\columnwidth]{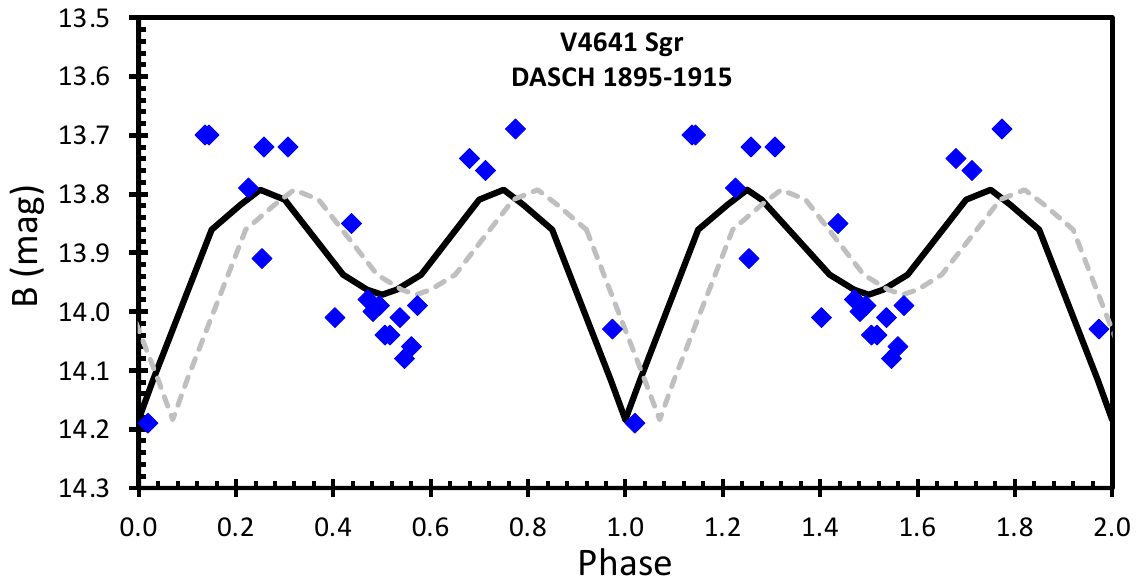}
    \caption{The V4641 Sgr folded light curve for the HCO data from 1895.0--1915.0.  The magnitudes were measured multiple times with my by-eye measures and with DASCH, with the averages plotted here.  The 20 Johnson $B$ magnitudes (blue diamonds) are sparse, yet nevertheless adequate to define the phase of the primary and secondary eclipses.  The light curve is folded twice into phase, with adopted period 2.81728 days, and the zero phase taken from the best fit time of minimum at HJD 2415407.2711 (in 1901.1).  The template light curve (thick black curve) is defined from later observations with many magnitudes.  As seen in the $O-C$ curve, the concave-down parabola shape is largely determined by the one measured time from 1901.1.  The HCO times of minimum are very reliable to within their quoted error bars.  A natural question for the V4641 $O-C$ plot is to wonder whether a zero-$\dot{P}$ solution is reasonable.  Without the old plates, the best straight-line would be close to 0.20 days later than the measured value in 1901.1, with this being  a 2.6-$\sigma$ deviation.  In this anti-historical case, the template would be shifted right by 0.07 in phase, as shown by the thin grey template.  Such a shift is clearly wrong, as can be seen in the figure where all the magnitudes at the two maxima are $<$0.25 and $<$0.75 in phase, the faintest magnitude is moved significantly away from the template minimum, and the secondary minimum magnitudes are all $<$0.5 in phase.  This points to the need for the $O-C$ curve to have a concave-down shape, which is to say $\dot{P}$$<$0.  }
\end{figure}

V4641 Sgr is an intermediate-mass X-ray binary that has frequent fast-rise exponential-decay outbursts (MacDonald et al. 2014, Mu\~{n}oz-Darias, Torres, \& Garcia 2018).  The first discovered outburst was in 1979 with optical light alone, then a large outburst in 1999 was discovered by two X-ray satellites, and there have been eight outbursts from 1999--2023.  The companion is a B9 {\rm III} star already evolved off the main sequence, showing ellipsoidal variations with amplitude near 0.40 mag.  A RV curve shows a 2.817 day period, for which the compact object has a mass of 6.4$\pm$0.6 M$_{\odot}$, so it must be a black hole.  The companion has a mass of 2.9$\pm$0.4 M$_{\odot}$.

The system is fairly bright in optical light, with $V$ hovering around 13.5 mag in quiescence.  This means that many data sets can get adequate optical light curves to define the times of minimum light in the ellipsoidal modulation, and that an $O-C$ curve is possible with many data points going far back in time.  For this end, the best work is from the SMARTS telescope at Cerro Tololo observatory, operating out of Yale, with ten years of wonderful light curves and spectra (MacDonald et al. 2014).  Their work shows a well-formed and stable ellipsoidal light curve, with small additions while in an active state.  I have also collected light curves from the Harvard plate archives (see Figure 15), the Maria Mitchell Observatory (MMO) plate archives, the AAVSO, ASAS, Pan-STARRS, ZTF, and from four papers in the literature.  With a light curves template to represent the ellipsoidal modulation, I have performed chi-square fits to all the light curves so as to derive the times of the primary minima (see Table 4).

With my 20 times of conjunction and their $O-C$ values, I have constructed a plot of the $O-C$ curve (Fig. 16).  From 1989 to 2023, there is a narrow envelope of the measures with high accuracy, and this runs significantly downward as time goes on.  If $\dot{P}$=0, then either this narrow envelope will be violated or the early time observations from the Harvard plates will be violated, so it looks like there is significant downward curvature.  But a zero curvature case is rejected only at a moderate significance level.  My parabola fits give $\dot{P}$=($-$4.7$\pm$3.4)$\times$10$^{-10}$.

\begin{table}
	\centering
	\caption{V4641 Sgr Primary Minimum Times}
	\begin{tabular}{llrrl}
		\hline
		Minimum Time (HJD)   &   Year   &   $N$   &   $O-C$$^a$ & Ref$^f$ \\
		\hline
2415407.2711	$\pm$	0.0695	&	1901.1	&	-14584	&	-0.0454	&	[1]	\\
2426690.5978	$\pm$	0.0323	&	1932.0	&	-10579	&	0.0749	&	[2]	\\
2437289.1331	$\pm$	0.0226	&	1961.0	&	-6817	&	0.0029	&	[3]	\\
2447707.4865	$\pm$	0.0038$^b$	&	1989.5	&	-3119	&	0.0548	&	[4]	\\
2448000.4509	$\pm$	0.0050$^c$	&	1990.3	&	-3015	&	0.0221	&	[5]	\\
2451443.2273	$\pm$	0.0520$^d$	&	1999.7	&	-1793	&	0.0824	&	[6]	\\
2451764.3372	$\pm$	0.0188	&	2000.6	&	-1679	&	0.0223	&	[4]	\\
2452896.8757	$\pm$	0.0143	&	2003.7	&	-1277	&	0.0143	&	[7]	\\
2453057.4241	$\pm$	0.0016	&	2004.1	&	-1220	&	-0.0223	&	[8]	\\
2453156.0822	$\pm$	0.0100$^e$	&	2004.4	&	-1185	&	0.0310	&	[9]	\\
2453635.0011	$\pm$	0.0052	&	2005.7	&	-1015	&	0.0123	&	[10]	\\
2455928.2843	$\pm$	0.0183	&	2012.0	&	-201	&	0.0296	&	[11]	\\
2456057.9252	$\pm$	0.0320	&	2012.4	&	-155	&	0.0756	&	[12]	\\
2456494.5280	$\pm$	0.0029	&	2013.6	&	0	&	0.0000	&	[13]	\\
2456874.8614	$\pm$	0.0029	&	2014.6	&	135	&	0.0006	&	[14]	\\
2457117.1494	$\pm$	0.0030	&	2015.3	&	221	&	0.0025	&	[15]	\\
2458421.5110	$\pm$	0.0277	&	2018.8	&	684	&	-0.0365	&	[16]	\\
2458708.8762	$\pm$	0.0259	&	2019.6	&	786	&	-0.0339	&	[17]	\\
2459072.3422	$\pm$	0.0100	&	2020.6	&	915	&	0.0030	&	[18]	\\
2459579.4765	$\pm$	0.0201	&	2022.0	&	1095	&	0.0269	&	[19]	\\
		\hline
	\end{tabular}	
	
\begin{flushleft}	
\
$^a$ $O-C$ is in units of days, with the fiducial linear ephemeris with a period of 2.81728 days and an epoch of HJD 2456494.5280.\\
$^b$ Minimum time calculated and reported by Orosz et al. (2001).\\
$^c$ The published epoch was for a time greatly after the end of the data, so as to allow easy comparison with later data, so I have converted the epoch back to a time near the middle of the original data with their adopted period of 2.81728 days.\\
$^d$ Epoch shifted by 0.25 in phase to get to the conjunction time of the deepest photometric minimum.\\
$^e$ Epoch shifted by 0.5 in phase to get to the conjunction time of the deepest photometric minimum.\\
$^f$ References:  
{\bf [1]}~HCO 1895--1914,~
{\bf [2]}~HCO 1918--1942,~
{\bf [3]}~Maria Mitchell Observatory 1956--1972,~
{\bf [4]}~Goranskij (2001),~
{\bf [5]}~Barsukova et al. (2014),~
{\bf [6]}~Orosz et al. (2001),~
{\bf [7]}~AAVSO visual light curve 2000--2007,~
{\bf [8]}~AAVSO CR light curve 2002--2005,~
{\bf [9]}~Lindstr{\o}m et al. (2005),~
{\bf [10]}~ASAS 2001--2009,~
{\bf [11]}~AAVSO visual light curve 2008--2015,~
{\bf [12]}~Pan-STARRS,~
{\bf [13]}~SMARTS 2013,~
{\bf [14]}~SMARTS 2014,~
{\bf [15]}~SMARTS 2015,~
{\bf [16]}~AAVSO visual light curve 2016--2023,~
{\bf [17]}~ZTF {\it zg} 2018--2020,~
{\bf [18]}~AAVSO V light curve 2016--2023,~
{\bf [19]}~ZTF {\it zg} 2021--2022.\\

\end{flushleft}

\end{table}

\begin{figure}
	\includegraphics[width=1.0\columnwidth]{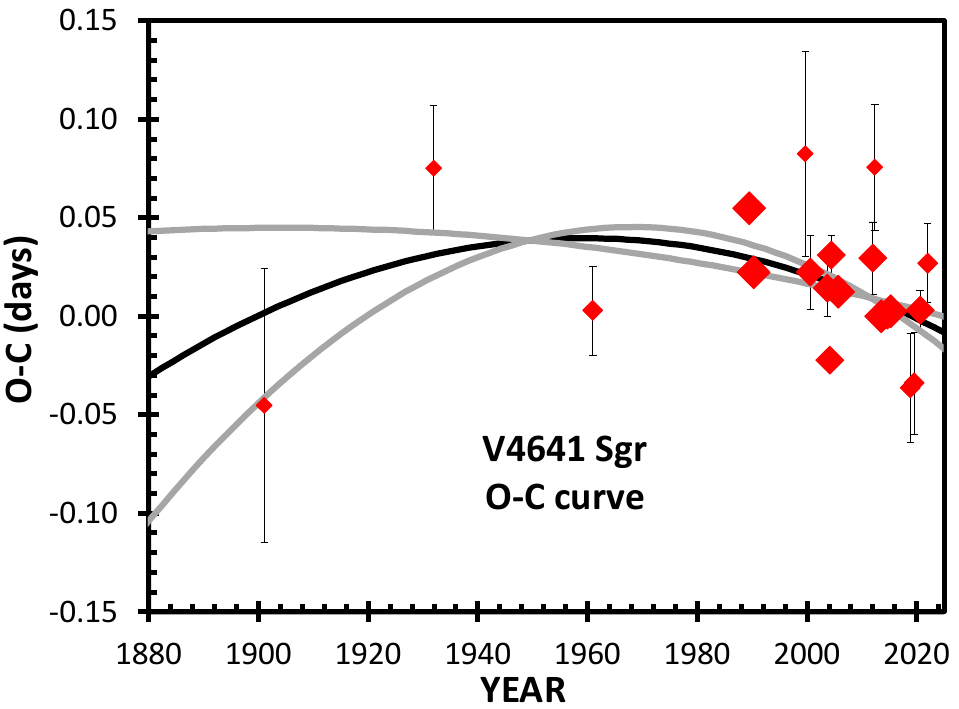}
    \caption{$O-C$ curve for V4641 Sgr.  This $O-C$ curve is constructed from the 20 times of primary minimum using data from 1894 to 2023 (see Table 4).  The size of the red diamonds is connected to the size of the error bars, with small diamonds representing values with relatively large uncertainties.  The best-fitting parabola (the thick black curve) is concave-down, so the period is decreasing.  Nevertheless, the uncertainties (as shown by the 1-sigma limits depicted as thick gray curves) are fairly large, with the $\dot{P}$=0 case (i.e., a straight line in the plot) being not much past the 1-sigma limit.}
\end{figure}



\subsection{V884 Sco = 4U 1700-377}

\begin{figure}
	\includegraphics[width=1.0\columnwidth]{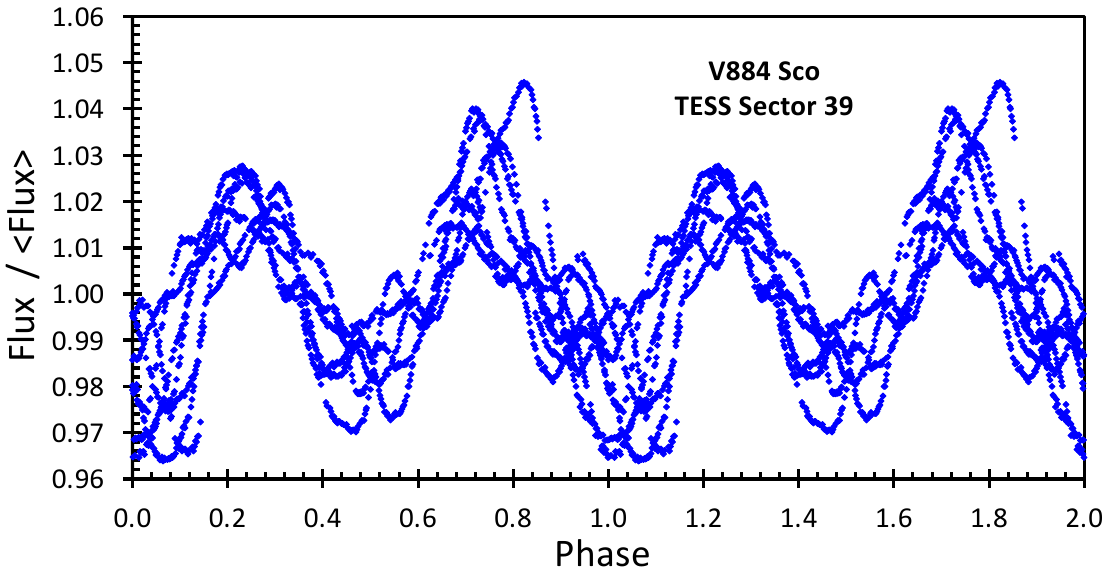}
    \caption{The V884 Sco folded light curve for {\it TESS} Sector 39 from 2021.4.  This unfiltered CCD light curve covers 24.8 days, with few gaps, with 2417 integrations each 600 seconds in duration.  The folding uses a period of 3.41165 days and the zero phase is at BJD 2459376.5484.  The individual fluxes have been normalized by the average-flux, and are shown as the small blue diamonds.  The photometric error bars are smaller than the plot symbols.  Strings of blue indicate the light curve for a given night.  The light curve shows large variability in shape, most particularly with a wide variation in the times of maxima and minima.  This means that the {\it TESS} light curve cannot provide any accurate measure of the minimum times, and all the sparse light curves (like with ASAS) will have relatively large measurement errors.  In the end, the optical-minimum $O-C$ diagram is not useful to constrain $\dot{P}$, whereas the X-ray eclipse $O-C$ curve provides a nice measure of $\dot{P}$.}
\end{figure}

\begin{figure}
	\includegraphics[width=1.0\columnwidth]{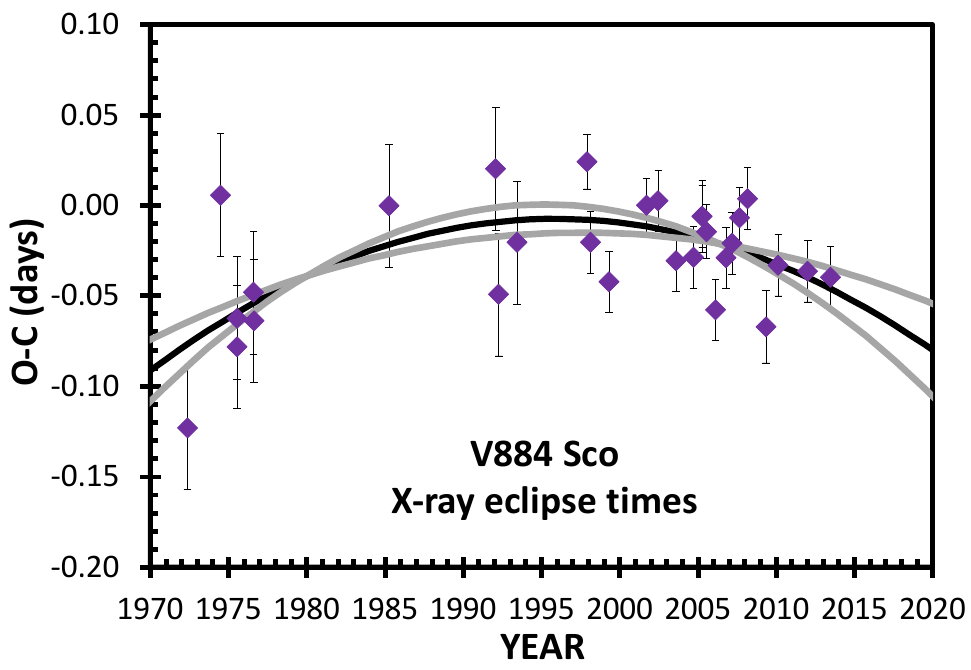}
    \caption{$O-C$ curve for V884 Sco.  This $O-C$ curve is constructed from the 29 times of mid-eclipse using X-ray data from 1972 to 2013, with $P_{\rm O-C}$=3.411650 days and $E_{\rm O-C}$=2452176.079.  The best-fitting parabola (the thick black curve) is concave-down, so the period is decreasing.  The 1-sigma limits depicted as thick gray curves) flank the best-fitting parabola.  I find $\dot{P}$=($-$6.4$\pm$2.4)$\times$10$^{-9}$.}
\end{figure}

V884 Sco appears as a bright ($V$=6.5) blue supergiant (O6 {\rm I}afpe) just off the sting of the Scorpion (Islam \& Paul 2016 and references therein).  In optical light, the system shows a typical ellipsoidal modulation with the primary minimum having an amplitude around 0.1 mag, while the secondary minimum has a somewhat smaller amplitude, all with substantial orbit-to-orbit variations.  In X-ray light, the system shows modulation on the same period, while a sharp-edged total eclipse becomes visible for energies above 10 keV, with the eclipse duration of 20 per cent of the orbit.  The $P$ is 3.412 days.  No X-ray pulsations or X-ray bursts are seen.  The compact object is likely a NS, as based on the X-ray spectrum, plus a possible detection of a cyclotron scattering resonance feature.  The accretion on to the NS is from a heavy wind from the supergiant, with a terminal wind velocity of 1900 km s$^{-1}$ and a wind mass loss rate of 10$^{-5.6}$ M$_{\odot}$ yr$^{-1}$ (Mart\'{i}nez-Chicharro et al. 2021).

V884 Sco shows significant ellipsoidal modulations in optical light, and these are tied to the orbital period.  So I have measured times of the primary minimum from a wide variety of data sets, hoping to get a useable $O-C$ curve to test the X-ray $\dot{P}$.  Unfortunately, the light curve shape varies by up to 50 per cent of the full amplitude from orbit-to-orbit.  This is best seen in the {\it TESS} light curves (for Sectors 12 and 39), where the phases of primary minima vary within a range larger than 0.25 in phase (see Figure 17).  Further, the average {\it TESS} primary minima are at a phase of 0.56 with the X-ray eclipse ephemeris, which means that the optical phasing has offsets from times of conjunction that are of unknown origin.  Further, most of the optical light curves plotted or tabulated in the literature are missing critical data for determining the times of minimum light, and this leads to large error bars for putting into an $O-C$ curve.  Other than for {\it TESS} and {\it ASAS}, all the optical light curves that I know of are too sparse, or have unusably large photometric errors.  In the end, I did construct an optical $O-C$ curve, but I have little confidence that my 7 optical times (1973--2021) are good measures of the orbit, and my resultant curve has real error bars that are greatly larger than any possible sagitta for a parabola shape.  So, reluctantly, we have no $\dot{P}$ from the optical light curves.

V884 Sco shows total eclipse in hard X-rays, as the NS passes behind the supergiant star.  Previously published values of $\dot{P}$ have ranged over a factor of 7, but all these were only derived from subsets of the available data.  To get the best value for $\dot{P}$ from the X-ray eclipses, I have used all 29 times collected by Falanga et al. (2015) and Islam \& Paul (2016), with these ranging from 1972--2013.  These times are plotted in an $O-C$ diagram in Figure 18.  With these 29 X-ray eclipse times, I derive a $\dot{P}$ of ($-$6.4$\pm$2.4)$\times$10$^{-9}$.

\subsection{QV Nor = 4U 1538-522}

QV Nor is a HMXB with a 526.8 s X-ray pulsar (Hemphill et al. 2019 and references therein).  The NS is in a 3.728 day orbit with an eccentricity of 0.18, which is fine for wind-fed accretion.  A total X-ray eclipse is visible above 10 keV with a duration of 0.143 in phase.  Hemphill et al. (2019) reports one new eclipse time from 2016 and adds in 10 eclipse times from the literature (1972--2009) to construct an $O-C$ curve (see their Figure 3) that shows a clear parabolic curvature.  They derive $\dot{P}$ to be ($-$9.7$\pm$3.8)$\times$10$^{-9}$.

\subsection{SMC X-1}

SMC X-1 is an HMXB in the Small Magellanic Cloud (Brumback et al. 2022 and references therein).  The primary star is an X-ray pulsar with spin period of 0.71 s.  The companion star is a B0 {\rm I} supergiant star, with an estimated mass around 18 M$_{\odot}$.  The orbital period is 3.892 days, as determined from periodic X-ray eclipses.  In the $V$-band, the binary shows ellipsoidal modulation with variations over a range 13.05$<$$V$$<$13.18.

The orbital period (and its changes) can be measured from times of the X-ray eclipses.  Falanga et al. (2015) has measured and collected 20 times from 1971 to 2005, while Brumback et al. (2022) has added two more times from 2017 and 2018.  Both show $O-C$ curves with a highly significant and well-formed parabola with concave-down curvature.  Falanga et al. get $\dot{P}$=($-$3.776$\pm$0.002)$\times$10$^{-8}$.  Brumback et al. do not quote a revised value, but show that their new times match nicely to the earlier fit.



\subsection{V1357 Cyg = Cyg X-1}

\begin{figure}
	\includegraphics[width=1.0\columnwidth]{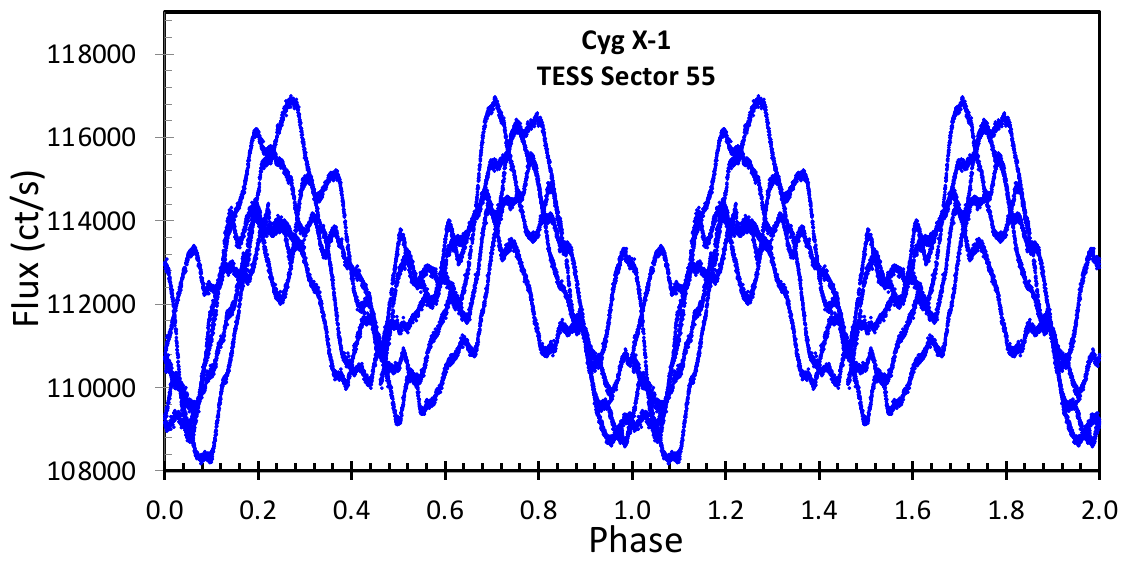}
    \caption{The Cyg X-1 folded light curve for {\it TESS} Sector 55 from 2022.6.  This unfiltered CCD light curve covers 27.2 days, with few gaps, with 18880 integrations each 120 seconds in duration.  The folding uses a period of 5.599848 days and the zero phase is at BJD 2459813.8888.  Strings of blue indicate the light curve for a given night.  The light curve shows the basic variability with a primary and secondary eclipse.  But the shape changes greatly orbit-to-orbit.  Importantly, the times of minima (and maxima) shift around by a third of the orbit.  This means that the intrinsic variability in Cyg X-1 will produce a large jitter, even for the magnificent {\it TESS} light curves.  Further, the sparse light curves (including those from AAVSO and KWS) must have large measurement errors.  This makes for a large scatter in the $O-C$ curve.}
\end{figure}

Cyg X-1 is the prototype XRB with a BH, famous as the first reasonable case for the existence of BHs, even though its low mass function allowed for long running doubts (Jiang 2024 and references therein).  The companion star is an O9.7 {\rm I}ab supergiant, which displays ellipsoidal modulations with a range of 8.72$<$$V$$<$8.93.  The ellipsoidal modulations show $P$ to be 5.600 days, and this is confirmed with RV curves.  Analyses of the radial velocities, astrometry, and allied information (like the absence of eclipses) yield the mass for the BH over a wide range, including 14.8$\pm$1.0 and 21.2$\pm$2.2 M$_{\odot}$, with the corresponding masses of the supergiant companion as 19.2$\pm$1.9 and 40.6$\pm$7.5 M$_{\odot}$ (Orosz et al. 2011, Miller-Jones et al. 2021).

There are two ways to construct an $O-C$ curve for Cyg X-1, and the first is to use the time when the velocity increases through the $\gamma$-velocity to serve as markers for the conjunction of the orbit.  Due to its importance for measuring the BH, many RV curves have been measured, but only for the years 1971--2000.  I have collected 17 RV curves from the literature (see Table 5).  In almost all cases, the individual measured times and velocities are listed, and I have used these in my own fits to a sinusoidal model.  These fits give the times of conjunction and the one-sigma measurement error bar. 

Of prime importance are the two RV measures from 1939 and 1940, with their importance arising from their extending the $O-C$ curve backwards in time by over 30 years.  LaSala et al. (1998) chose not to consider these two measures because of concerns that the averaging of lines of several different species would make for considerable uncertainties.  But such measures are perfectly fine, as long as appropriate error bars are used.  From detailed comparison between RVs for a variety of lines, the deviations are small, so this is not a substantial problem.  I have assigned an error bar of $\pm$10 km s$^{-1}$, even though this is likely greatly too large.  We are left with good times and RVs, good to within the stated error bars.  I have used these two RV measures in a sinewave fit, where I have had to adopt a global average $\gamma$-velocity, K amplitude, and $P$.  This sinewave fit to two points yields the time of conjunction, albeit with a formal uncertainty (from the propagation of the various uncertainties) of $\pm$0.114 days.  The cycle count back to 1939 is accurately known, while any alternative counts force an implausibly large period change soon before 1970.  The resultant $O-C$ value from 1939 is good, good to within the quoted errors, and must be used.

The second way to construct an $O-C$ curve for Cyg X-1 is to use times of the photometric minima in the optical ellipsoidal variations.  A substantial unrealized problem is that the actual shape of the ellipsoidal variations varies greatly from orbit-to-orbit.  This is best seen with the exquisite {\it TESS} light curves from Sectors 14, 54, and 55.  The depths of the secondary minima varies randomly from 0.5--1.0 times the depths of the adjacent primary minima, and the durations of the primary minimum vary by a factor of 3.  Importantly for the question of the accuracy of $O-C$ measures, the times of the faintest light in the primary minima varies with extremes from $-$0.10 to $+$0.10 in phase.  The RMS scatter of minima times is approximately 0.15 days.  This intrinsic jitter, from Cyg X-1 itself, cannot be corrected in any way, so the time of conjunction, as estimated from the time of photometric minimum, will suffer an additional uncertainty of $\pm$0.15 days for one orbit.  This is in addition to the measurement uncertainty, as taken from the chi-square fit.  For a light curve that samples $N_{\rm cycles}$, the real total uncertainty in the measure of the time of conjunction will be the addition in quadrature of the measurement error and $0.15/\sqrt{N_{\rm cycles}}$ days.  For example, each {\it TESS} sector has excellent measurement accuracy ($\pm$0.003 days) but only 5 orbits worth of data, so the real uncertainty for measuring the instant of conjunction is $\pm$0.067 days.  A similar result applies to the RV times of conjunction.

\begin{figure}
	\includegraphics[width=1.0\columnwidth]{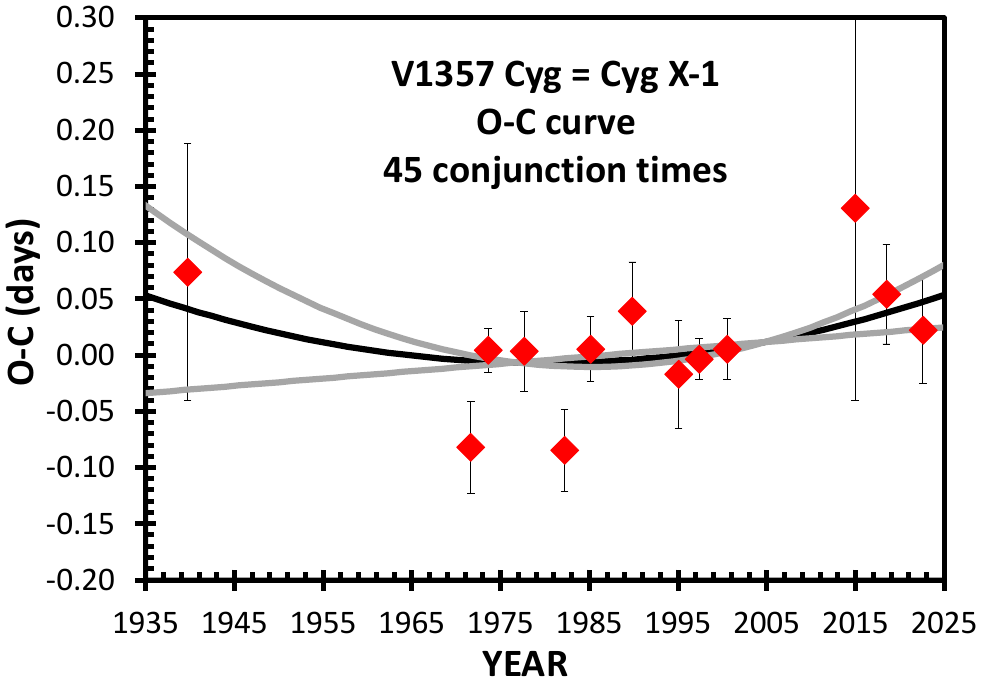}
    \caption{$O-C$ curve for Cyg X-1.  This $O-C$ curve is constructed from the 45 times of conjunction (27 times of primary minima and 18 times when the RV increases past the $\gamma$-velocity) from 1939--2022, as given in Table 5.  For display purposes only, the $O-C$ values have been formed into 4-year bins by weighted least squares.  The best-fitting parabola (thick black curve) has a positive curvature (with the period {\it increasing} over the years), for $\dot{P}$=($+$2.5$\pm$2.4)$\times$10$^{-9}$.  The positive-curvature case is to be preferred, but the $\dot{P}$=0 case (i.e., a straight line in the plot) is slightly outside the 1-sigma limits (the gray curves).}
\end{figure}

As with the RV measures, photometric light curves of Cyg X-1 were made by many workers from 1971--1998, but none has been published for observations after 1998.  (I am disappointed that I cannot pull out any useable phase information from the Harvard plates.)  I have collected light curves from the literature, performed chi-square fits to an ellipsoidal template, and derived the times of primary minimum light, along with its one-sigma measurement error.  Further, I have used unpublished light curves from {\it TESS} (e.g., see Figure 19), {\it Hipparcos}, AAVSO, and the Kamogata/Kiso/Kyoto Wide-field Survey (KWS) so as to get more times of conjunction.  These light curves from all-sky surveys provide the only information on the orbit of Cyg X-1 after the year 2000.  All the times of primary minima are given in Table 5.

The RV and photometric $O-C$ values show no systematic offset from each other, nor is any offset expected.  So I have used both equally in a single fit to the $O-C$ curve.  From 1971--2000, the many $O-C$ values overlap confusingly in an $O-C$ plot, so for plotting purposes only (see Fig. 20), I have binned them together as weighted averages over 4-year time bins.  I have fitted a parabola to the full list of 45 times of conjunction (see Table 5) that form the $O-C$ curve (see Fig. 20).  Without the 1939 measure, all the other $O-C$ measures reasonably fit a straight line (i.e., $\dot{P}$=0), with this making the 1939 measure into a two-sigma deviation.  With good confidence in the 1939 measure, the existence of a {\it positive} $\dot{P}$ is preferred, although not to the extent of high confidence.  The formal fit for a parabola gives $\dot{P}$ equal to ($+$2.5$\pm$2.4)$\times$10$^{-9}$.  With this, the $\dot{P}$=0 case is slightly more than one-sigma from the best fit.

\startlongtable
\begin{deluxetable}{lllrrl}
\tablecaption{Cyg X-1 Conjunction Times}
\tablewidth{600pt}
\tabletypesize{\scriptsize}
\tablehead{
Conjunction (HJD)   &   Year  &  \colhead{Type}   &   $N$   &   $O$-$C$$^a$ & Ref$^b$          
}
\startdata
2429535.520	$\pm$	0.114	&	1939.7	&	RV	&	-2475	&	0.074	&	[1]	\\
2441166.220	$\pm$	0.060	&	1971.6	&	Phot	&	-398	&	-0.110	&	[2]	\\
2441211.076	$\pm$	0.056	&	1971.7	&	RV	&	-390	&	-0.053	&	[3]	\\
2441317.512	$\pm$	0.058	&	1972.0	&	RV	&	-371	&	-0.014	&	[4]	\\
2441356.734	$\pm$	0.040	&	1972.1	&	RV	&	-364	&	0.009	&	[5]	\\
2441457.331	$\pm$	0.088	&	1972.4	&	Phot	&	-346	&	-0.192	&	[6]	\\
2441485.587	$\pm$	0.066	&	1972.5	&	RV	&	-341	&	0.065	&	[7]	\\
2441491.052	$\pm$	0.065	&	1972.5	&	Phot	&	-340	&	-0.070	&	[8]	\\
2441675.957	$\pm$	0.065	&	1973.0	&	RV	&	-307	&	0.040	&	[9]	\\
2441698.461	$\pm$	0.095	&	1973.0	&	Phot	&	-303	&	0.145	&	[10]	\\
2441871.986	$\pm$	0.057	&	1973.5	&	RV	&	-272	&	0.075	&	[11]	\\
2441888.691	$\pm$	0.065	&	1973.6	&	Phot	&	-269	&	-0.020	&	[6]	\\
2441944.636	$\pm$	0.048	&	1973.7	&	Phot	&	-259	&	-0.073	&	[12]	\\
2441944.698	$\pm$	0.041	&	1973.7	&	RV	&	-259	&	-0.011	&	[5]	\\
2442297.545	$\pm$	0.063	&	1974.7	&	Phot	&	-196	&	0.045	&	[13]	\\
2442314.416	$\pm$	0.056	&	1974.7	&	Phot	&	-193	&	0.117	&	[6]	\\
2442353.516	$\pm$	0.036	&	1974.8	&	RV	&	-186	&	0.018	&	[14]	\\
2442622.245	$\pm$	0.050	&	1975.6	&	Phot	&	-138	&	-0.046	&	[15]	\\
2442678.271	$\pm$	0.060	&	1975.7	&	Phot	&	-128	&	-0.018	&	[6]	\\
2442913.596	$\pm$	0.067	&	1976.4	&	Phot	&	-86	&	0.113	&	[16]	\\
2443047.959	$\pm$	0.063	&	1976.7	&	Phot	&	-62	&	0.080	&	[6]	\\
2443053.481	$\pm$	0.036	&	1976.8	&	RV	&	-61	&	0.002	&	[5]	\\
2443394.932	$\pm$	0.048	&	1977.7	&	Phot	&	0	&	-0.138	&	[15]	\\
2443770.299	$\pm$	0.048	&	1978.7	&	Phot	&	67	&	0.039	&	[16]	\\
2444145.375	$\pm$	0.042	&	1979.7	&	RV	&	134	&	-0.075	&	[5]	\\
2444878.866	$\pm$	0.067	&	1981.8	&	Phot	&	265	&	-0.164	&	[15]	\\
2445035.779	$\pm$	0.042	&	1982.2	&	RV	&	293	&	-0.046	&	[17]	\\
2445242.976	$\pm$	0.079	&	1982.7	&	Phot	&	330	&	-0.044	&	[16]	\\
2445713.420	$\pm$	0.018	&	1984.0	&	RV	&	414	&	0.013	&	[17]	\\
2446525.383	$\pm$	0.064	&	1986.3	&	RV	&	559	&	-0.002	&	[18]	\\
2447247.862	$\pm$	0.065	&	1988.2	&	Phot	&	688	&	0.097	&	[19]	\\
2448423.715	$\pm$	0.059	&	1991.5	&	Phot	&	898	&	-0.019	&	[20]	\\
2449538.018	$\pm$	0.073	&	1994.5	&	RV	&	1097	&	-0.085	&	[21]	\\
2449958.143	$\pm$	0.041	&	1995.7	&	Phot	&	1172	&	0.051	&	[22]	\\
2450238.059	$\pm$	0.052	&	1996.4	&	RV	&	1222	&	-0.025	&	[23]	\\
2450322.070	$\pm$	0.032	&	1996.7	&	Phot	&	1237	&	-0.012	&	[22]	\\
2450652.512	$\pm$	0.042	&	1997.6	&	RV	&	1296	&	0.039	&	[24]	\\
2450669.273	$\pm$	0.038	&	1997.6	&	Phot	&	1299	&	0.000	&	[22]	\\
2451033.243	$\pm$	0.037	&	1998.6	&	Phot	&	1364	&	-0.020	&	[22]	\\
2451733.249	$\pm$	0.027	&	2000.5	&	RV	&	1489	&	0.005	&	[25]	\\
2457014.031	$\pm$	0.171	&	2015.0	&	Phot	&	2432	&	0.131	&	[26]	\\
2457904.308	$\pm$	0.058	&	2017.4	&	Phot	&	2591	&	0.031	&	[27]	\\
2458693.932	$\pm$	0.067	&	2019.6	&	Phot	&	2732	&	0.077	&	[28]	\\
2459785.802	$\pm$	0.067	&	2022.6	&	Phot	&	2927	&	-0.023	&	[29]	\\
2459813.892	$\pm$	0.067	&	2022.6	&	Phot	&	2932	&	0.067	&	[30]	\\
\enddata
\begin{flushleft}	
$^a$ $O-C$ is in units of days, with the fiducial linear ephemeris with a period of 5.599848 days and an epoch of HJD 2443395.070.\\
$^b$ References:~~
~~{\bf [1]}~	Seyfert \& Popper (1941)
~~{\bf [2]}~	Khalliullin (1975)
~~{\bf [3]}~	Webster \& Muirden (1972)
~~{\bf [4]}~	Brucato \& Kristian (1973)
~~{\bf [5]}~	Gies \& Bolton (1982)
~~{\bf [6]}~	Walker \& Quintanilla (1978)
~~{\bf [7]}~	Smith et al. (1973)
~~{\bf [8]}~	Lyutyi et al. (1973)
~~{\bf [9]}~	Mason et al. (1974)
~~{\bf [10]}~	Hilditch \& Hill (1974)
~~{\bf [11]}~	Brucato \& Zappala (1974)
~~{\bf [12]}~	Cherepashchuk et al. (1974)
~~{\bf [13]}~	Lester et al. (1976)
~~{\bf [14]}~	Abt et al.  (1977)
~~{\bf [15]}~	Lyutyi (1985)
~~{\bf [16]}~	Kemp et al. (1987)
~~{\bf [17]}~	Ninkov et al. (1987)
~~{\bf [18]}~	Sowers et al.  (1998)
~~{\bf [19]}~	Rossiger \& Luthardt (1988)
~~{\bf [20]}~	{\it Hipparcos}
~~{\bf [21]}~	Canalizo et al. (1995)
~~{\bf [22]}~	Karitskaya et al. (2001)
~~{\bf [23]}~	LaSala et al. (1998)
~~{\bf [24]}~	Brocksopp et al. (1999)
~~{\bf [25]}~	Gies et al. (2003)
~~{\bf [26]}~	AAVSOnet
~~{\bf [27]}~	Kamogata/Kiso/Kyoto Wide-field Survey \url{http://kws.cetus-net.org/~maehara/VSdata.py?object=V1357+Cyg&plot=1}
~~{\bf [28]}~	{\it TESS} 14
~~{\bf [29]}~	{\it TESS} 54
~~{\bf [30]}~	{\it TESS} 55  \\
\end{flushleft}	
	
\end{deluxetable}



\subsection{GP Vel = Vela X-1}

Vela X-1 is a well-studied bright HMXB (Kretschmar et al. 2021 and references therein).  The primary star is an X-ray pulsar, with a spin period of around 283 seconds, although this suffers unexplained variations at larger than the parts-per-thousand level on long and short time-scales.  In the X-ray regime, the system displays deep total eclipses with a period 8.964 days.  The companion star is a B0.5 {\rm I}ae supergiant star, emitting a substantial stellar wind that provides the material for accretion on to the NS.

GP Vel, the optical counterpart of Vela X-1, shows low amplitude modulations over the range 6.76$<$$V$$<$6.99 on the orbital period.  This can provide an opportunity for tracking the orbit with old and recent light curves.  The best representation of the light curve on time-scales of less than a month is with the {\it TESS} data from Sectors 8, 9, and 35.  Unfortunately, by eye, I cannot see any signature of any periodicity, orbital or otherwise, rather, light curve peaks are apparently chaotically separated by 2--14 days.  A Fourier transform of all the {\it TESS} data shows a peak at near the half-period, but a fold on the half-period shows a sloppy sinewave where the minima spread over half a cycle.  I could formally fit an ellipsoidal light curve to the {\it TESS} light curve, but the uncertainty in the time of minimum is near 2 days, and is completely useless for $O-C$ purposes.  Similarly, the many light curves reported in the literature and sky surveys are not useful.  In all, I have found only one light curve source that is adequate, and that is the ASAS database with 747 magnitudes from 2001--2009, showing a phase-averaged light curve with a distinct light curve shape (substantially different from any ellipsoidal modulation).  But with only one useable light curve, I cannot construct an $O-C$ curve of optical times of minimum light.

Vela X-1 has X-ray eclipses that can be used to track the changes of $P$.  Falanga et al. (2015) has collected 24 X-ray eclipse times from the years 1972 to 2009, for construction of an $O-C$ curve.  They find no significant curvature.  Formally, their measure is that $\dot{P}$ is ($-$2.4$\pm$7.4)$\times$10$^{-8}$.  The one-sigma range covers both positive and negative values, but the error bars are small enough that there might be some utility is carrying the measure in Table 1 and Fig. 1.

\subsection{EXO 1722-363}

EXO 1722-363 is an X-ray pulsar with a spin period of 413.8 seconds, in orbit around a high mass (13.6 M$_{\odot}$) supergiant star of spectral class B0--B1 {\rm I}a (Mason et al. 2010).  The star is an eclipsing binary with $P$=9.740 days.  Falanga et al. (2015) collected 11 eclipse times from 1998--2006.  They fitted their $O-C$ curve to a parabola, finding $\dot{P}$ equal to ($-$5.6$\pm$3.7)$\times$10$^{-7}$.  The uncertainty is large due to having only an 8 year interval.  The $\dot{P}$ is not far from zero, and its error bar is sufficiently large that it is unclear whether this measure is useful.

\subsection{OAO 1657-415}

OAO 1657-415 is an X-ray binary with the primary being a wind-fed X-ray pulsar with a 37 second spin period (Jenke et al. 2012 and references therein).  The companion star is an Ofpe/WN9 supergiant, with a mass near 16 M$_{\odot}$.  The orbital period is 10.447 days, and the system displays eclipses lasting around 1.7 days.  The system has a Galactic latitude of $+$0.32$\degr$, so the extinction is high, and the counterpart has only been barely visible in the infrared.  Falanga et al. (2015) has collected 22 X-ray eclipse times from the literature and added four new times.  The $O-C$ curve from 1991--2011 shows a good parabola, concave down.  Their derived $\dot{P}$ is ($-$9.73$\pm$0.43)$\times$10$^{-8}$.

\section{COLLECTING $\dot{P}$ MEASURES}

\begin{figure*}
	\includegraphics[width=2.0\columnwidth]{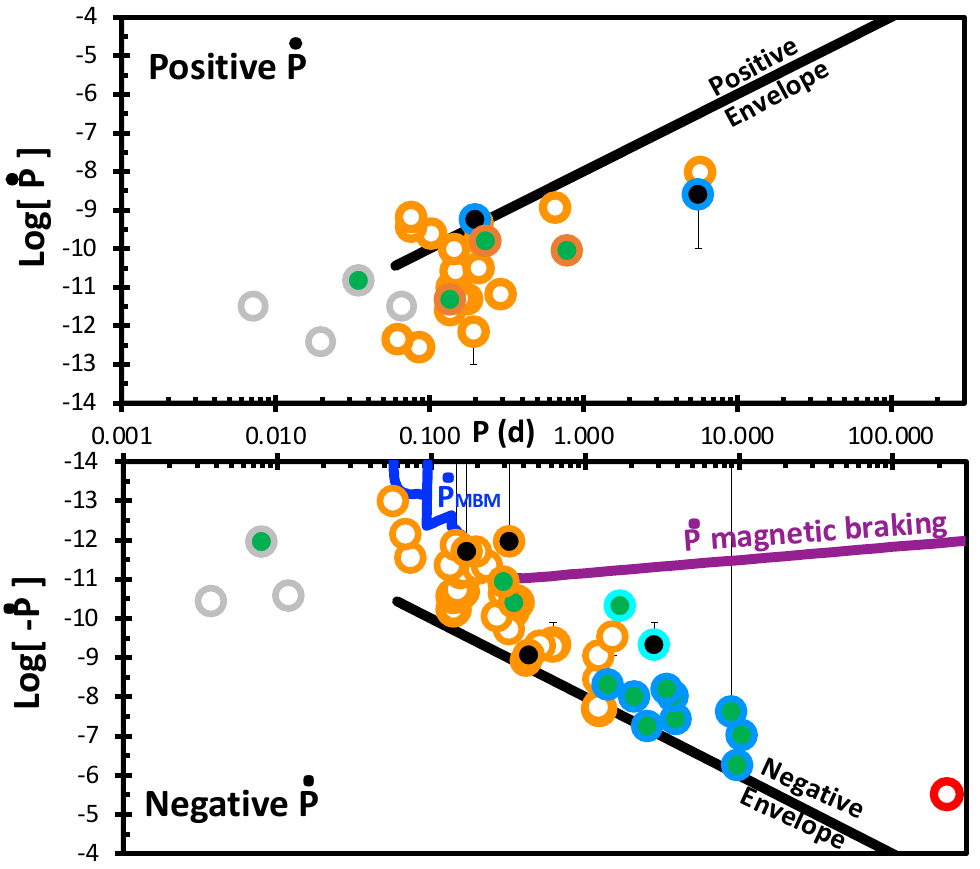}
    \caption{$\dot{P}$ versus $P$ for XRBs and CVs.  This figure shows the reality of binary evolution.  This plot provides the test of model prediction for the driver of the binary evolution.  This plot includes the 23 XRBs and the 49 CVs that have parabolic $\dot{P}$ measures, as a quick pictorial overview of the situation.  The details of this plot are the same as for Figure 1.  The primary take-aways from looking at this plot are: {\bf (1)}~The $\dot{P}$ values are mostly inside an envelope defined by $\dot{P} = \pm (P/10000)^2$, with $P$ measured in units of days (shown as the two black lines for the Positive Envelope and the Negative Envelope).  {\bf (2)}~The HMXBs with NSs (the symbols with a green core and a blue outer ring) are all strongly clustered just inside the Negative Envelope.  This suggests that those systems have their real dominant AML mechanism being proportional to $-P^2$.  {\bf (3)}~Other than the NS/HMXBs, the $\dot{P}$ values have no systematic effects based on the nature of the compact objects or the nature of the companion star.  That is, the distribution of $\dot{P}$ in the plot is the same for black hole, neutron star, and white dwarf primaries.  The NS HMXBs are separate due to their wind accretion, rather than RLOF accretion.  Similarly, the binaries with highly-magnetized primary stars have the same distribution as the other stars.  {\bf (4)}~Roughly one-third of the XRBs and CVs have {\it positive} $\dot{P}$.  This is impossible with MBM.  {\bf (5)}~The MBM predicts a single unique evolutionary track for all interacting binaries, with the period change $\dot{P}_{\rm MBM}$ for $P$$<$6 hours depicted in the blue curve, and the $\dot{P}$ from magnetic braking alone depicted by the purple line.  Even for the negative-$\dot{P}$ XRB systems, all-but-one have greatly larger sizes (more negative), while 13-out-of-16 are more than a factor of 10$\times$ in deviation from the MBM predictions.  {\bf (6)}~The primary point from this figure is that the observed measures are almost always greatly in disagreement with the MBM predictions.  The essence of MBM is its `recipe' for AML, yet this `recipe' is wrong by orders of magnitude, so no calculation of evolution with the MBM can be used. }
\end{figure*}

In the previous Section, I have measured, collected, improved, and updated the $\dot{P}$ for 25 XRBs.  These are tabulated in Table 1, ordered by $P$.  These are also displayed in Figure 21.  For comparison and generality, Figure 21 has added in my measures of 52 CVs (Schaefer 2024).  (This figure plots 10 points for 3 recurrent novae with measures over multiple inter-eruption intervals.  Not plotted are any points for the 3 CVs and 2 XRBs with large non-parabolic variations.)  The XRBs are distinguished by the circles with a black central area for XRBs with black holes, and with a green central region for XRBs with neutron stars.  CVs are the represented by the symbols with a white central area for white dwarfs.  This plot covers a range of a factor of 61,000$\times$ in $P$, so the horizontal axis can only be usefully made with a logarithmic scale.  The vertical axis for $\dot{P}$ also covers a range of over 7 orders-of-magnitude for the positive $\dot{P}$ values, and covers a range of over 9 orders-of-magnitude for the negative $\dot{P}$ values.  With both positive and negative $\dot{P}$ values, a simple logarithmic plot is not possible.  My best solution for giving a single overview plot is to present a stacked pair of plots, with the top panel for the positive values and the bottom plot for the negative values.  With this, each panel can be a logarithmic plot so as to cover the full range of measures without crowding.  In such a plot, the zone between the panels is for the near-zero measures.

My overview plot of $\dot{P}$ versus $P$ can also report on various model predictions.  In particular, the revised-MBM prediction from K2011 are presented as a blue curve for $P$$<$6 hours.  The revised-MBM predictions for the magnetic braking mechanism alone are depicted as a purple curve along the top of the bottom panel.  For the $P$$>$6 hour systems, this purple curve should be close to the total predicted $\dot{P}_{\rm MBM}$ for the revised-MBM, because the magnetic braking effects are thought to dominate over all other mechanisms.  So the blue curve and the purple curve represent the revised-MBM prediction, with the testing of these being one of the primary purpose of this paper.  

One of the striking properties of Fig. 21 is that the systems appear to be contained inside an envelope.  The Positive Envelope is displayed as a thick black slanted line in the upper half of Fig. 21, while the Negative Envelope is a mirror-image line in the lower half of the figure.  There appears to be a fairly `hard' edge for the $\dot{P}$ distribution of my 25+52 systems.  The width of the envelope appears to be increasing with $P$ as the square of the period.  To be specific, the extremes of the $\dot{P}$ range are $\pm$$(P/10000{\rm d})^2$.  I have 72 systems going into these limits, and the limits are fairly sharp with a strong dependence on $P$, and these suggest that the dominant AML mechanism has its strength varying as something like $P^2$.

Figure 21 also displays a lack of systems with small values of $\vert$$\dot{P}$$\vert$ (i.e., near zero close to the horizontal axis in the center between the halves of the plot).  This effect is purely an illusion due to my use of logarithmic scales on the vertical axes.  For a uniform $\dot{P}$ distribution between the envelopes, the number of systems close to the axis must be relatively small because the range of $\dot{P}$ is very small.  To illustrate with a specific case, for systems with $P$ near 1.0 days, the uniform distribution will be from $-$10$^{-8}$ to $+$10$^{-8}$.  In this case, 90 per cent of the systems will appear within 1.0 of the envelope, with this being a small distance on the plot.  So most systems will cluster close to the envelope lines.  Only 0.1 per cent of the systems will have $\vert$$\dot{P}$$\vert$$<$10$^{-11}$, which is to say that the center of the graph will have few if any systems.  This result applies to all the $P$ values, so the conclusion is that the distribution of $\dot{P}$ for a given $P$ does not have a minimum for small values, but rather that the distribution is fairly uniform between the envelopes.

Figure 21 shows us that the HMXBs with NSs (the symbols with the blue outer circle and the green center) are strongly clustered near the negative envelope, with no exception.  This implies that the dominant AML mechanism for the NS HMXBs has a strength proportional to $P^2$.  It appears that the HMXBs all have a simple and single mechanism that controls their orbital evolution.

Figure 21 allows us to see the $\dot{P}$ distribution for systems whose compact star is a BH (symbols with black centers), a NS (green centers), and a WD (white centers).  Excepting the HMXBs with NSs, all of the $\dot{P}$ distributions are the same for the BH systems, the NS systems, and the WD systems.  That is, the WD/NS/BH systems are not all positive-or-negative, not clumped anywhere in the diagram, and are not systematically offset from other systems.  Rather, the BH and NS distribution in the figure is the same as for WDs, following along the strips just inside the positive and negative envelopes.  The distribution of $\dot{P}$ for a given period range is indistinguishable for BH, NS, and WD systems.  This implies that the dominant AML mechanism does not depend on the nature of the compact star.

Does the unknown mechanism controlling the $\dot{P}$ depend on the magnetic fields of the compact stars?  The XRBs with strong magnetic fields are identified in Table 1, and shown with square symbols in Figure 1.  Strikingly, all the HMXBs are X-ray pulsars (with high magnetic fields) that are in a small clump.  This isolated clump might be due to the high-$M_{\rm prim}$ and the necessarily long-$P$, or it might be due to the magnetic field.  We only have three other high-field systems in Figure 1 (V691 CrA, AX J1745.6-2901, and Her X-1).  These three magnetic-systems have a similar distribution in Figure 1 as do the non-magnetic systems.  Similarly, the magnetic CVs have a distribution indistinguishable from the non-magnetic CVs.  So, other than whatever is going on with HMXBs, the $\dot{P}$ does not appear to have anything to do with the magnetic fields on the compact stars. 

Figure 21 shows the reality of how the population of XRBs and CVs evolve.  That is, the evolution is measured and driven by $\dot{P}$, and Fig. 21 shows us the actuality of evolution.

\section{TESTING THE PREDICTIONS OF MBM}

A primary task of this paper is to test the MBM predictions with the XRB systems.  The MBM prediction of $\dot{P}_{\rm mb}$ is its most fundamental prediction because the period changes are what drives and measures the evolution.  This one core prediction of a single evolutionary track for all systems is a strong requirement of MBM.  For XRBs and CVs, this core prediction has been untested until the present work.  Further, the MBM has {\it indirect} predictions for $P_{\rm min}$, $P_{\rm gap-}$, $P_{\rm gap+}$, and $\dot{M}$ as a function of P, and these should be given a modern analysis.  Here, I will use the XRBs to test these predictions, and I will add in the results from my 52 CVs.

The MBM was originally formulated to explain the mass transfer in XRBs (Rappaport et al. 1983).  However, there is no expectation that the magnetic braking mechanism or the MBM AML `recipe' are applicable to the XRBs that have WD companions or high-mass companion stars.  For the systems with WD companions, there will be no stellar wind and no reason to think that the real AML will follow either the MBM presumption or the same relation as systems with main sequence companions.  For the eleven systems with high-mass companions, the large binary separations, the massive winds, and the short lifetimes point to the magnetic braking mechanism as likely being negligibly small.  I am unclear as to whether the two IMXB systems should be thought to obey the MBM AML `recipe'.  For the XRBs used in this paper, this leaves only the ten XRB systems with lower-mass main-sequence companions (for periods from 0.08 to 0.8 days) that can be used to test the MBM.

\subsection{Testing the core prediction of $\dot{P}$}

K2011 states repeatedly and emphatically that ``Theoretically, all CVs with initially unevolved donors are expected to quickly join onto a {\it unique} evolution track, whose properties are determined solely by the mechanism for AML from the system (Paczynski \& Sienkiewicz 1983; Ritter \& Kolb 1992; Kolb 1993; Stehle et al. 1996).''  So we have a definite and unique prediction to test.

For the ten XRBs with MS companions, the majority violate the MBM predictions:  {\bf 1.} XTE J1710-281, V691 CrA, and Sco X-1 have positive $\dot{P}$, with this being impossible for MBM.  {\bf 2.} GU Mus has an observed $\dot{P}$ that is 90$\times$ more-negative than predicted by MBM.  AX J1745.6-2901 has its observed value 4.6$\pm$0.1 times the MBM prediction.  {\bf 3.} V4580 Sgr and UY Vol have fast changes in the $O-C$ diagram, for both positive and negative $\dot{P}$, with such being impossible for the MBM.  

Three of the ten XRBs have $\dot{P}$ possibly consistent with the MBM predictions.  Two of these cases (KV UMa and V616 Mon) have agreement only because their error bars are large.  Perhaps a better evaluation is to acknowledge that these two systems are not a useful test of the MBM predictions due to their large measurement uncertainties.  With this, only one (V2134 Oph) out of eight XRBs can be said to obey the MBM predictions.  The MBM is refuted for XRBs with main-sequence companion stars because $\frac{7}{8}$ have their measured $\dot{P}$ greatly different from the theory predictions.

The same conclusion is reached, with much the same details, for the 52 CVs reported in Schaefer (2024).  With this, the MBM predictions have a grand failure, usually by orders-of-magnitude, for its one most fundamental property.

\subsection{Testing the indirect prediction of $P_{\rm min}$}

Historically, the MBM prediction of the minimum orbital period for binaries with non-degenerate companions had been a primary reason for broad acceptance of the theory.  The `standard-MBM' predicts $P_{\rm min}$ to be 72 minutes (K2011).  But a `revised-MBM' models the AML by increasing the effect of gravitational radiation by a factor of 2.47$\pm$0.22, with the explicit motivation so as to predict a minimum period of 82.4 minutes (K2011).  

These predictions can be compared to the observed $P_{\rm min}$ of 82.4$\pm$0.7 minutes for CVs (G\"{a}nsicke et al. 2009).  The period distribution of XRBs can be seen from the $P$ values catalogued in Liu, van Paradijs, \& van den Heuvel (2007).  For systems with non-degenerate companion stars, HETE J1900.1-2455 has the shortest period at 83.4 minutes.  The XRB distribution around $P_{\rm min}$ is greatly sparser than for CVs.  Nevertheless, it appears that the XRBs share the same minimum period as do CVs.  The XRB+CV minimum period is thus 82.4$\pm$0.7 minutes.

If we are testing the standard-MBM, then the prediction (72 minutes) greatly disagrees with the observations (82.4$\pm$0.7 minutes).  If we are testing the revised-MBM, there is no $P_{\rm min}$ prediction because the AML law was adjusted by-hand to match the observed value.  So in all cases, there is no successful MBM prediction.

\subsection{Testing the indirect prediction of the Period Gap}

Historically, the MBM prediction of the range of the Period Gap (from $P_{\rm gap-}$ to $P_{\rm gap+}$) is perceived as a success, and a substantial reason to accept the model.  K2011 analyzed a group of CVs with superhumps to observe a Gap from 2.15$\pm$0.03 to 3.18$\pm$0.04 hours.  The XRB $P$ distribution is sparse around the period Gap, but there are only two known systems from 2.1--3.2 hours (Liu et al. 2007), and is consistent with the CV Gap.

The standard-MBM predicts a Gap from 2.24-to-3.52 hours (K2011).  With this, the standard-MBM prediction is a failure.  The revised-MBM had the AML adjusted by-hand to best match the observations, for a Gap from 2.24--3.24 hours (K2011).  So the revised-MBM did not make a {\it prediction} of the Gap.  In all cases, there is no successful MBM prediction.

The situation is actually much worse for the MBM, because the theory requires that XRBs and CVs of all types have the exact same Period Gap.  This requirement is contradicted by the various types of systems having greatly different Gaps (Schaefer 2022a, 2023).  The Gap for novae is from 1.70--2.66 hours, the Gap for nova-like CVs is from 1.94--3.14 hours, the Gap for superhump CVs is 2.15--3.18 hours, the Gap for dwarf novae is from 2.22--3.38 hours, the Gap for the low-luminosity Sloan CVs is 2.45--3.18 hours, while polars do not even have any Period Gap.  That is, these subsets all have greatly different Gaps, with some systems have little overlap in the Gap with other systems.  So  $P_{\rm gap-}$ to $P_{\rm gap+}$ vary by large factors.  This is a simple and sure refutation of the MBM requirement that all CVs and XRBs follow a single unique evolutionary track.

\subsection{Testing the indirect prediction of the accretion rate}

MBM predicts that all CVs and XRBs must have a single unique $\dot{M}$ as a function of $P$.  Contrarily, since the 1980s, it is well-known that systems with any given $P$ have a range of accretion that spans several orders of magnitude (Patterson 1984; Warner 1987; K2011).  The frequently-invoked reconciliation is to speculate that $\dot{M}$ for every single system varies by orders-of-magnitude on timescales of millennia, or longer, such that the time-average on million-year timescales somehow returns the MBM prediction.  This attempt at reconciliation has the problems that the physical mechanism is unknown, the effect has never been observed, and there is no reason to expect that the $\dot{M}$ variations will average to the MBM value.  Further, with the accretion rate varying by orders-of-magnitude, any magnetic braking effect is negligibly small, and evolution is dominated by un-modeled mechanisms.  It does not matter what the excuse is, the bottom line is that the MBM prediction fails.

From all of Section 4, we see that the MBM predictions all fail, usually by orders-of-magnitude.  This is true for the XRBs, the CVs, and for both taken together.  This is true for both the direct and critical prediction as well as for the historically-important indirect predictions.  With these broad and deep failures, we can only conclude that the MBM is refuted.

\section{ACCOUNTING FOR MASS TRANSFER AND GR}

The physics of period changes is standard, see, for example Frank, King, \& Raine (2002).  The observed $\dot{P}$ measures (see Table 1) have contribution from three mechanisms, General Relativity gravitational wave emission (`GR'), ordinary mass transfer by the companion star (`mt'), and by other angular momentum losses from the orbit (`AML').  These effect are additive, with
\begin{equation}
\dot{P} = \dot{P}_{\rm AML} + \dot{P}_{\rm GR} + \dot{P}_{\rm mt}.
\end{equation}
Both $\dot{P}_{\rm GR}$ and $\dot{P}_{\rm mt}$ are surely known, or at least to within the measurement uncertainties for the stellar masses and mass-loss rates.  Within the MBM, the critical assumption was that $\dot{P}_{\rm AML}$=$\dot{P}_{\rm mb}$ is a particular power law in $P$, but the MBM has now been thoroughly refuted.  So now, the primary issue for the large fields of XRB and CV evolution and demographics is entirely on understanding the true function for $\dot{P}_{\rm AML}$.  Thus, an important task is to correct the observed $\dot{P}$ for the effects of GR and mass transfers.  That is, the AML should be isolated by taking out the GR and mass transfer contributions, 
\begin{equation}
\dot{P}_{\rm AML} = \dot{P} - \dot{P}_{\rm GR} - \dot{P}_{\rm mt}.
\end{equation}
The terms on the right-hand-side of Eq. 2 are known for each of my 25 XRBs and 52 CVs, so the effects of the AML alone can be calculated.

$\dot{P}_{\rm AML}$ is simply the period change remaining after the known GR and mass transfer effects have been taken out.  By taking out the variable and confounding effects of $\dot{P}_{\rm GR}$ and $\dot{P}_{\rm mt}$, we are left with the unknown mechanism(s), isolated and visible.  $\dot{P}_{\rm AML}$ represents the unknown effects that are dominating binary evolution.  $\dot{P}_{\rm AML}$ has long been equated to the magnetic braking prediction, but this is refuted.  Perhaps $\dot{P}_{\rm AML}$ is produced by a mechanism that transfers the spin angular momentum of the compact star to the disk and then to the orbit.  Perhaps $10^{42.5}$ erg Superflares (like on V2487 Oph) are stealing energy from the orbit.  $\dot{P}_{\rm AML}$ might arise from multiple mechanisms, say, effects of mass lost by the binary plus accelerations from a planet orbiting the binary.  Nobody knows the mechanism(s) that create $\dot{P}_{\rm AML}$.  The purpose of this paper is to characterize and quantify its relation to the binary properties so as to hopefully allow a recognition of the unknown mechanism(s).  Even if the mechanism is not named or recognized, the empirical relations to system properties will allow a useful quantification and parameterization that will serve the needs of theorists calculating models, population synthesis, and demographics. 

To get equations for $\dot{P}_{\rm GR}$ and $\dot{P}_{\rm mt}$, we should start with the definition of the orbital angular momentum as
\begin{equation}
J = M_{\rm prim} M_{\rm comp} \sqrt{\frac{Ga}{M_{\rm total}}}.
\end{equation}
Here, subscripts `prim' refers to the accreting primary star (either a WD, NS, or BH), `comp' refers to the donor companion star, and `total' refers to the both stars together, so $M_{\rm total}$=$M_{\rm prim}$$+$$M_{\rm comp}$.  The usual mass ratio is $q$=$M_{\rm comp}$/$M_{\rm prim}$.  The binary semi-major axis, $a$, is taken from Kepler's Law,
\begin{equation}
a^3 = (G/4\pi^2) M_{\rm total} P^2.
\end{equation}
Equations 3 and 4 can be combined by eliminating $a$, the natural log of the equation can then be easily differentiated with respect to time to get the basic relation between the time derivatives of $P$, $J$, and the stellar masses,
\begin{equation}
\frac{\dot{P}}{P} = 3\frac{\dot{J}_{\rm AML}}{J} + 3\frac{\dot{J}_{\rm GR}}{J} + \left( \frac{\dot{M}_{\rm total}}{M_{\rm total}} - 3\frac{\dot{M}_{\rm prim}}{M_{\rm prim}} - 3\frac{\dot{M}_{\rm comp}}{M_{\rm comp}} \right)
\end{equation}
Here, the rate of change of the orbital angular momentum has been divided into the AML and GR components, so $\dot{J}=\dot{J}_{\rm AML}+\dot{J}_{\rm GR}$.  Now we can identify the terms in Eq. 1,
\begin{equation}
\dot{P}_{\rm AML} = 3P\frac{\dot{J}_{\rm AML}}{J}, 
\end{equation}
\begin{equation}
\dot{P}_{\rm GR} = 3P\frac{\dot{J}_{\rm GR}}{J}, 
\end{equation}
\begin{equation}
\dot{P}_{\rm mt} = P \left(\frac{\dot{M}_{\rm total}}{M_{\rm total}} - 3\frac{\dot{M}_{\rm prim}}{M_{\rm prim}} - 3\frac{\dot{M}_{\rm comp}}{M_{\rm comp}} \right).
\end{equation}
So the task then comes to evaluating $\dot{P}_{\rm GR}$ and $\dot{P}_{\rm mt}$, so as to calculate $\dot{P}_{\rm AML}$ from Eq. 2 for many systems, then looking for correlations that can point to the real dominant AML law.

The GR contribution is known exactly, for the given masses and periods, with
\begin{equation}
\dot{P}_{GR} = \frac{-96}{5} (2\pi)^{\frac{8}{3}} \frac{G^{\frac{5}{3}}}{c^5} M_{\rm prim} M_{\rm comp} (M_{\rm prim} + M_{\rm comp})^{\frac{-1}{3}} P^{\frac{-5}{3}}.
\end{equation}
This contribution is always small, see the $G^{\frac{5}{3}}$ in the numerator and the $c^5$ in the denominator.  This contribution is always negligibly small except in the case for systems below the Period Gap.

In practice, $\dot{P}_{\rm mt}$ will be one of two simple cases.  The first case is Roche lobe overflow (RLOF), for which conservative mass transfer is close to the real situation.  The mass accretion rate is usually taken to be a positive quantity labeled like $\dot{M}_{\rm RLOF}$.  With this, $\dot{M}_{\rm comp}$=$-$$\dot{M}_{\rm RLOF}$, $\dot{M}_{\rm prim}$ = $\dot{M}_{\rm RLOF}$, $\dot{M}_{\rm total}$=0, and 
\begin{equation}
\dot{P}_{\rm mt} = 3P(1-q)\frac{\dot{M}_{\rm RLOF}}{M_{\rm comp}}. 
\end{equation}
For essentially all RLOF systems, $q<1$, this means that the usual XRB or CV with RLOF will have $\dot{P}_{\rm mt}$ positive, which is to say that mass transfer will {\it increase} $P$ and the binary separation.

The second case is for wind accretion, when the companion star is ejecting a substantial stellar wind, of which only a fraction ($\epsilon$) is intercepted and accreted onto the primary star.  The remainder escapes from the binary and is lost to the system.  I will notate the strength of the wind as the positive quantity $\dot{M}_{\rm wind}$.  With this, $\dot{M}_{\rm comp}$ = $-$$\dot{M}_{\rm wind}$, $\dot{M}_{\rm prim}$ = $\epsilon$$\dot{M}_{\rm wind}$, and $\dot{M}_{\rm total}$ = ($\epsilon$-1)$\dot{M}_{\rm wind}$.  Then we get
\begin{equation}
\dot{P}_{\rm mt} = P\frac{\dot{M}_{\rm wind}}{M_{\rm comp}}  \left( \frac{3+2q-2\epsilon q -3\epsilon q^2}{1+q} \right). 
\end{equation}
This reduces to Eq. 10 for the case with $\epsilon$=1.  For the relevant cases with wind accretion, i.e., the HMXBs, with modestly small $\epsilon$ and $q$$>$2 or so, the factor in parentheses can be either positive or negative.  With $\epsilon$ being only poorly known from theory, the $\dot{P}_{\rm mt}$ for HMXBs will have substantial uncertainties.

So now we have a prescription for calculating $\dot{P}_{\rm AML}$ from the observed $P$, $\dot{P}$, $M_{\rm prim}$, $M_{\rm comp}$, either $\dot{M}_{\rm RLOF}$ or $\dot{M}_{\rm wind}$, and possibly an $\epsilon$ from theory.  This would be to use Equations 2, 9, and either 10 or 11.

To produce these calculated $\dot{P}_{\rm AML}$, I have assembled the system properties into Table 6.  The individual values are mostly taken from references already cited for the individual sources, while others are taken as judicious averages from several papers, while others are my estimates based on the typical values for similar sources.  These input measures often have substantial uncertainties, and the asymmetric limits make for skewed error bars.  For the acceptable ranges of the input parameters, I have made an evaluation of the acceptable range of $\dot{P}_{\rm AML}$, as shown in the last column of Table 6.

The $P$ measures are always known with high accuracy and confidence for the 77 systems.  The sources for the XRBs are the cited papers, while the sources for the CVs are in Schaefer (2023, 2024) and citations therein.  The $\dot{P}$ measures are often known with exquisite accuracy following perfect parabolas, and are sometimes known with only poor accuracy.  The sources and calculations for each of the 77 systems are extensively given for each individual XRB in this paper and for each individual CV in Schaefer (2024).  Note that the reported $\dot{P}$ measures for DQ Her, V394 CrA, IM Nor, and U Sco were previously published with units days/cycle, and these are here expressed in dimensionless units of days/day or s/s.  For several of the CVs, I report small updates based on recent eclipse timings.  For the systems with their $O-C$ curves greatly departing from any parabola (OY Car, Z Cha, V4580 Sgr, SW Sex, and UY Vol), I will take $\dot{P}$ to be zero with an uncertainty corresponding to the maximum possible curvature hiding in the observed wiggles of the $O-C$ curve.  For the RNe T Pyx, CI Aql, U Sco, and T CrB, I calculate $\dot{P}_{\rm AML}$ for each of their 2--5 inter-eruption intervals with an independent measured $\dot{P}$.

The stellar masses are often known from radial velocity curves, and for some cases this produces accurate masses, while for other (e.g., when the inclination is not known from eclipses and the companion star is not seen) the accuracy is not high.  For some systems, no direct measure is known, so I have estimated the masses based on typical values for the system.  Thus, an unmeasured NS is taken to be 1.4 M$_{\odot}$, RNe will be 1.2--1.4 M$_{\odot}$, and D- and J-class CNe will be 0.75--1.0 M$_{\odot}$.  Fairly accurate estimates of $M_{\rm comp}$ can come from radial velocity curves, from the surface temperatures plotting onto an HR diagram, or from the Roche lobe size derived from $P$.

The accretion rates are often poorly known, with the real uncertainties typically like a factor of ten.  Further, accretion rates often change by large factors, and estimating the appropriate average has large uncertainties.  For RLOF, $\dot{M}_{\rm RLOF}$ usually comes from some model of the system, usually with the critical calculation trying to reproduce the optical or X-ray luminosity of the accretion disk and its boundary layer.  Such measures usually have a distance-squared dependency, so it is important to correct the published estimates to the distances from the latest $Gaia$ measures.  A particularly useful limit is that the dividing line between nova-like CVs and DNe is around $\sim$3$\times$$10^{-9} $M$_{\odot}$ yr$^{-1}$, for stars just above the Period Gap, and this cannot be violated (Dubus, Otulakowska-Hypka, \& Lasota 2018).  It is important to not use a published accretion rate that is based on some model for the observed $\dot{P}$.  

For the HMXBs, $\dot{M}_{\rm wind}$ is usually taken from some canonical theoretical value based on the observed spectral class on the companion.  These wind strengths are $\gtrsim$3$\times$10$^{-7}$ M$_{\odot}$ yr$^{-1}$ (Falanga et al. 2015) or $\gtrsim$10$^{-6}$ M$_{\odot}$ yr$^{-1}$ (Walter et al. 2015, Fornasini, Antoniou, \& Dubus 2023).  For systems that underfill their Roche lobes, the companion's stellar wind can result in accretion on the primary star either through Bondi-Hoyle capture (direct wind accretion) or as disk wind accretion mediated through an accretion disk created from the stellar wind (Fornasini et al. 2023).  Direct wind accretion is very inefficient at transferring mass from the wind to the primary, with typical efficiencies of $\epsilon$ ranging from 0.0001 to 0.01 (Davidson \& Ostriker 1973, Lamers et al. 1976), with this still being adequate to power the X-ray luminosity.  For the relatively low-velocity wind of RedG stars, $\epsilon$ ranges from $\lesssim$0.001 to 0.2, depending on the binary separation (Perets \& Kenyon 2013).  For the high-velocity winds of HMXBs, any formation of a disk around the primary star can only be with the Bondi-Hoyle mechanism, with the efficiencies 0.0001 to 0.01.  There is a further efficiency for what fraction of the material in the wind-formed disk will actually make it to the accreting star.  For this extra efficiency factor, Perets \& Kenyon calculate that from 40--90\% of the disk gas actually makes it down to the WD, with the primary dependency on the binary separation.  For the calculation of $\dot{P}_{\rm mt}$, for HMXBs with accretion from a wind, I will adopt $\epsilon$ values from 0.0001 to 0.01 for large-$P$ to small-$P$ with no disk intermediary, and I will adopt 0.004 to 0.00004 respectively for cases where there is a disk formed as an intermediary for the accretion.

With the properties in Table 6, I can now calculate $\dot{P}_{\rm GR}$, $\dot{P}_{\rm mt}$, and $\dot{P}_{\rm AML}$.  The period change values are cast in dimensionless units of $10^{-12}$, as this makes for the comparisons easy without keeping track of the exponents.  The $\dot{M}$ listings are for either $\dot{M}_{\rm RLOF}$ or $\dot{M}_{\rm wind}$, in units of $10^{-8}$  M$_{\odot}$ yr$^{-1}$.  The $\dot{P}$ entries have the notation ``k'' used to indicate a multiplication by 1000$\times$, solely to keep the column widths manageable.  The $\dot{P}_{\rm AML}$ column includes the calculated value plus my estimated full range of plausible values in parentheses, in dimensionless units of $10^{-12}$. The $\dot{P}_{\rm AML}$ values for each of my 25 XRBs and 52 CVs are displayed in Fig. 22.  This plot has the identical format as in Fig. 1, allowing easy comparisons.

\begin{longrotatetable}
\begin{deluxetable*}{lllllllllllll}
\tablecaption{Measured $\dot{P}_{\rm AML}$ for XRBs and CVs}
\tablewidth{600pt}
\tabletypesize{\scriptsize}
\tablehead{
Star   &   Class$^a$  &  $P$  &  Prim.  &  Comp.  &  Accretion  & $M_{\rm prim}$  &  $M_{\rm prim}$  &  $\dot{M}_{-8}$  &  $\dot{P}$  &  $\dot{P}_{\rm GR}$  &  $\dot{P}_{\rm mt}$ &   $\dot{P}_{\rm AML}$    \\
  &     &  (days)  &    &    &    & ($M_{\odot}$)  &  ($M_{\odot}$)  &  ($10^{-8}$)  &  ($10^{-12}$)  &  ($10^{-12}$)  &  ($10^{-12}$) &   ($10^{-12}$)      
}
\startdata
HM Cnc	&	AM CVn	&	0.00372	&	WD	&	WD	&	RLOF	&	1.0	&	0.17	&	1	&	 $-$36.57 $\pm$ 0.01	&	-39.37	&	4.5	&	 $-$1.7 ($-$21 to $+$17)	\\
V407 Vul	&	AM CVn	&	0.00660	&	WD	&	WD	&	RLOF	&	0.47	&	0.24	&	5	&	 $+$3.17 $\pm$ 0.10	&	-11.87	&	17	&	 $-$1.5 ($-$74 to $+$19)	\\
ES Cet	&	AM CVn	&	0.00717	&	WD	&	WD	&	RLOF	&	0.7	&	0.06	&	5	&	 $+$3.18 $\pm$ 0.11	&	-3.76	&	135	&	 $-$128 ($-$200 to $+$4)	\\
4U 1820-30	&	LMXB	&	0.00793	&	NS	&	WD	&	RLOF	&	1.58	&	0.07	&	1.2	&	 $-$1.131 $\pm$ 0.029	&	-6.47	&	32	&	 $-$27 ($-$140 to $+$5)	\\
AM CVn	&	AM CVn	&	0.0119	&	WD	&	WD	&	RLOF	&	0.71	&	0.13	&	0.71	&	 $-$27 $\pm$ 3	&	-3.43	&	13	&	 $-$37 ($-$42 to $-$27)	\\
YZ LMi	&	AM CVn	&	0.0197	&	WD	&	WD	&	RLOF	&	0.85	&	0.035	&	0.0009	&	 $+$0.39 $\pm$ 0.05	&	-0.47	&	0.12	&	 $+$0.7 ($-$0.5 to $+$0.9)	\\
V1405 Aql	&	LMXB	&	0.0347	&	NS	&	WD	&	RLOF	&	1.40	&	0.07	&	0.076	&	 $+$14.6 $\pm$ 0.3	&	-0.51	&	8.8	&	 $+$6 ($-$5 to $+$14)	\\
WZ Sge	&	DN	&	0.0567	&	WD	&	MS	&	RLOF	&	1.0	&	0.07	&	0.0052	&	 $-$0.102 $\pm$ 0.014	&	-0.18	&	0.97	&	 $-$0.9 ($-$7 to 0)	\\
V2051 Oph	&	DN	&	0.0624	&	WD	&	MS	&	RLOF	&	0.43	&	0.11	&	0.013	&	 $+$0.44 $\pm$ 0.20	&	-0.13	&	1.4	&	 $-$0.8 ($-$1.2 to $+$0.3)	\\
OY Car	&	DN	&	0.0631	&	WD	&	MS	&	RLOF	&	0.88	&	0.093	&	0.0067	&	 0 $\pm$ 3	&	-0.18	&	1.0	&	 $-$0.8 ($-$3.8 to $+$2.2)	\\
EX Hya	&	DN	&	0.0682	&	WD	&	MS	&	RLOF	&	0.79	&	0.11	&	0.0039	&	 $-$0.72 $\pm$ 0.05	&	-0.17	&	0.51	&	 $-$1 ($-$32 to 0)	\\
HT Cas	&	DN	&	0.0736	&	WD	&	MS	&	RLOF	&	0.57	&	0.09	&	0.0017	&	 $-$2.85 $\pm$ 0.18	&	-0.10	&	0.29	&	 $-$3.0 ($-$12 to $-$2.5)	\\
Z Cha	&	DN	&	0.0745	&	WD	&	MS	&	RLOF	&	0.80	&	0.15	&	0.0011	&	 0 $\pm$ 3	&	-0.20	&	0.11	&	 $+$0.1 ($-$2.9 to $+$3.1)	\\
T Pyx	&	RN P(62)	&	0.0762	&	WD	&	MS	&	RLOF	&	1.30	&	0.20	&	10	&	 $+$649 $\pm$ 7	&	-0.36	&	790	&	 $-$145 ($-$550 to $+$72)	\\
T Pyx	&	RN P(62)	&	0.0762	&	WD	&	MS	&	RLOF	&	1.30	&	0.20	&	5	&	 $+$367 $\pm$ 27	&	-0.36	&	400	&	 $-$30 ($-$270 to $+$130)	\\
V4580 Sgr	&	LMXB	&	0.0839	&	NS	&	MS	&	RLOF	&	1.40	&	0.05	&	0.1	&	 0 $\pm$ 2	&	-0.08	&	40	&	 $-$40 ($-$250 to $+$1)	\\
DV UMa	&	DN	&	0.0859	&	WD	&	MS	&	RLOF	&	0.9	&	0.14	&	0.015	&	 $+$0.28 $\pm$ 0.08	&	-0.16	&	1.9	&	 $-$1.5 ($-$9 to 0)	\\
IM Nor	&	RN P(80)	&	0.103	&	WD	&	MS	&	RLOF	&	1.21	&	0.20	&	15	&	 $+$245 $\pm$ 7	&	-0.21	&	1600	&	 $-$1340 ($-$4400 to $-$390)	\\
V482 Cam	&	Nlike (SW)	&	0.134	&	WD	&	MS	&	RLOF	&	0.8	&	0.3	&	0.7	&	 $-$4.5 $\pm$ 1.5	&	-0.15	&	48	&	 $-$52 ($-$100 to $-$19)	\\
SW Sex	&	Nlike (SW)	&	0.135	&	WD	&	MS	&	RLOF	&	0.44	&	0.30	&	1.9	&	 0 $\pm$ 10	&	-0.09	&	67	&	 $-$67 ($-$140 to 0)	\\
DW UMa	&	Nlike (SW)	&	0.137	&	WD	&	MS	&	RLOF	&	0.82	&	0.3	&	2.28	&	 $+$2.49 $\pm$ 0.73	&	-0.14	&	160	&	 $-$160 ($-$210 to $-$120)	\\
XTE J1710-281	&	LMXB	&	0.137	&	NS	&	MS	&	RLOF	&	1.40	&	0.4	&	0.1	&	 $+$4.7 $\pm$ 0.3	&	-0.28	&	6.0	&	 $-$1 ($-$27 to $+$4)	\\
TT Tri	&	Nlike (SW)	&	0.140	&	WD	&	MS	&	RLOF	&	0.8	&	0.27	&	0.4	&	 $-$60.2 $\pm$ 4.3	&	-0.12	&	34	&	 $-$94 ($-$100 to $-$85)	\\
V1315 Aql	&	Nlike (SW)	&	0.140	&	WD	&	MS	&	RLOF	&	0.73	&	0.30	&	0.5	&	 $+$9.74 $\pm$ 0.43	&	-0.13	&	34	&	 $-$24 ($-$44 to $-$10)	\\
V1500 Cyg	&	CN S(4)	&	0.140	&	WD	&	MS	&	RLOF	&	1.15	&	0.20	&	0.2	&	 $-$27 $\pm$ 10	&	-0.12	&	28	&	 $-$55 ($-$70 to $-$41)	\\
RR Pic	&	CN J(122)	&	0.145	&	WD	&	MS	&	RLOF	&	0.95	&	0.40	&	10	&	 $+$95.8 $\pm$ 3.4	&	-0.19	&	520	&	 $-$420 ($-$1300 to $-$150)	\\
PX And	&	Nlike (SW)	&	0.146	&	WD	&	MS	&	RLOF	&	0.73	&	0.3	&	0.4	&	 $-$1.37 $\pm$ 1.37	&	-0.12	&	28	&	 $-$30 ($-$58 to $-$23)	\\
V1024 Cep	&	Nlike (SW)	&	0.149	&	WD	&	MS	&	RLOF	&	0.8	&	0.3	&	0.4	&	 $+$25.6 $\pm$ 6.7	&	-0.12	&	31	&	 $-$4.8 ($-$10 to $+$2.8)	\\
HS 0220+0603	&	Nlike (SW)	&	0.149	&	WD	&	MS	&	RLOF	&	0.80	&	0.3	&	4	&	 $-$18.8 $\pm$ 2.7	&	-0.12	&	310	&	 $-$320 ($-$400 to $-$250)	\\
BP Lyn	&	Nlike (SW)	&	0.153	&	WD	&	MS	&	RLOF	&	0.80	&	0.3	&	0.5	&	 $+$19.6 $\pm$ 3.9	&	-0.12	&	39	&	 $-$19 ($-$51 to $-$3.8)	\\
BH Lyn	&	Nlike (SW)	&	0.156	&	WD	&	MS	&	RLOF	&	0.7	&	0.3	&	0.4	&	 $-$7.073 $\pm$ 0.039	&	-0.10	&	29	&	 $-$36 ($-$65 to $-$25)	\\
IP Peg	&	DN	&	0.158	&	WD	&	MS	&	RLOF	&	1.16	&	0.55	&	0.04	&	 $-$24.2 $\pm$ 1.8	&	-0.25	&	1.5	&	 $-$25 ($-$27 to $-$23)	\\
LX Ser	&	Nlike	&	0.158	&	WD	&	MS	&	RLOF	&	0.60	&	0.30	&	0.4	&	 $+$5.2 $\pm$ 0.3	&	-0.09	&	26	&	 $-$21 ($-$27 to $-$11)	\\
UY Vol	&	LMXB	&	0.159	&	NS	&	MS	&	RLOF	&	2.0	&	0.4	&	0.04	&	 0 $\pm$ 10	&	-0.28	&	3.1	&	 $-$3 ($-$14 to 6)	\\
UU Aqr	&	Nlike	&	0.164	&	WD	&	MS	&	RLOF	&	0.67	&	0.43	&	0.5	&	 $-$20.8 $\pm$ 1.1	&	-0.12	&	17	&	 $-$38 ($-$43 to $-$18)	\\
V1776 Cyg	&	Nlike (SW)	&	0.165	&	WD	&	MS	&	RLOF	&	0.8	&	0.43	&	0.8	&	 $-$6.1 $\pm$ 2.4	&	-0.14	&	35	&	 $-$41 ($-$93 to $-$19)	\\
KV UMa	&	LMXB	&	0.170	&	BH	&	MS	&	RLOF	&	7.30	&	0.18	&	0.00079	&	 $-$2 $\pm$ 11	&	-0.28	&	0.18	&	 $-$1.9 ($-$13 to $+$9)	\\
U Gem	&	DN	&	0.177	&	WD	&	MS	&	RLOF	&	1.17	&	0.44	&	0.07	&	 $+$4.8 $\pm$ 0.7	&	-0.17	&	4.3	&	 $+$0.6 ($-$1.8 to $+$2.7)	\\
DQ Her	&	CN D(100)	&	0.194	&	WD	&	MS	&	RLOF	&	0.80	&	0.50	&	0.2	&	 $+$0.7 $\pm$ 0.6	&	-0.12	&	7.2	&	 $-$6.3 ($-$26 to $+$0.3)	\\
UX UMa	&	Nlike	&	0.197	&	WD	&	MS	&	RLOF	&	0.78	&	0.47	&	1	&	 $-$2 $\pm$ 1	&	-0.11	&	41	&	 $-$43 ($-$67 to $-$24)	\\
Cyg X-3	&	HMXB	&	0.199	&	BH	&	HiMass	&	Wind	&	2.4	&	10	&	1000	&	 $+$563 $\pm$ 2	&	-3.33	&	3500	&	 $-$2900 ($-$6500 to $-$1200)	\\
T Aur	&	CN D(84)	&	0.204	&	WD	&	MS	&	RLOF	&	0.80	&	0.50	&	1	&	 $-$5.4 $\pm$ 2.4	&	-0.11	&	38	&	 $-$43 ($-$81 to $-$24)	\\
V617 Sgr	&	CBSS	&	0.207	&	WD	&	MS	&	RLOF	&	1.3	&	0.5	&	40	&	 $+$460 $\pm$ 8	&	-0.16	&	2500	&	 $-$2050 ($-$3100 to $-$920)	\\
EX Dra	&	DN	&	0.210	&	WD	&	MS	&	RLOF	&	0.75	&	0.56	&	0.079	&	 $+$30.5 $\pm$ 3.2	&	-0.11	&	1.9	&	 $+$29 ($+$25 to $+$32)	\\
HR Del	&	CN J(231)	&	0.214	&	WD	&	MS	&	RLOF	&	0.67	&	0.55	&	8	&	 $+$180 $\pm$ 40	&	-0.10	&	137	&	 $+$43 ($-$370 to $+$170)	\\
RW Tri	&	Nlike	&	0.232	&	WD	&	MS	&	RLOF	&	0.70	&	0.6	&	1.5	&	 $-$4.6 $\pm$ 0.8	&	-0.10	&	20	&	 $-$25 ($-$40 to $-$4.5)	\\
V691 CrA	&	LMXB	&	0.232	&	NS	&	MS	&	RLOF	&	1.69	&	0.5	&	0.05	&	 $+$157 $\pm$ 13	&	-0.16	&	4.0	&	 $+$153 (140 to 166)	\\
1RXS J0644	&	Nlike (SW)	&	0.269	&	WD	&	MS	&	RLOF	&	0.8	&	0.7	&	1	&	 $-$89 $\pm$ 15	&	-0.09	&	11.8	&	 $-$101 ($-$120 to $-$87)	\\
EM Cyg	&	Nlike	&	0.291	&	WD	&	MS	&	RLOF	&	1.0	&	0.77	&	0.1	&	 $+$6.6 $\pm$ 3.4	&	-0.11	&	2.1	&	 $+$4.6 ($+$0.3 to $+$8.0)	\\
V2134 Oph	&	LMXB	&	0.296	&	NS	&	MS	&	RLOF	&	1.48	&	0.8	&	0.019	&	 $-$11.8 $\pm$ 0.6	&	-0.15	&	0.80	&	 $-$12.4 ($-$13 to $-$10)	\\
AC Cnc	&	DN	&	0.300	&	WD	&	MS	&	RLOF	&	0.76	&	0.77	&	0.2	&	 $-$22.9 $\pm$ 3.4	&	-0.08	&	$-$0.25	&	 $-$22.6 ($-$26 to $-$19)	\\
V363 Aur	&	Nlike (SW)	&	0.321	&	WD	&	MS	&	RLOF	&	0.90	&	0.88	&	1.3	&	 $-$189.9 $\pm$ 1.2	&	-0.09	&	2.6	&	 $-$192 ($-$210 to $-$180)	\\
V616 Mon	&	LMXB	&	0.323	&	BH	&	MS	&	RLOF	&	6.6	&	0.8	&	0.01	&	 $-$1.1 $\pm$ 1.3	&	-0.39	&	0.87	&	 $-$1.6 ($-$3.3 to $-$0.3)	\\
BT Mon	&	CN F(182)	&	0.334	&	WD	&	MS	&	RLOF	&	1.04	&	0.87	&	0.09	&	 $-$68 $\pm$ 3	&	-0.10	&	1.39	&	 $-$69 ($-$83 to $-$66)	\\
AX J1745.6-2901	&	LMXB	&	0.348	&	NS	&	MS	&	RLOF	&	1.40	&	0.80	&	0.3	&	 $-$40.3 $\pm$ 3.2	&	-0.11	&	13.8	&	 $-$54 ($-$82 to $-$45)	\\
QZ Aur	&	CN 	&	0.358	&	WD	&	MS	&	RLOF	&	0.98	&	0.93	&	3	&	 $-$39 $\pm$ 14	&	-0.09	&	14.5	&	 $-$53 ($-$70 to $-$39)	\\
WX Cen	&	CBSS	&	0.417	&	WD	&	MS	&	RLOF	&	0.9	&	0.6	&	30	&	 $-$1200 $\pm$ 100	&	-0.04	&	1710	&	 $-$2900 ($-$4600 to $-$1800)	\\
GU Mus	&	LMXB	&	0.433	&	BH	&	MS	&	RLOF	&	8	&	0.9	&	0.17	&	 $-$890 $\pm$ 170	&	-0.31	&	17.6	&	 $-$910 ($-$1100 to $-$740)	\\
V Sge	&	V Sge	&	0.514	&	WD	&	MS	&	Runaway	&	0.85	&	3.3	&	2300	&	 $-$506 $\pm$ 6	&	-0.12	&	33k	&	 $-$33000 ($-$62000 to $+$30000)	\\
CI Aql	&	RN P(32)	&	0.618	&	WD	&	MS	&	RLOF	&	1.21	&	0.85	&	10	&	 $-$470 $\pm$ 600	&	-0.04	&	530	&	 $-$1000 ($-$1600 to $-$400)	\\
CI Aql	&	RN P(32)	&	0.618	&	WD	&	MS	&	RLOF	&	1.21	&	0.85	&	10	&	 $-$516 $\pm$ 45	&	-0.04	&	530	&	 $-$1050 ($-$1600 to $-$860)	\\
QR And	&	CBSS	&	0.660	&	WD	&	MS	&	RLOF	&	1.0	&	0.4	&	15	&	 $+$1130 $\pm$ 40	&	-0.02	&	3700	&	 $-$2500 ($-$14000 to $-$1300)	\\
Sco X-1 	&	LMXB	&	0.787	&	NS	&	MS	&	RLOF	&	1.40	&	0.7	&	0.6	&	 $+$87 $\pm$ 51	&	-0.02	&	83	&	 $+$4 ($-$510 to $+$130)	\\
U Sco	&	RN PP(3)	&	1.231	&	WD	&	subG	&	RLOF	&	1.36	&	1.0	&	9.8	&	 $-$3600 $\pm$ 2100	&	-0.02	&	780	&	 $-$4300 ($-$5800 to $-$2000)	\\
U Sco	&	RN PP(3)	&	1.231	&	WD	&	subG	&	RLOF	&	1.36	&	1.0	&	9.9	&	 $-$900 $\pm$ 900	&	-0.02	&	800	&	 $-$1700 ($-$2900 to $-$720)	\\
U Sco	&	RN PP(3)	&	1.231	&	WD	&	subG	&	RLOF	&	1.36	&	1.0	&	12.9	&	 $-$21k $\pm$ 3100	&	-0.02	&	1030	&	 $-$22k ($-$25k to $-$19k)	\\
U Sco	&	RN PP(3)	&	1.231	&	WD	&	subG	&	RLOF	&	1.36	&	1.0	&	8.2	&	 $-$17k $\pm$ 5600	&	-0.02	&	660	&	 $-$17.7k ($-$12k to $-$6k)	\\
U Sco	&	RN PP(3)	&	1.231	&	WD	&	subG	&	RLOF	&	1.36	&	1.0	&	9.5	&	 $-$18700 $\pm$ 7k	&	-0.02	&	760	&	 $-$19k ($-$26k to $-$12k)	\\
LMC X-4	&	HMXB	&	1.408	&	NS	&	HiMass	&	Wind	&	1.57	&	18.0	&	24	&	 $-$4970 $\pm$ 40	&	-0.13	&	300	&	 $-$5300 ($-$6200 to $-$5100)	\\
V394 CrA	&	RN P(5)	&	1.515	&	WD	&	subG	&	RLOF	&	1.34	&	0.94	&	7	&	 $+$300 $\pm$ 600	&	-0.01	&	830	&	 $-$530 ($-$3900 to $+$240)	\\
Her X-1	&	IMXB	&	1.700	&	NS	&	IntMass	&	RLOF	&	1.4	&	2.2	&	0.6	&	 $-$48.5 $\pm$ 1.3	&	-0.02	&	$-$65	&	 $+$17 ($-$16 to $+$49)	\\
Cen X-3	&	HMXB	&	2.087	&	NS	&	HiMass	&	RLOF	&	1.57	&	24	&	53	&	 $-$10189 $\pm$ 6	&	-0.08	&	$-$16k	&	 $+$6k ($-$7k to $+$20k)	\\
M82 X-2	&	HMXB	&	2.533	&	NS	&	HiMass	&	RLOF	&	1.4	&	8	&	240	&	 $-$57k $\pm$ 2400	&	-0.02	&	$-$88k	&	 $+$31k ($-$47k to $+$950k)	\\
V4641 Sgr	&	IMXB	&	2.817	&	BH	&	IntMass	&	RLOF	&	6.4	&	2.9	&	2	&	 $-$470 $\pm$ 340	&	-0.03	&	260	&	 $-$730 ($-$1500 to $-$390)	\\
V884 Sco	&	HMXB	&	3.412	&	NS	&	HiMass	&	Wind	&	1.96	&	46	&	250	&	 $-$6400 $\pm$ 2400	&	-0.07	&	2050	&	 $-$8.4k ($-$15k to $-$6k)	\\
QV Nor	&	HMXB	&	3.728	&	NS	&	HiMass	&	Wind	&	1.02	&	16	&	83	&	 $-$9700 $\pm$ 380	&	-0.02	&	2500	&	 $-$12k ($-$16k to $-$10k)	\\
SMC X-1	&	HMXB	&	3.892	&	NS	&	HiMass	&	Wind	&	1.21	&	18	&	150	&	 $-$37730 $\pm$ 20	&	-0.02	&	5030	&	 $-$43k ($-$55k to $-$39k)	\\
Cyg X-1	&	HMXB	&	5.600	&	BH	&	HiMass	&	Wind	&	21.2	&	40.6	&	260	&	 $+$2500 $\pm$ 2400	&	-0.27	&	6800	&	 $-$4300 ($-$11k to $-$130)	\\
V1017 Sgr	&	CN	&	5.786	&	WD	&	subG	&	RLOF	&	1.10	&	0.7	&	0.4	&	 $+$9500 $\pm$ 2900	&	0.00	&	300	&	 $+$9200 ($+$6300 to $+$12k)	\\
GP Vel	&	HMXB	&	8.964	&	NS	&	HiMass	&	Wind	&	2.12	&	26	&	90	&	 $-$24k $\pm$ 74k	&	-0.01	&	5300	&	 $-$29k ($-$100k to $+$ 45k)	\\
EXO 1722-363	&	HMXB	&	9.740	&	NS	&	HiMass	&	Wind	&	1.91	&	18	&	90	&	 $-$560k $\pm$ 370k	&	-0.01	&	8400	&	 $-$570k ($-$940k to $-$200k)	\\
OAO 1657-415	&	HMXB	&	10.45	&	NS	&	HiMass	&	Wind	&	1.74	&	17.5	&	40	&	 $-$97k $\pm$ 2800	&	0.00	&	4100	&	 $-$101k ($-$110k to $-$97k)	\\
T CrB	&	RN S(6)	&	227	&	WD	&	RG	&	RLOF	&	1.35	&	0.81	&	6.4	&	 $-$3.1kk $\pm$ 1.6kk	&	0.00	&	177k	&	 $-$3.3kk ($-$4.9kk to $-$1.7kk)	\\
\enddata
\begin{flushleft}	
\
$^a$ XRB and CV classes:\\
{\bf AM CVn~} AM CVn with a WD primary and a degenerate dwarf companion, \\
{\bf DN~} Dwarf Nova, \\
{\bf CN~} Classical Nova, light curve class (S, P, O, D, J, F) and $t_3$ in parentheses, \\
{\bf RN~} Recurrent Nova, light curve class (S, P, O, D, J, F) and $t_3$ in parentheses, \\
{\bf Nlike~} Nova-like CV \\
{\bf Nlike (SW)~} Nova-like CV of the SW Sex class \\
{\bf CBSS~} Close-binary supersoft source \\
{\bf LMXB~} Low-mass X-ray Binary (NS or BH primary, plus low-mass main-sequence companion) \\
{\bf IMXB~} Intermediate-mass X-ray Binary (NS or BH primary, plus $\sim$3 $M_{\odot}$ companion) \\
{\bf HMXB~} High-mass X-ray Binary (NS or BH primary, plus $\gtrsim$8 $M_{\odot}$ companion) \\

\end{flushleft}	
	
\end{deluxetable*}
\end{longrotatetable}


\begin{figure*}
	\includegraphics[width=2.0\columnwidth]{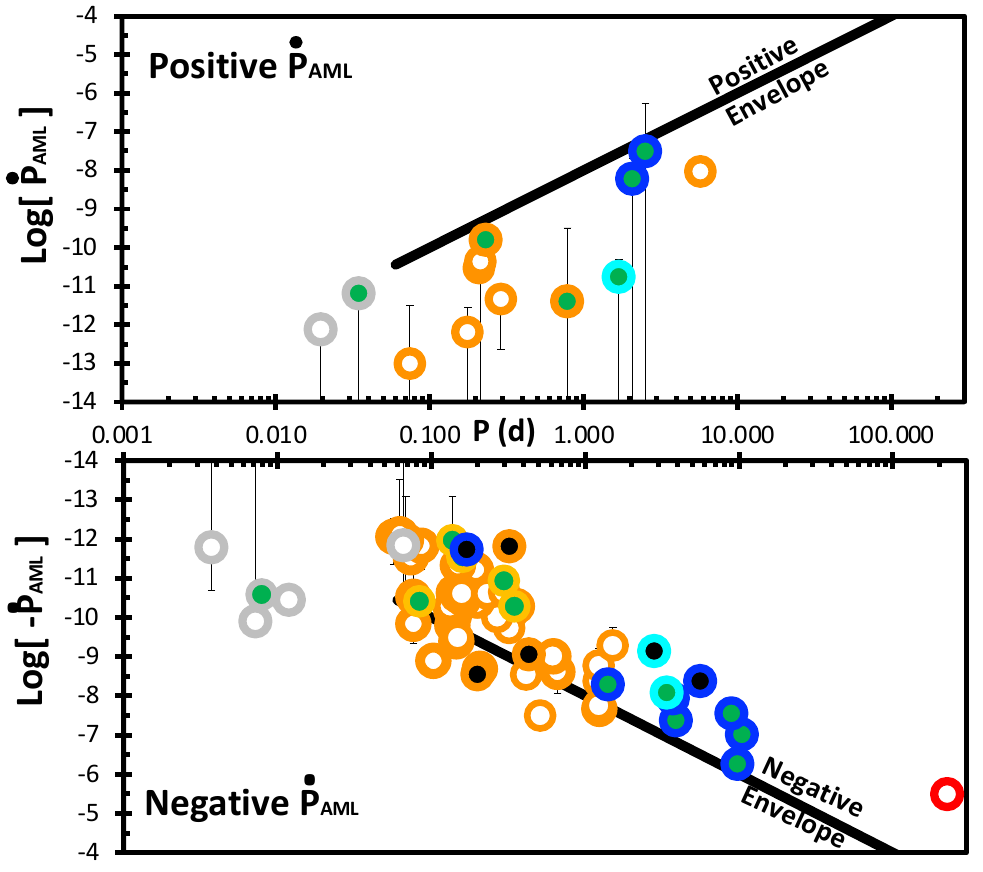}
    \caption{$\dot{P}_{\rm AML}$ versus $P$ for XRBs and CVs.  This figure shows the specific properties that can reveal the real dominant AML mechanism.  The format and descriptions for this plot are identical to those in Figure 1.  {\bf (1.)~}Most binaries lineup just inside the Negative Envelope (the thick black line in the lower panel for $-$$P/10000$), with this correlation being highly significant.  This property holds for periods from 0.0567 to 227 days.  {\bf (2.)~}This gives an AML law of $\dot{P}_{\rm AML} \propto - P^{\alpha}$, for $P$ in units of days, and $\alpha$ somewhere between 1.4 and 2.0.  The scatter about this relation is roughly one order-of-magnitude, due to variations in other binary properties, chiefly the accretion rate.  {\bf (3.)~}Roughly one third of the systems appear in the upper panel of Fig. 21 (i.e., positive $\dot{P}$), but most of these have moved to the lower panel in this figure (i.e., negative $\dot{P}_{\rm AML}$) due to mass transfer effects always being positive (i.e., $\dot{P}_{\rm mt}$$>$0).  Most of the systems remaining in the upper panel have very large error bars such that the acceptable range extends into the lower panel.  Only three stars appear to have significantly positive $\dot{P}_{\rm AML}$, EX Dra, V1017 Sgr, and V691 CrA.   }
\end{figure*}

Figure 22 shows that most XRBs and CVs follow a line just inside the Negative Envelope line.  This suggests an AML law of 
\begin{equation}
\dot{P}_{\rm AML} \propto - P^{\alpha},
\end{equation}
for periods given in days.  By itself, this AML law would be a great improvement over all prior models, including the MBM.  The power-law index, $\alpha$, appears to be from roughly 1.4--2.0.  But the scatter in Figure 22 is still around one order-of-magnitude.  This variance will be shown to arise from the other fundamental system parameters $M_{\rm prim}$, $M_{\rm comp}$, and especially $\dot{M}$.  The best AML law will be a function of of all of these system properties.

\section{EXAMPLES OF $\dot{P}_{\rm AML}$}

{\bf AM CVn} is the prototype star for its CV class.  After an epic program, Patterson et al. (2019) have measured the steady orbital period change with 10 per cent accuracy.  The observed $\dot{P}$ is substantially more negative than $\dot{P}_{\rm GR}$.  In this case, with $\dot{P}_{mt}$ being necessarily positive, there is no possibility that $\dot{P}_{\rm AML}$ can be near zero, for any accretion rate.  So here we have a confident case that CVs below the Period Gap must have some unidentified mechanism that is steadily reducing the $P$ at a rate that is 4$\times$ the GR effect.  This could well be connected with the observation (Schaefer 2024) that its minimum period (i.e., the bounce period) for AM CVn is substantially smaller than the CV period spike of G\"{a}nsicke et al. (2009).
 
{\bf T Pyx} is a recurrent nova with a unique history.  After a long wait as an ordinary CV, except with a high-mass WD, T Pyx erupted around the year 1866 as a {\it classical} nova, as proven by the high-mass and low-velocity of its ejecta, now seen closely by {\it HST} (Schaefer, Pagnotta, and Shara 2010).  The 1866 CN event excited a very high accretion rate onto the WD, which lead to a recurrent nova eruption, which kept the accretion high, so another RN eruption, and so on with fast repeating RN events.  Observed RN eruptions are in the slowing sequence of 1890, 1902, 1920, 1944, 1967, and 2011.  The slowing of the recurrence intervals is explained by a substantial dropping of the quiescent brightness (from $B$=13.8 {\it before} the 1890 eruption to $B$=16.3 in 2023) and the corresponding accretion rate drop by nearly a factor of 10$\times$.  The slowing down of the induced accretion makes for the slowing of the eruptions, with T Pyx even now transitioning back to its pre-1866 state.  The entire RN phase only produced $\sim$8 RN eruptions over 150 years or so.  The pre-2011 $\dot{P}$ was extremely large-positive (even being above the Positive Envelope), while the post-2011 $\dot{P}$ was also extremely large-positive, yet still 2$\times$ lower than before 2011.  Presumably the sharp drop in $\dot{P}$ is associated with the sharp drop in $\dot{M}$.  This can be tested by correcting the observed $\dot{P}$ for the reasonably-accurate $\dot{P}_{\rm mt}$ measure.  The resultant $\dot{P}_{\rm AML}$ values are consistent with zero.  That is, the ordinary mass-transfer effects from the very-high $\dot{M}$ provides a good explanation for the changes in the large-positive $\dot{P}$ measures.  Further, the T Pyx case shows no need for any period-change mechanism below the Period Gap.
 
{\bf IM Nor} is a recurrent nova regarded as a sister system to T Pyx.  However, IM Nor shows no evidence for the complex secular evolution similar to that arising from the 1866 regular-nova eruption of T Pyx that kickstarted the century-long episode of high accretion (Schaefer et al. 2010).  As an RN, IM Nor must have an accretion rate around 10$^{-7}$ $M_{\odot}$ yr$^{-1}$ and $M_{\rm WD}$$>$1.20 $M_{\odot}$ so as to support its fast recurrence timescale.  Based on the optical luminosity of the disk alone, Patterson et al. (2022) estimated that the accretion rate is 1.0$\times$10$^{-7}$ $M_{\odot}$ yr$^{-1}$, where I have corrected their estimate by 2.0$\times$ to account for the better distance in Schaefer (2022b).  The recurrence timescale and the $M_{\rm WD}$ force the accretion rate to be (1.5$\pm$0.5)$\times$10$^{-7}$ $M_{\odot}$ yr$^{-1}$.  IM Nor has a large and positive $\dot{P}$ that is accurately measured.  We can get the zero-case ($\dot{P}_{\rm AML}$=0) for an accretion rate of 2.3$\times$10$^{-8}$ $M_{\odot}$ yr$^{-1}$.  But this low rate is not possible for a recurrent nova, so the zero-case is rejected.  Hence, IM Nor is a CV below the Period Gap where some additional and unknown mechanism is operating to smoothly decay the orbit.

{\bf V482 Cam} is an ordinary SW Sex class nova-like CV with a 0.1336 day period just above the Period Gap.  The observed $\dot{P}$ has the formal uncertainty that is one-third of the measure, but this relatively large error bar will not contribute much change in the $\dot{P}_{\rm AML}$.  V482 Cam has an absolute magnitude of $+$7.0 so the accretion rate is somewhere around 7$\times$10$^{-9}$ $M_{\odot}$ yr$^{-1}$, and has no DN events so the accretion rate must be $>$3$\times$10$^{-9}$ $M_{\odot}$ yr$^{-1}$.  Given the period, $M_{\rm comp}$ is 0.30 $M_{\odot}$, for which I will allow the range 0.25--0.35 $M_{\odot}$.  For SW Sex stars just above the Gap, $M_{\rm WD}$ is in the range 0.6--1.0 $M_{\odot}$ (Rodr\'{i}guez-Gil et al. 2004), and I will take 0.8 $M_{\odot}$ as the best estimate.  With these best estimates, $\dot{P}_{\rm AML}$ is $-$52 in the dimensionless units of 10$^{-12}$.  Pushing all the input to their acceptable extremes, $\dot{P}_{\rm AML}$ ranges from $-$100 to $-$19 in units of 10$^{-12}$.  The modeled magnetic braking effect is close to $-$1.6 in units of 10$^{-12}$ (eq. 8 of Paxton et al. 2015), so the MBM is rejected by over one order-of-magnitude.

{\bf U Gem} is the prototypical DN, with well measured properties.  Echevarria et al. (2007), Dubus et al. (2018), and Pala et al. (2022) agree that $M_{\rm WD}$=1.17$\pm$0.03 $M_{\odot}$, $M_{\rm comp}$=0.44$\pm$0.04 $M_{\odot}$, and $\dot{M}$=(7.0$\pm$0.08)$\times$10$^{-10}$ $M_{\odot}$ yr$^{-1}$.  The best value for $\dot{P}_{\rm AML}$ is $+$0.64 in units of 10$^{-12}$.  When all the inputs are simultaneously pushed to their extremes, the range is $-$1.8 to $+$2.7 in units of 10$^{-12}$.   This is very small, and is even consistent with zero.  For the case of magnetic braking alone, the requirement is that the AML equals $-$3.1, and this is significantly more negative than is possible for U Gem.  Nevertheless, the measured range is sufficiently close to the MBM prediction such that we could imagine some large error capable of reconciling the two.

{\bf V617 Sgr} is one of only three known CBSS stars in our Milky Way, with these being characterized as WDs with a very high accretion rate supporting steady hydrogen burning.  These three stars (V617 Sgr, WX Cen, and QR And) all have their mass ratio $q<\frac{2}{3}$, which gives then completely different physics, evolution, and properties from V Sge\footnote{The three CBSS stars were originally labeled as `V Sge stars', based on some modest spectral similarities to the unique V Sge.  However, the stark and critical difference in mass ratio ($q<\frac{2}{3}$ versus $q=3.9$) makes everything in completely different modes, so the three CBSS stars must not be classed as V Sge stars.}.  For calculating $\dot{P}_{\rm AML}$, a potentially significant problem is the ambiguous evidence for either a jet or wind coming from the WD.  I have not found any estimate of the strength of any WD-wind or jet, and such might be small.  In such a case, I can only proceed to calculate $\dot{P}_{\rm AML}$ with the default value for simple RLOF.  Given the required high $\dot{M}_{\rm RLOF}$ and the well-measured $O-C$ curve parabola, the $\dot{P}_{\rm AML}$ is calculated to be $-$2100 (with an acceptable range from $-$3100 to $-$910) in dimensionless units of 10$^{-12}$.

{\bf V Sge} is the prototype for the `V Sge' class of CVs, although it is the only known member.  V Sge is unique as being a CV with the mass ratio $q$=3.9$\gg$1 driving a runaway accretion.  The very high $\dot{M}_{\rm RLOF}$ is driving a fast and strong wind that is ejecting a large fraction of the mass falling onto the accretion disk.  The observational measures have the RLOF at (0.6--3)$\times$10$^{-5}$ $M_{\odot}$ yr$^{-1}$ and the wind coming from the WD at (0.55--2.3)$\times$10$^{-5}$ $M_{\odot}$ yr$^{-1}$.  The wind must be smaller than the RLOF, although both strengths appears to be comparable.  For calculating $\dot{P}_{\rm mt}$ from Eq. 11, this case is the same as if $\epsilon$ equals 0.1 or so (and with $\dot{M}_{\rm wind}$ replaced by $\dot{M}_{\rm RLOF}$).  The uncertainty in $\dot{P}_{\rm AML}$ is dominated by the uncertainty in $\epsilon$, so the quoted range (-62,000 to $+$30,000 in units of 10$^{-12}$) is for $\epsilon$ varying from 0.01 to 0.3.  In the end, the AML effect has such a large error bar (including both large-positive and large-negative values) as to be not-helpful for understanding the AML mechanism, and to be not-helpful for understanding the nature of the V Sge wind.

{\bf Sco X-1} is the prototypical LMXB.  The observed $O-C$ curve (Fig. 12) shows a noisy parabola with a positive curvature (but only at the 1.6-$\sigma$ level), with the curvature greatly smaller than for all other systems with similar period.  In this case, the $\dot{P}_{\rm AML}$ measure is nearly the negative of $\dot{P}_{\rm mt}$.  The realistic uncertainties in $M_{\rm prim}$ (ranging from 1.2 to 1.4 $M_{\odot}$), $M_{\rm comp}$ (0.4--0.9 $M_{\odot}$), $\dot{M}_{\rm RLOF}$ (1.0--15$\times$10$^{-9}$ $M_{\odot}$ yr$^{-1}$), and $\dot{P}$ ($+$36 to $+$138 times 10$^{-12}$) contribute from $\pm$15 to $\pm$90 (in units of $10^{-12}$) when each input is varied over its acceptable range.  When all four inputs are pushed to their limit in the same direction, the extreme range of $\dot{P}_{\rm AML}$ is from -510 to +130 times $10^{-12}$, with the best estimate close to zero.  This is a useful result, because it shows the unknown AML mechanism has a near-zero effect for the Sco X-1 case.

{\bf Cyg X-1} is the prototypical BH HMXB, the first system with a plausible case for containing a black hole.  This can serve as an example for HMXBs, with this case having comparatively well-observed system parameters.  Despite huge and important efforts, the system parameters still have substantial uncertainties, for example with the BH mass having published values from excellent studies ranging by a factor of 2.  Fortunately, $\dot{P}_{\rm AML}$ has only small dependence over the range for $M_{\rm prim}$ and only modest dependences on the uncertainties of $M_{\rm comp}$ and $\dot{P}$.  The dominant uncertainty in $\dot{P}_{\rm AML}$ is from $\epsilon$$\dot{M}_{\rm wind}$.  I am taking the best estimates as $0.004$ times $2.6\times 10^{-6}$ $M_{\odot}$ yr$^{-1}$ (Miller-Jones et al. 2021) or $10^{-8}$ in the same units.  The quoted value of $\dot{M}_{\rm wind}$ has a small quoted error bar, and the range of possible values is relatively small (Gies et al. 2003), yet I expect the real uncertainty is substantially larger.  For wind strengths over the range (1--5)$\times 10^{-6}$ $M_{\odot}$ yr$^{-1}$, $\dot{P}_{\rm AML}$ changes from $-11,000$ to $-130$ in dimensionless units of $10^{-12}$.  As for all the other HMXBs, the range of acceptable AML effects is large, but the values are certainly negative, hugely-negative.

\section{COMPARISONS OF $\dot{P}_{\rm AML}$}

Table 6 lists my 84 derived $\dot{P}_{\rm AML}$ for the 25 XRBs and 52 CVs.  These can be compared with each other in Fig. 22.  This figure has the same format and legend as Fig. 21.  Whereas Fig. 21 shows the reality of what the binary evolution actually is, Fig. 22 shows the properties of the unknown AML mechanism(s).  In the comparison of the two figures, the most obvious difference is that the many systems with certainly-positive $\dot{P}$ have been lowered to near-zero or negative values, after the effects of mass transfer and GR have been removed.  That is, to within the quoted error bars, almost all of the binaries are between the horizontal zero-line and the Negative Envelope.  This is satisfying because astronomers have strongly known since the 1970s that binary evolution must have a negative AML so as to evolve the systems from long-$P$ to short-$P$.  

\subsection{Outliers}

Figure 22 shows most systems to be between the zero-line and the Negative Envelope (i.e., appearing in the lower panel), but there are a few systems that have their formal $\dot{P}_{\rm AML}$ value as positive.  A question is whether any system has significantly positive $\dot{P}_{\rm AML}$.  For M82 X-2, Cen X-3, Her X-1, YZ LMi, U Gem, HR Del, T CrB, V1405 Aql, and Sco X-1, the uncertainties are so large as to have the acceptable range including zero, so these are not significant outliers.  For EM Cyg, the zero value is barely outside the acceptable range, so it is easier to make a case that my acceptable range is slightly too narrow, rather than to make a case than the binary must have a positive value.  This leaves us with 3 out of 77 systems that confidently have positive-$\dot{P}_{\rm AML}$.

EX Dra has well-measured system parameters, with an acceptable range for $\dot{P}_{\rm AML}$ running from $+$25 to $+$32 in units of $10^{-12}$.  I do not know of any plausible dodge or mechanism that could change the value to being negative.

V691 CrA has the best estimate $\dot{P}_{\rm AML}$ at $+$153 in units of $10^{-12}$, and a full range of $+$140 to $+$166.  This seems to be a case where the AML effect is positive.  I am not able to impeach this result, because the two $O-C$ curves are in close agreement and the parabolas are finely measured, while the stellar masses are known to within modest error bars.  The accretion rate is known from the X-ray luminosity for a NS at the accurate {\it Gaia} distance.  The values of $\dot{P}_{\rm GR}$ and $\dot{P}_{\rm mt}$ are both greatly smaller than $\dot{P}$, so $\dot{P}_{\rm AML}$ is simply close to the well-measured $\dot{P}$ for all plausible system parameters.  If $\dot{M}$ is increased by a factor of 40$\times$ (from an estimated 5$\times 10^{-10}$ $M_{\odot}$ yr$^{-1}$ up to 2$\times 10^{-8}$ $M_{\odot}$ yr$^{-1}$), then the AML effect will be zeroed.  But this is accretion at the Eddington limit for this LMXB, and this violates the observed X-ray luminosity.  So I can only conclude that V691 CrA has a significantly {\it positive} AML mechanism at work.

V1017 Sgr has measured system parameters known with moderate accuracy, producing an acceptable range for $\dot{P}_{\rm AML}$ running from $+$6300 to $+$12000 in units of $10^{-12}$, so it appears that this is a confident case of a positive $\dot{P}_{\rm AML}$.  However, if a worker wanted to force a negative value, there is a possible way to achieve this.  The problem is that the $O-C$ curve has a long gap from the eruption in 1919 up to the light curve measured by A. Landolt starting in 1975.  During this gap, we only have sparse data (27 B-magnitudes from the Harvard plates from 1923--1950).  The trick is that the cycle count between minima around the eruption (fairly well determined from the pre-eruption Harvard plates) to minima in the late 1970s might be off by 1 cycle.  The cycle count was determined by fitting the phased light curve to a broken parabola, with one particular cycle count giving a substantially better chi-square.  However, it is possible that the cycle count is off by 1 cycle, and such would change the sign of the post-eruption $\dot{P}$, making it significantly negative, and making the $\dot{P}_{\rm AML}$ value to also be negative.  So the best information is that V1017 Sgr is an outlier with a significantly positive value, but at some unknown confidence level, the $\dot{P}_{\rm AML}$ might be negative.  At this late date, the only way to resolve this ambiguity is to use {\it TESS} in the upcoming Sectors 91 and 92 and to use the best plate-scale archival photographs at the Sonneberg Observatory.

In the end, only three out of the 77 XRBs and CVs have good cases for any positive-$\dot{P}_{\rm AML}$.  The other systems plotted in the upper panel of Fig. 22 all have large error bars, where the acceptable range includes negative values.  I do not know whether these three outliers represent some additional AML mechanism in operation, or whether the outliers have some unrecognized large error.

\subsection{Comparing to Magnetic Braking}

Fig. 1 and the analysis in Section 4.1 are testing the Magnetic Braking Model versus the observed $\dot{P}$, with both including the effects of GR and mass-transfer.  The MBM has failed utterly.  A related-but-different question is testing whether the specific prediction for the magnetic braking mechanism alone ($\dot{P}_{\rm mb}$) equals observed $\dot{P}_{\rm AML}$.  To derive the predicted period change from the magnetic braking mechanism alone, I use the formulation of equation 8 in Paxton et al. (2015).  This prediction is shown in Figures 1 and 21 as the purple line at the top of the lower panel.  Figures 1, 21, and 22 have identical axes, so the positioning of the purple line is the same in each.  The magnetic braking mechanism is zero below the Period Gap and might or might-not be applicable to HMXBs and IMXBs\footnote{Rappaport, Joss, and Webbink (1982) constructed the first paper proposing the MBM as a way to explain XRBs.}, so this leaves me with 50 measures for 44 CV and LMXB systems of all types.  For these systems, the central 80\% have $\dot{P}_{\rm mb}$ between $-$0.8 and $-$11 in units of $10^{-12}$.  The comparison to magnetic braking can be quantified by the parameter $\dot{P}_{\rm AML}$/$\dot{P}_{\rm mb}$.  K2011 emphatically state several times that all CVs and LMXBs must quickly and closely track along the singular path where $\dot{P}_{\rm AML}$/$\dot{P}_{\rm mb}$=1.0.  This prediction can be tested with my 50 measures.

The required prediction fails.  The average $\dot{P}_{\rm AML}$/$\dot{P}_{\rm mb}$ is 1700, while the median value is 15.  Of my 50 measured acceptable ranges, 9 ranges include $\dot{P}_{\rm mb}$, while five of these cases are only due to the acceptable range being very large, so the consistency is useless.  Three of the systems have the entire acceptable range being positive, which is impossible for magnetic braking.  From this large sample, 54\% of the systems have deviations from the magnetic braking predictions by more than a factor of 10$\times$, while 32\% deviate by more than a factor of 100$\times$, and 16\% deviate by more than a factor of 1000$\times$.  By any of these criteria, the magnetic braking mechanism fails in its predictions by orders-of-magnitude.

The magnetic braking mechanism might still be operating, but its effects are necessarily being dwarfed by some other mechanism.  In almost all of my systems, any contribution of $\dot{P}_{\rm mb}$ must be negligibly small.  As such, it is inconsequential as to whether magnetic braking is included or not-included.

\subsection{Comparing WDs versus NSs versus BHs}

In principle, the AML law might depend on the nature of the compact star.  Perhaps the incredibly high magnetic fields possible around NSs drive dynamics inside the accretion disk so as to govern the angular momentum loss, or perhaps the BH event horizon swallowing up half the accretion energy changes the irradiation environment so as to change the AML.  

The possibility that the AML law depends on the WD/NS/BH nature in the binary can be tested by looking at the distributions of the measured $\dot{P}_{\rm AML}$ in Figure 22.  For this, it is easy to spot the 6 systems with the black central regions and compare these to the the other NS and WD systems.  We see that the BH systems are overlapping all the other systems, following along a line parallel and inside the Negative Envelope.  The BH systems are above and below the Negative Envelope just like for the WD and NS systems.  Similarly, the NS binaries follow the same distribution as do the BH and WD binaries.  The NSs follow the line just inside the Negative Envelope, with a scatter comparable to those of the BHs and WDs.  Six NS systems appear in the upper panel of Figure 22, but just as for the WD systems in the upper panel, the measured values have their uncertainties consistent with zero.  The distributions of $\dot{P}_{\rm AML}$ are indistinguishable between the WD, NS, and BH binaries.  So I conclude that the AML law does not depend on the nature of the accreting star.

\subsection{Comparing $\dot{P}_{\rm AML}$ for five U Sco inter-eruption intervals}

I discovered the eclipses of U Sco back in 1989 and I have observed 63 eclipse times up until 2016.  The {\it Kepler} spacecraft K2 mission provided 59 eclipse times in 2014.  Since 2019, the 39 measured eclipse times have been from just one observer, G. Myers (AAVSO).  Figure 23 shows the complete $O-C$ curve with seasonal averages (from 173 eclipses outside of eruptions) from 1989 until one week ago.  This records the sudden period changes ($\Delta P$) across four RN eruptions, plus the $\dot{P}$ for 5 inter-eruption intervals.  To within the uncertainties, the $\dot{P}$ was the same low value for inter-eruption intervals 1987--1996 and 1999--2010, but then suddenly changed to a value 20$\times$ larger for the intervals 2010--2016, 2016--2022, and 2022--2024.  So something happened in the U Sco system around the time of the 2010 eruption so as to increase the $\dot{P}$ by a factor of 20$\times$.

Whatever mechanism is governing the AML law, it suffered a fast change in under one year.  Such a fast change is impossible for the magnetic braking mechanism, because it operates to change $\dot{P}$ on a timescale no faster than the companion star's rotational-synchronization timescale, which is 280 years for U Sco.  Such is also impossible with the conjectural `Applegate mechanism' (Applegate 1992).  Many of the other speculative period-change mechanisms (planets in orbit, globular cluster accelerations, and so on) also fail to allow the fast U Sco $\dot{P}$ change, so we know that $\dot{P}$ in U Sco is dominated by none of these conjectured mechanisms.

We learned from the case of T Pyx that both the $\dot{P}$ and $\dot{M}$ changed together suddenly across the RN event, so we have to wonder whether U Sco also has its fast changing $\dot{P}$ governed by some mechanism tied to the accretion rate.  For U Sco, the accretion rate is accurately measured from the out-of-eclipse brightness, for which we have a long record since 1969 (Schaefer 2022c).  I can update table 5 of Schaefer (2022c) with $\langle V \rangle$ from the post-2022 inter-eruption intervals, with the average-$V$ equaling 18.01 mag from 2023.0--2024.4.  With small variations, the brightness level (and hence the $\dot{M}$) is nearly constant.  Most critically, there is no accretion level change associated with the huge and sudden change in $\dot{P}$ in 2010.

The sudden change in $\dot{P}$ in 2010 must be somehow related to the 2010 RN eruption.  But this eruption was identical in details with all previous eruptions, both photometrically and spectroscopically, so it is a mystery as to why U Sco changed its $\dot{P}$.

\begin{figure}
	\includegraphics[width=1.0\columnwidth]{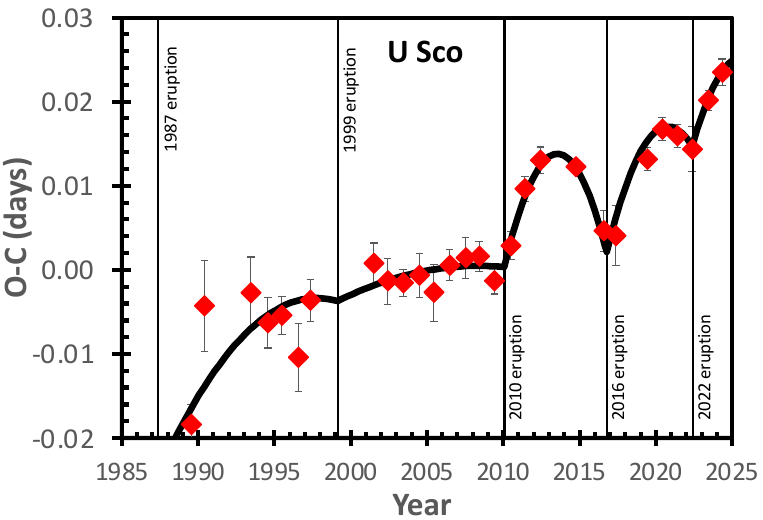}
    \caption{$O-C$ curve for U Sco.  This updated $O-C$ curve is constructed from seasonal averages of 173 eclipse times from 1989--2024.  This plot shows the kinks (i.e., sudden period changes by $\Delta P$) across the four RN eruptions in 1999, 2010, 2016, and 2022.  The $\Delta P$ changed from near zero in 1999, to large positive values in the last three eruptions.  The large $\Delta P$ requires that the binary separation suddenly increased, as caused by the nova eruption, although the physical mechanism remains a mystery.  The $\dot{P}$ between eruptions is represented by the fitted broken parabolas, shown as a thick black curve.  The first two inter-eruption intervals displays a relatively small and negative $\dot{P}$, consistent with no change in $\dot{P}$ before the 2010 eruption.  After the 2010 eruption, U Sco shows a nearly-constant $\dot{P}$, with a high value that is significantly higher than before-2010.  So the 2010 eruption somehow caused a sudden change in the $\Delta P$ and $\dot{P}$ behavior of the binary.  The 2010 eruption was then the all-time best observed nova eruption, and this eruption was indistinguishable from all the other eruptions, both photometrically and spectroscopically.  And with extensive photometry during quiescence from 1969--2024, the brightness level does not change greatly, so in particular, the period changes cannot be readily tied to any changes in accretion.  While the causes for the $\Delta P$ and $\dot{P}$ variations are still mysterious, they can be used to confidently rule out models (the MBM and the Hibernation model) as well as many particular physical mechanisms (including magnetic braking and the Applegate conjecture).}
\end{figure}

\subsection{Comparing sister systems}

Many pairs or groups of systems have closely similar configuration, and these can be used to see if there is consistency of $\dot{P}_{\rm AML}$ within each sisterhood.

AM CVn and YZ LMi are sisters, with nearly identical properties.  Yet AM CVn certainly has a substantial negative $\dot{P}_{\rm AML}$ (ranging from $-$42 to $-$27 in units of $10^{-12}$), while YZ LMi must be within a tight range centered on zero ($-$0.5 to $+$0.9).  What is different between the two systems to account for their greatly different $\dot{P}_{\rm AML}$?  My best explanation is to point to YZ LMi as having an accretion rate that is 355$\times$ smaller than for AM CVn, and similar factors for the other three AM CVn stars.  This could provide information that the AML law under the Period Gap has the effect being something like proportional to $\dot{M}$.  That is, YZ LMi has the AML effect so close to zero because it has minuscule accretion.  The other AM CVn stars all have $\dot{M}$$\sim$10$^{-8}$ $M_{\odot}$ yr$^{-1}$, with acceptable ranges consistent with the AML effect being, say, proportional to the accretion rate.

T Pyx and IM Nor are sister RNe below the Period Gap, sharing many of the same fundamental properties.  But T Pyx has $\dot{P}_{\rm AML}$ consistent with zero (for both before and after the 2011 eruption, with its large change in $\dot{M}$), while IM Nor has a definite and large negative-$\dot{P}_{\rm AML}$.  Their acceptable ranges have small overlap, so if a worker is desirous of making them equal, they can selectively push various system parameters to extremes so as to achieve the match.  Nevertheless, most workers will consider the systems' period changes to be greatly different, and wonder what condition makes for this difference.  The only answer that I can think of is that T Pyx is suffering a turn-off of its high accretion after its $\sim$1866 CN eruption, while there is no indication of such secular changes in IM Nor.  For this, I have no idea for how the turn-off makes to zero-out the $\dot{P}_{\rm AML}$ in T Pyx.

RR Pic, DQ Her, and T Aur form a sisterhood of three similar CNe.  They have similar periods, similar slow-$t_3$  D- and J-class light curves (pointing to $M_{\rm prim}$$\sim$0.8 $M_{\odot}$), and similar $M_{\rm comp}$.  Yet they have $\dot{P}_{\rm AML}$ values of $-$420, $-$6.3, and and $-$43, respectively, in units of $10^{-12}$.  How can these three sisters with similar binary properties return $\dot{P}_{\rm AML}$ values over a range of 100$\times$?  I think that the answer is that modest differences in $P$ and $\dot{M}$ combine to make the observed variation.  From Figure 22 we get $\dot{P}_{\rm AML}$$\propto$$P^2$, while from the next section we get $\dot{P}_{\rm AML}$$\propto$$\dot{M}$.  So a rough prediction is that the values should scale as $\dot{M}$$P^2$.  And indeed, this prediction is a reasonable match for the observed values.  The ratio of observed-to-predicted $\dot{P}_{\rm AML}$ is 1.6, 0.6, and 0.8, respectively.  So the observed scatter by a factor of 100$\times$ is reduced to a scatter of near 2$\times$.  So these three sisters have their wide variations in $\dot{P}_{\rm AML}$ accounted for by their modest differences in $P$ and $\dot{M}$.

The three CBSS systems (V617 Sgr, WX Cen, and QR And) are sisters with similar properties, often discussed together in papers.  In this case, the $\dot{P}_{\rm AML}$ measures are closely similar for all three sisters, being $-$2100, $-$2900, and $-$2500 in the usual dimensionless units, respectively.  For these stars, modest variations in $P$ (over a range of just over 3$\times$) are mirrored by modest variations in $\dot{M}$ (over a range a bit under 3$\times$), with these offsetting to produce the small scatter in $\dot{P}_{\rm AML}$.  That is, the variation of $\dot{M}$$P^2$ is relatively small.  Specifically, the ratio of observed-to-predicted values is 1.7, 0.8, and 0.5, respectively.  So for the three CBSS systems, the lack of variation in $\dot{P}_{\rm AML}$ goes along with the lack of variation in $\dot{M}$$P^2$.

\subsection{Comparing classes}

All of the members of the various classes of XRBs and CVs (AM CVns, ... HMXBs) share similar binary properties and are expected to share similar mechanisms and strengths of period-changes.  Indeed, each class has a scatter that is smaller than expected from the general population (tempered by the sometimes large error bars).  So we can meaningfully compare the properties between the classes.

\begin{table*}
	\centering
	\caption{Median properties for binary classes}
	\begin{tabular}{lrrrrcrr}
		\hline
		Binary class   &   $\#$   &   $\langle P \rangle$   &   $\langle M_{\rm prim}\rangle$ & $\langle M_{\rm comp}\rangle$  &  $\langle\dot{M}_{-8}\rangle$  &  $\langle\dot{P}\rangle$  &  $\langle\dot{P}_{\rm AML}\rangle$  \\
		   &      &   (days)   &   ($M_{\odot}$) & ($M_{\odot}$)  &  (10$^{-8}$ $M_{\odot}~ {\rm yr}^{-1}$)  &  ($10^{-12}$)  &  ($10^{-12}$)  \\
		\hline
AM CVn	&	5	&	0.007	&	0.71	&	0.08	&	1	&	0.4	&	-37	\\
LMXB below Gap	&	3	&	0.035	&	1.40	&	0.07	&	0.1	&	0.0	&	-27	\\
DN below Gap	&	7	&	0.07	&	0.80	&	0.11	&	0.007	&	-0.1	&	-1	\\
RN under Gap	&	3	&	0.08	&	1.30	&	0.20	&	10	&	367	&	-268	\\
SW Sex Novalike	&	13	&	0.15	&	0.80	&	0.30	&	0.7	&	-4.5	&	-57	\\
LMXB just above Gap	&	3	&	0.16	&	2.00	&	0.40	&	0.04	&	0.0	&	-2	\\
DN above Gap	&	4	&	0.19	&	0.96	&	0.56	&	0.075	&	-9	&	-11	\\
Novalike above Gap	&	5	&	0.20	&	0.70	&	0.47	&	0.5	&	-2.0	&	-25	\\
CN above Gap	&	7	&	0.20	&	0.93	&	0.50	&	1	&	-5.4	&	-43	\\
LMXB $P$$>$0.2 day	&	6	&	0.34	&	1.59	&	0.80	&	0.11	&	-5.7	&	-32	\\
CBSS	&	3	&	0.42	&	1.00	&	0.50	&	30	&	460	&	-2600	\\
RN $P$$\sim$1 day	&	8	&	1.2	&	1.36	&	1.00	&	9.8	&	-2200	&	-2900	\\
IMXB	&	2	&	2.3	&	3.90	&	2.6	&	0.013	&	-260	&	-360	\\
HMXB	&	10	&	3.8	&	1.83	&	18	&	90	&	-9900	&	-10400	\\
		\hline
	\end{tabular}	
\end{table*}

I have constructed median properties and period-changes for each of the classes (Table 7).  Some of the classes have been broken down by period range.  I have not included T CrB, V Sge, V1017 Sgr, or M82 X2, as these are all in unique classes by themselves, and they happen to have uselessly large error bars on $\dot{P}_{\rm AML}$.  Each RN interval is treated separately.  I use the median for each property, as such is less sensitive to the poor estimates of binary properties (especially for $\dot{M}$), and is less sensitive to the systems with large error bars.  The period changes are cast into the dimensionless units of $10^{-12}$ so as to have the values like integers of modest size.  The classes are ordered by $P$.

These median properties show a strong correlation between $\dot{P}_{\rm AML}$ and $\dot{M}$.  We can see this in Table 7, where all the classes with high-$\dot{M}$ are exactly those that have the largest negative-amplitude of $\dot{P}_{\rm AML}$.  That is, the RNe, CBSSs, and the HMXBs are notorious for their very-high accretion and wind rates, and those are just the systems the most negative $\dot{P}_{\rm AML}$.  Further, the two classes with the smallest $\dot{P}_{\rm AML}$ values also have the smallest $\dot{M}$ measures.

I have made a log-log plot for $-$$\dot{P}_{\rm AML}$ versus $\dot{M}$ for these classes, as shown in Fig. 24.  We see a fairly tight power law relation, with a slope near unity.  That is to say $\dot{P}_{\rm AML}$$\propto$$\dot{M}$.  This correlation looks strong, tight, and simple.  This correlation is not the best.  Part of the trouble is that the $\dot{M}$ is also correlated with the $P$, and $\dot{P}_{\rm AML}$ is well-correlated to $P$, so we can have trade-offs between $P$ and $\dot{M}$ dependencies that have been overlooked.  Further, the lumping together of the systems into classes has lost much of the information arising from the variance of the changing stellar masses.  

That $\dot{P}_{\rm AML}$ is approximately proportional to $\dot{M}$ is telling us that the AML mechanism is driven by the physics of the accretion.  This immediately rejects most proposed physical mechanisms for the dominant cause of the steady AML seen in all the XRBs and CVs.  So the AML driving binary evolution is itself being driven by some physics from the accretion disk, or its edges.

A better analysis is to seek to explain the huge variations in the group $\dot{P}_{\rm AML}$ as varying proportional to $P^{\alpha} \dot{M}^{\delta}$.  Previously, preliminary values of $\alpha$=2 and $\delta$=1 have been suggested.  These power-law indices can be optimized by minimizing the scatter of $\log[\dot{P}_{\rm AML}/(P^{\alpha} \dot{M}^{\delta})]$.  The minimum RMS scatter is achieved when $\alpha$=0.4 and $\delta$=0.7.  The AML law would then be expressed as 
\begin{equation}
\dot{P}_{\rm AML} = -170\times10^{-12} P^{0.4} \dot{M}^{0.7}_{-8}.
\end{equation}
Here, $P$ is in units of day, $\dot{M}_{-8}$ is in units of $10^{-8}~M_{\odot} ~ {\rm yr}^{-1}$, and the period change is in dimensionless units.  The scatter is just 0.33 in the logarithm, which is a great improvement over the original 4 orders-of-magnitude variation.  This scatter is getting comparable to the measurement errors, which is to say that the real AML law is similar to Eq. 13, and this has applicability for all XRBs and CVs.

This analysis needs improvement.  For one issue, it lumps all the binaries in each class together, so information and variations within each class are lost.  For a second issue, information and correlations involving the stellar masses are lost.  For a third issue, the dominant AML mechanisms are likely to be greatly different for binaries below the Period Gap, for LMXBs and CVs with periods above the Gap up to around 1.0 days, and for long period binaries like the HMXBs.  So what is needed is an analysis that keeps track of all the fundamental binary properties (i.e., $P$, $M_{\rm prim}$, $M_{\rm comp}$, and $\dot{M}$), uses all the individual $\dot{P}_{\rm AML}$ measures, and is calculated separately for each period range.

\begin{figure}
	\includegraphics[width=1.0\columnwidth]{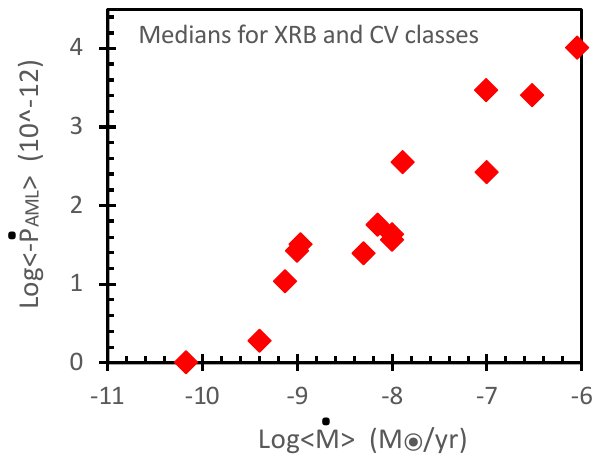}
    \caption{Median $\dot{P}_{\rm AML}$ versus $\dot{M}$ for XRB and CV classes.  For the classes and median properties in Table 7, we see that the values are closely and significantly correlated.  The slope in this log-log plot is close to unity, so $\dot{P}_{\rm AML}$$\propto$$\dot{M}$.  Most of the scatter in this relation is correlated with the $P$, indicating a more complicated relation for the AML law.  Considering $P$ and $\dot{M}$ together for these group medians, a better relation is $\dot{P}_{\rm AML}$$\propto$$P^{0.4}$$\dot{M}^{0.7}$.  Importantly, the dependency on $\dot{M}$ proves that the unknown AML mechanism is controlled by the accretion, which is to say that the physics of the angular momentum loss must arise somewhere in the accretion disk, the accretion stream, or maybe the boundary layer.}
\end{figure}

\section{NATURE OF THE AML}

The observed period changes of the CVs and XRBs certainly have contributions from mass transfer and GR, and most of the systems have an excess or extra component, denoted by $\dot{P}_{\rm AML}$.  This AML component certainly has no significant contribution from magnetic braking, and the dominant AML mechanism is not known.  With the strong need for understanding the evolution, demographics, and population synthesis of all close binaries, the identification and testing of the real AML component are now the forefront for stellar research.  For this, the path forward is to take the observed $\dot{P}_{\rm AML}$ values and try to deduce the properties of the real AML mechanism.

Many papers have calculated the equivalent of $\dot{P}_{\rm AML}$ for individual systems, and they often try to model the deviations from some expected value by invoking some new mechanism.  These mechanisms are of a wide variety and often rather imaginative.  New mechanisms, already published, to explain variations in the period include  {\bf (A)} dark matter winds (Pani 2015; G\'{o}mez \& Rueda 2017), {\bf (B)} multiple asteroid scatterings (Jain, Sharma, \& Paul 2022), {\bf (C)} circumbinary disks (Chen \& Li 2015), {\bf (D)} apsidal motion (Africano \& Wilson 1976, Antokhin \& Cherepashchuk 2019), {\bf (E)} acceleration from a globular cluster core (Chou \& Jhang 2023), {\bf (F)} stellar activity cycles changing the oblateness of the companion (Applegate 1992, Illiano et al. 2023), {\bf (G)} the Kozai effect (Prodan \& Murray 2012), {\bf (H)} non-conservative mass transfer involving accretion disks (Chou \& Jhang 2023), {\bf (I)} acceleration from a planet in orbit around the binary (Iaria et al. 2021), {\bf (J)} linear proper motion (Shklovskii 1970, Pajdosz 1995), {\bf (K)} unipolar inductors (Carvalho et al. 2022), {\bf (L)} flyby of a ``dark remnant'' (Peuten et al. 2014), and {\bf (M)} the old default of magnetic braking of the companion's rotation (K2011).  This litany of conjectured mechanisms is impressive.  The multitude of greatly different mechanism (and their limited applicability) further impresses me that none of the mechanisms has much of a chance of being the general answer for AML.  Well, some of the mechanisms are not `angular momentum loss', so I have to acknowledge that the $\dot{P}_{\rm AML}$ more-generally refers to all the mechanisms other than mass transfer and GR.  This paper is about the measured $\dot{P}$ and $\dot{P}_{\rm AML}$, so the existence and validity of any of these new mechanisms does not change  any of the results of this paper.  Rather, this paper is looking at the reality of period changes so as to try to extract the properties of the dominant AML mechanism(s).

In this section, I will use all 84 $\dot{P}_{\rm AML}$ measures from all 77 XRBs and CVs to derive the properties of the AML.  For this, I will be fitting the data to a model with
\begin{equation}
\dot{P}_{\rm AML} = - C P^{\alpha} M_{\rm prim}^{\beta}  M_{\rm comp}^{\gamma}  \dot{M}_{-8} ^{\delta},
\end{equation}
where $P$ is in units of days, the stellar masses are in units of $M_{\odot}$, and the accretion rate is in units of $10^{-8}~M_{\odot} ~ {\rm yr}^{-1}$.  This functional form is keying off the theory effects having multiplicative dependencies on the binary parameters, and this makes for power laws.  And this formulation involves the fundamental binary properties, so the equation is likely to provide a good representation of any dominant AML mechanism.  The model can be fit against the observed values in the usual chi-square sense, for deriving the best values of $C$, $\alpha$, $\beta$, $\gamma$, and $\delta$.  The goal here is that this set of parameters will describe an empirical model that can be used for modelers of individual binaries, for binary demographers, and for population syntheses.  The goal is also to provide a set of parameters that can be used to inspire and test physical models for the unknown dominant AML mechanism. 

A simple expectation is that the basic AML law for XRBs and CVs will be different for systems below the Period Gap, systems with $P$ from 0.13--1.0 days, and for HMXBs.  So I will first consider these groups separately, before considering the entire group of 77 binaries.

\subsection{Systems Below the Period Gap}

For systems below the Period Gap, with $P$$<$0.13 days or so, standard theory says that magnetic braking is not operational, nor is any other AML mechanism, so the evolutionary period change is only by mass transfer and GR (K2011).  In this case, $\dot{P}_{\rm AML}$ should equal 0.  However, there is some evidence for non-zero AML, with the revised-MBM model of Knigge et al. (2011) requiring that $\dot{P}_{\rm AML}$=1.4$\dot{P}_{\rm GR}$ so as to get the predicted minimum period ($P_{min}$) equal to the observed value.  For the real nature of the AML below the Period Gap, we can only look at the measured $\dot{P}_{\rm AML}$.

We can start with the four systems OY Car, Z Cha, T Pyx, and V4580 Sgr.  All four show highly-significant and fast kinks in their $O-C$ curves.  The short-duration events with large and fast $\dot{P}_{\rm AML}$ changes can be called `jerks'.  The 2010 jerk for T Pyx is certainly caused by the 2010 RN eruption, although the specific mechanism is still unknown.  However, the jerks for OY Car, Z Cha, and V4580 Sgr are certainly not from any nova event, because any such eruption would have been blatant in the tight on-going coverage of their light curves.  The jerks cannot arise from any conjectural magnetic braking or stellar activity cycle, as both mechanism can only change the $P$ on the rotational synchronization timescale, which is near 140 years for these four systems.  The jerks cannot be from any third body dynamics, because the effects are not periodic.  The jerks are not correlated with any brightening or dimming of the stars, either in the X-ray or optical, so they are not associated with any mass transfer events.  So the physical mechanism for the jerks is mysterious.  Nevertheless, jerks occur once every decade or two on 4 of my 18 systems below the Period Gap.  And these jerks are dominating the shape of the $O-C$ curves.  So the jerks prove that roughly $\frac{4}{18}$ of interacting binaries below the Period Gap have their evolution controlled by some unknown AML mechanism.

Further, the two systems AM CVn and IM Nor certainly have significantly negative $\dot{P}_{\rm AML}$.  Both stars have well-measured $O-C$ curves, so their $\dot{P}$ values are accurate and confident.  The uncertainty on the $\dot{P}_{\rm GR}$ contribution is negligibly small, and the entire GR contribution to $\dot{P}_{\rm AML}$ is negligibly small, so there is no possibility of zeroing the AML effect by varying the GR contribution.  So the only way to speculatively zero the $\dot{P}_{\rm AML}$ is to change $\dot{P}_{\rm mt}$, and the only way to try this is to decrease the $\dot{M}$.  But the limit on the acceptable range for AM CVn only moves slightly even for a zero-$\dot{M}$, and 
IM Nor has a strict lower limit of $\dot{M}$ based on the requirement that the accretion must be high so as to make a fast recurrence timescale.  So I see no possibility of AM CVn or IM Nor having a zero-$\dot{P}_{\rm AML}$, or anything near to zero.

So 6 of my 18 systems below the Period Gap certainly have some significant and unidentified AML mechanism dominating the evolution.  But what about the other 12 systems?  For these, they are all consistent with zero-$\dot{P}_{\rm AML}$.  (Well, HT Cas has its acceptable range coming close to zero, so a modest increase in the acceptable range of $\dot{M}$ can easily zero the value.)  For these 12 systems, the straight average is $-$50$\pm$6 (in units of $10^{-12}$), being dominated by the two systems with by far the largest error bars (T Pyx and ES Cet).  The weighted average is $+$0.34$\pm$0.23 in the same units, being dominated by the one system with by far the smallest error bar (YZ LMi).  The median equals -1.7, with this being less sensitive to outliers and to poor estimated $\dot{M}$ values.  The number of these systems that are consistent with a single value is maximized for $\dot{P}_{\rm AML}$=0.  With these conflicting criteria, I am unclear as to whether the 12 systems have zero AML.  For the 12 systems, the general $\dot{P}_{\rm AML}$ might be either zero, or perhaps some small negative value.

For the certainly-non-zero cases, can we recognize any system property that could be associated with the deviations from zero?  But V4580 Sgr has a sister system in V1405 Aql, so none of the visible system properties can explain the jerks.  OY Car and Z Cha are indistinguishable from the other DNe below the Gap, so the cause for their jerks is not apparent in the primary system parameters.  AM CVn is the prototype with properties similar to the other 4 AM CVn stars.  IM Nor is the sister of T Pyx, so we have no apparent cause for their difference in $\dot{P}_{\rm AML}$.  This is all discouraging for identifying the mechanism that is driving the non-zero AML evolution below the Period Gap.

From this analysis, below the Period Gap, 6-out-of-18 systems certainly have non-zero $\dot{P}_{\rm AML}$, while all the remainder have average values that are consistent with zero or with some small negative value.  The existence of these effects means that these systems cannot be used as effective tests for GR.

The 18 binaries with $P$ below the Period Gap can be used collectively to determine the AML law as in Equation 14.  The task is to vary the parameters until the chi-square between the observed $\dot{P}_{\rm AML}$ and the model $\dot{P}_{\rm AML}$ is minimized.  For this, the $\sigma$ is taken to be the addition in quadrature of the measurement error (see the last column in Table 6) and a systematic error.  A systematic error of 0.25 in the logarithm of the model $\dot{P}_{\rm AML}$ (as adopted for all binary groups) returns a reduced chi-square well below unity, which means that I have over-estimated the size of the measurement errors.  The one-sigma error bars on the best fit parameters are found from the region in parameter space over which the chi-square is within 1.0 of the minimum value.  The best fit model parameters are $\alpha$=$+$0.50$^{+0.54}_{-0.24}$, $\beta$=$-$0.3$^{+1.2}_{-1.8}$, $\gamma$=$-$0.50$\pm$1.4, and $\delta$=$+$0.87$\pm$0.20 (see Table 8).  The median for the model values is $-$12 in dimensionless units of $10^{-12}$, which is comparable to the one-system total-sigma.  When this best fit is compared to the $C$=0 case, the chi-square increases by a factor of 4.5$\times$, with this being a proof that binaries below the Period Gap do not have zero-$\dot{P}_{\rm AML}$.

The best parameters for this fit are presented in Table 8.  The reduced chi-square is around 0.4, which indicates that my adopted measurement error range are somewhat too large.  This makes for the quoted error bars to be somewhat larger than they should be.  The error bars on $\beta$ and $\gamma$ are large, mainly because there is only a small range for the stellar masses.  This makes for small overall uncertainty when the the relation is applied to stellar masses in the usual range for binaries below the Period Gap.  To be specific, the AML law for systems below the Period Gap is
\begin{equation}
\dot{P}_{\rm AML} = -110\times10^{-12} P^{0.50} M_{\rm prim}^{-0.3}  M_{\rm comp}^{-0.5}  \dot{M}^{0.87}_{-8}.
\end{equation}
$P$ is in units of days, $M_{\rm prim}$ and $M_{\rm comp}$ are in units of $M_{\odot}$, $\dot{M}_{-8}$ is in units of $10^{-8}~M_{\odot} ~ {\rm yr}^{-1}$, and $\dot{P}_{\rm AML} $ is in dimensionless units.  This empirical equation is now the best AML law for all contact binaries below the Period Gap.

\begin{table*}
	\centering
	\caption{The best empirical AML laws}
	\begin{tabular}{lccrrrrr}
		\hline
		Binaries   &   Number   &   \colhead{$C$}   &   \colhead{$\alpha$} & \colhead{$\beta$}  &  \colhead{$\gamma$}  &  \colhead{$\delta$}  &  $\chi^2$  \\
		  &      &   \colhead{$(10^{-12}$)}   &    \colhead{($P$ index)} &  \colhead{($M_{\rm prim}$ index)}  &   \colhead{($M_{\rm comp}$ index)}  &   \colhead{($\dot{M}_{-8}$ index)}  &    \\
		\hline
Below Gap	&	18	&	$110^{+1100}_{-88}$			&	$+0.50^{+0.54}_{-0.24}$			&	$-0.3^{+1.2}_{-1.8}$			&	$-$0.5	$\pm$	1.4	&	$+$0.87	$\pm$	0.20	&	5.6	\\
0.13$<$$P$$<$1.0 d	&	43	&	$1500^{+1630}_{-520}$			&	$+1.29^{+0.31}_{-0.84}$			&	$+2.75^{+1.45}_{-0.95}$			&	$-$1	$\pm$	0.45	&	$+$0.43	$\pm$	0.11	&	39.9	\\
HMXBs \& IMXBs	&	13	&	$10500^{+40000}_{-2500}$			&	$+$1.43	$\pm$	0.54	&	$-0.9^{+1.4}_{-0.8}$			&	$-0.4^{+1.2}_{-1.8}$			&	$+0.1^{+0.5}_{-0.9}$			&	9.4	\\
All	&	84	&	$1800^{+580}_{-430}$			&	$+$1.66	$\pm$	0.21	&	$+$2.24	$\pm$	0.61	&	$-$0.75	$\pm$	0.22	&	$+$0.50	$\pm$	0.09	&	83.9	\\
		\hline
	\end{tabular}	
\end{table*}

\subsection{CVs and LMXBs above the Period Gap}

The CVs and LMXBs above the Period Gap are at the core of the important issues involving binary models, evolution, and demographics.  These are the binaries for which the MBM was applied.  But the MBM does not work, so we have the question as to the distribution and properties of $\dot{P}_{\rm AML}$, hoping for clues to the real mechanism of AML.  

The general fit to the empirical power-law model of Equation 14 has been performed for the 43 XRBs and CVs with periods from 0.13 to 1.0 days.  The resulting best fit parameters are $\alpha$=$+$$1.29^{+0.31}_{-0.84}$, $\beta$=$+$$2.75^{+1.45}_{-0.95}$, $\gamma$=$-$1.00$\pm$0.45, and $\delta$=$+$0.43$\pm$0.11 (as placed into Table 8).  With the systematic error of $\pm$0.25 in the logarithm of $\dot{P}_{\rm AML}$, the best fit reduced chi-square is close to unity.  This is an impressive good fit for the empirical power-law model.  The fit has reduced the 5 orders-of-magnitude range of the observed $\dot{P}_{\rm AML}$ (just considering the negative measures) down to just $\frac{1}{4}$ orders-of-magnitude, being an improvement of 20$\times$ in the exponent.  This relatively small scatter shows that the empirical power-law model describes all the details of the observations with reasonably good accuracy.

The old-default, now-refuted, MBM predicts a specific formulation close to $\alpha$=$+$$\frac{1}{3}$, $\beta$=$-$$\frac{1}{2}$, $\gamma$=$+$$\frac{7}{6}$, and $\delta$=0 (see equation 8 of Paxton et al. 2015).  When compared to the fit in the previous paragraph, the chi-square has risen by 16.3, which implies something like a 4-sigma rejection of the MBM.

Another published AML model is named `consequential angular momentum loss', or CAML (Schreiber, Zorotovic, \& Wijnen 2016).  The core of this model is the bold conjecture that the dominant AML law has its formulation nearly that of Equation 14 with $\alpha$=$+$1, $\beta$=$-$2, $\gamma$=$+$1, and $\delta$=1 (see equation 3 of Schreiber et al. 2016).  No physical mechanism and no input data lead to this CAML conjecture.  Starting with this as a postulate, CAML is applied to population synthesis calculations to predict various CV demographic quantities.  The predicted CAML populations are then seen to solve the WD mass problem, plus they agree with the observed orbital period distribution and the space density of CVs.  When confronted by the collected $\dot{P}$ measures (i.e., using the model exponents in a fit to Equation 14), the CAML fails with a chi-square that is 27.7 larger than the best fit.

The conjectured power-law exponents of CAML were tweaked to make an alternative prescription for the AML law, with this revised model being named `empirical CAML', or e-CAML.  The basis for this tweak is to improve the predicted distribution of WD masses.  The e-CAML case has $\alpha$=$+$1, $\beta$=$-$1, $\gamma$=$-$1, and $\delta$=$+$1 (see equation 5 of Schreiber et al. 2016).  With these postulated parameters, the chi-square is 27.7 larger than the chi-square of the best fit, and the e-CAML model is strongly rejected.

These three models have exponents varying over a wide range, and this is telling us that theoretical models have no useful confidence, at least currently.  Further, the theory models are effectively picking variously widely-separated points in parameter space, which is to say that there is no connection to the reality of binary evolution.  In this situation, we can only ignore conjectured AML laws.  Instead of testing conjectures against tertiary data (like CV densities or solar-mass pre-main-sequence star slow-downs), we should measure the AML directly against the many measures of the most fundamental property that drives evolution (i.e., the $\dot{P}$ measures).

To be specific, the best fit AML law is
\begin{equation}
\dot{P}_{\rm AML} = -1500\times10^{-12} P^{1.29} M_{\rm prim}^{2.75}  M_{\rm comp}^{-1.00}  \dot{M}^{0.43}_{-8}.
\end{equation}
Here, $P$ is in units of days, the stellar masses are in units of $M_{\odot}$, the accretion rate is in units of $10^{-8}~M_{\odot} ~ {\rm yr}^{-1}$, and the AML period change is in dimensionless units (s/s).  This empirical power-law model for the CVs and LMXRBs above the Period Gap  is now the best representation of the dominant AML mechanism.  As such, this should replace all other formulations as used by modelers, as used for binary demography, and as used for all future evolution calculations. 

\subsection{HMXBs and IMXBs}

The HMXBs and IMXBs have greatly different natures from the CVs and LMXBs for the companion star, the accretion mode, and the compact star.  So we strongly expect that their AML mechanism will be completely different from that considered in the previous subsections.  To understand the reality of how HMXBs and IMXBs evolve, we can only consider the cases as shown in Fig. 1.  To understand the reality of the AML mechanism for HMXBs, we can only consider the cases as shown in Fig. 22.

I have collected and derived full information on 11 HMXBs and 2 IMXBs.  Unfortunately, the derived $\dot{P}_{\rm AML}$ for Vela X-1, Cen X-3, and M82 X-2 have such a large uncertainty so as to be useless.  Their acceptable ranges are poor only because the system properties ($M_{\rm prim}$, $M_{\rm comp}$, $\dot{M}$, and $\epsilon$) cannot be usefully constrained.  For the remaining 8 HMXBs, the constraints on the system parameters are known well enough to produce large-but-useful error bars.  So for seeking correlations and comparisons, only 7 NS systems and 3 BH systems contribute substantial constraints.

The best fit parameters for all 13 systems to Eq. 15 are in Table 8.  The best fit has a reduced chi-square close to unity.  This means that the adopted systematic error ($\pm$0.25 in the logarithm of $\dot{P}_{\rm AML}$) is close to the real uncertainty in this relation reflecting reality.  Nevertheless, the uncertainties are large, partly from having only 10 useful measures, partly from having all-but-two $M_{\rm prim}$ values spanning a small range, and partly due to correlations amongst the fit parameters allowing for one parameter to be pushed to an extreme.  To be specific, the AML law for HMXBs and IMXBs is 
\begin{equation}
\dot{P}_{\rm AML} = -10500\times10^{-12} P^{1.43} M_{\rm prim}^{-0.9}  M_{\rm comp}^{-0.4}  \dot{M}^{0.1}_{-8},
\end{equation}
with the same units as the previous equations.  This empirical AML law should now be used for all calculations of models, demographics and evolution of HMXBs.

\subsection{All 75 XRBs and CVs}

The dominant AML mechanism is likely different for systems below the Gap, for LMXBs and CVs with periods from 0.13--1.0 days, and for HMXBs.  In this case, modelers should use Equations 15, 16, and 17 as appropriate.  However, it is possible that only one mechanism dominates the AML for all these systems.  Such a case might arise if the AML comes from some aspect of the accretion disk (say, where the gases necessarily must be shedding angular momentum somehow, simply so the gas can move inward), with accretion disks being the sole constant phenomenon throughout all 77 systems.  In this case, we had best make a fit of Equation 14 to all 84 measured $\dot{P}_{\rm AML}$ values from all 77 binaries.

The best fit parameters are $\alpha$=$+$1.66, $\beta$=$+$2.24, $\gamma$=$-$0.75, and $\delta$=$+$0.50 (see Table 8).  The reduced chi-square for this fit is near unity, for an adopted systematic uncertainty of 0.25 in the logarithm of $\dot{P}_{\rm AML}$ added in quadrature with the measurement error (from the last column of Table 6).  This means that the joint overall fit is equal in quality for explaining all 77 systems to the quality of the three individual fits.  This makes a plausible case that it is better to represent the AML law of contact binaries by this one overall relation.  That is, the 84 measured $\dot{P}_{\rm AML}$ are equally represented by three independent models involving 15 fit parameters and by one model involving 5 parameters, so the one overall fit would be preferred.

Nevertheless, the dominant AML mechanisms are likely different for each period group.  The fit parameters for each of the period-groups  are significantly different from each other and from the overall fit.  So the overall best fit cannot be best for any of the three period groups.  When the overall fit is applied to the 18 binaries below the Period Gap, the chi-square for these binaries is 21.6, which is nearly 4$\times$ worse than the fit optimized for the same systems, indicating that the specialized fit is better than the overall fit for the systems below the Gap.  When the overall fit is applied to the 0.13--1.0 day group, the chi-square is 41.5, versus 39.9 for the specialized fit, so the specialized fit is better and should be used.  The overall best fit is somewhat similar to the parameters for the 0.13--1.0 day group because most of the systems were used for the specialized fit.  When the overall fit is applied to the 13 HMXBs and IMXBs, the chi-square is 10.8, whereas the best fit for these 13 systems is 9.4, which is reasonably close.  But this low chi-square is only happenstance because the HMXB parameter error bars are large while the best fit parameters are significantly different.  Importantly, the three period-groups all have power-law indices that are substantially different from each other.  With this, the three AML laws (Equations 15, 16, and 17) should be applied separately.

What about the 10 $\dot{P}_{\rm AML}$ measures that were not included in the three period-groups?  These include 8 measures from 3 long-period RNe (U Sco, V394 Cra, and T CrB), the CN V1017 Sgr, plus the unique V Sge.  The $\dot{P}_{\rm AML}$ measures for V394 CrA, V Sge, and T CrB have such large error bars that they would fit most any model.  These 10 measures have an average chi-square of 0.9 when evaluated with the HMXB model.  So the  HMXB model (Eq. 17) is an adequate description for all the binaries with periods $>$1.0 days.

The three AML laws are applicable for all the three period ranges ($P$$<$01.3, 0.13$<$$P$$<$1.0, and $P$$>$1.0 days).  That is, my empirical AML law is now the best description for all 77 binaries.  My collection of 77 binaries spans all classes of XRBs and CVs.  So I can claim that the AML laws are `universal', being applicable to all accreting binaries involving compact stars.

\section{CONCLUSIONS}

Here, I have calculated and collected $\dot{P}$ measures for 25 XRBs.  These are combined with my 59 measures of $\dot{P}$ from 52 CVs of all types, with 3 RNe providing 10 measures for separate inter-eruption intervals.  This total sample of 84 $\dot{P}$ measures from 77 systems covers all types of interacting binaries.  These were collected to test the required and exacting predictions of the MBM.  Further, I have pulled out the effects of GR and mass transfer, so as to derive 84 measures of $\dot{P}_{\rm AML}$, which represents the angular momentum loss to the orbit from all other effects.  These $\dot{P}_{\rm AML}$ values are a direct measure of the binary evolution.  The dominant mechanism or mechanisms are not known, so the task for this paper is to quantify the real AML law.

{\bf (1.)}~The real AML law is not the venerable MBM.  This is demonstrated for the 25 XRBs collected in this paper.  The MBM is also refuted by the 52 CVs collected in this paper and its predecessor.  These are for both $\dot{P}$ and $\dot{P}_{\rm AML}$, with such explicitly testing the one fundamental postulate of the MBM, as these measure the period changes that drive the evolution.  The data show that the MBM is usually orders-of-magnitude in error in its predictions of the strength of evolution.  Further, we now realize that the traditional successes of the MBM (for the bounce period and the Period Gap edges) were put in by-hand, and so are not successful predictions.  Indeed, the wide variations in the Gap edges, the wide range of $\dot{M}$, and the huge variations in $\dot{P}$ disproves the required MBM ideal that all CVs follow a single unique evolution track.  Further, we have the realization that there never was any useable evidence to support the existence of the magnetic braking conjecture.  So we are left with the strong conclusion that the MBM is completely rejected, and that the magnetic braking mechanism cannot be more than negligible in strength.  With this, our community is faced with the dilemma that all the many past papers using the MBM for modeling, demographics, and evolution are now no longer valid, and all results making any critical use of the MBM must now be relegated.  For the future, all use of the MBM or magnetic braking must cease.

{\bf (2.)}~I do not know of any model or mechanism that can account for my $\dot{P}_{\rm AML}$ measures.  The MBM is thoroughly refuted, and the CAML and e-CAML conjectures do not agree with my $\dot{P}_{\rm AML}$ measures.  The dozen imaginative mechanisms itemized in the start of Section 8 all fail for a variety of reasons.  Some of the conjectures fail because the $O-C$ curves do not show the required periodicities.  For the Applegate mechanism, the stellar cycle must be between 4.6 to 14.9 years, so all the observed $O-C$ curves are too long to show only an apparently-parabolic segment.  For the effects of a hypothesized third body in orbit around the binary, detailed analysis shows that the sinewave $O-C$ curves can appear as a parabolic segment in only $<$0.6\% of the binaries (Schaefer 2024).  Many of the proposals have problems with being able to match the amplitude of the period changes, including the Applegate conjecture, the multiple asteroid scattering idea, planets in orbit around the binary, apsidal motion, linear proper motion, and the Kozai effect.  Some of the imaginative conjectures are highly improbable for even one binary, including the multiple-asteroid-collisions and the Kozai effect.  Some of the conjectured mechanisms have troubles with physics for application to real binaries, including the `dark matter winds', the unipolar inductors, and the Applegate mechanism.  Some of the proposed mechanisms require very special settings (including the accelerations by globular cluster cores and the change in radial velocity for linear proper motion of a nearby star), such that these can be relevant for at most one of my 77 binaries.  Most of the imaginative conjectures cannot account for changes in $\dot{P}$ on timescales faster than centuries.  For example, both the magnetic braking mechanism and the Applegate mechanism cannot operate faster than the rotational-synchronization timescale of usually several centuries.  For the sudden period changes ($\Delta P$) associated with novae, all published mechanisms have already been refuted (Schaefer 2023, 2024).  In the end, there is no plausible mechanism or model to explain the $\dot{P}_{\rm AML}$ or $\Delta P$ measures.

{\bf (3.)}~The general evolution of the binaries is expected to have a steady $\dot{P}$, resulting in parabolic $O-C$ curves, and this is indeed seen, often with high precision.  But many of the XRBs and CVs have transient fast period changes that dominate the evolution:  {\bf (A)}~Five systems (OY Car, Z Cha, V4580 Sgr, SW Sex, and UY Vol) have sudden period changes, with no associated brightness changes, and these `jerks' certainly are from a different mechanism than makes for the steady parabolas, and these jerks are dominating the evolution over any steady AML mechanism.  {\bf (B)}~Three RNe (T Pyx, U Sco, and T CrB) suffer sudden and large changes in their steady $\dot{P}$ between the pre-eruption and post-eruption intervals in quiescence.  The mechanism for fast changes in $\dot{P}$ must be different from the steady mechanism seen in many other systems, and for these systems the fast-change mechanism dominates.  {\bf (C)}~All of the CN systems have sudden changes in $P$ across each nova eruption ($\Delta P$), and the long-term effect makes for an evolutionary period change over the nova recurrence timescale of $\Delta P$/$\tau_{\rm rec}$ that dominates over the steady changes during quiescence.  This separate mechanism presumably also applies to all the nova-like CVs and also the DNe.  Past demographics and evolution calculations have ignored this effect.  In all, just over half of my 77 accreting binaries have the unknown fast-period-change mechanism(s) dominating the long-term evolution.  That is, for a substantial fraction of the systems, the long-term evolution is dominated by some fast-change mechanism, with this being separate from the common steady $\dot{P}$ mechanism.  So any realistic model or demographic calculation must involve some accounting for the fast changes in $P$.  And there is a strong imperative for theory to identify and prove the physical mechanism(s) for these fast changes.

{\bf (4.)}~The nature of the steady AML law can be recognized, in part, by the distributions of $\dot{P}_{\rm AML}$.  The distribution of systems in Figure 22 is same for binaries with WDs, NSs, and BHs, so the AML law should not have any dependence on the nature of the compact accreting star.  The distribution in Figure 22 for systems with a highly-magnetized primary is the same as for systems with no evidence for a magnetized primary, so the dominant AML mechanism is not related to magnetic effects.  That most systems are somewhat inside the Negative Envelope (see Figure 22) points to the AML law having a dependency like $\dot{P}_{\rm AML}$$\propto$$P^{\alpha}$ for $\alpha$ roughly 1.4--2.0.  When medianed by class (see Figure 24), the systems follow $\dot{P}_{\rm AML}$$\propto$$\dot{M}^{\delta}$, with $\delta$ close to 1.0.  The best power-law indices will differ somewhat, partly because trade-offs between correlated binary properties will skew the slopes, and partly because the best fit slopes will differ between the period groups.  From this, we have the clear picture that the dominant steady AML mechanism (as separate from the common fast-change AML mechanism) for all period classes does not depend on the nature or magnetism of the primary star, while the AML mechanism is associated-with and controlled-by the accretion process.

{\bf (5.)}~In the absence of any viable theoretical model or any knowledge of the real dominant mechanism, the only possible approach is empirical, to let the stars `tell' us the actual period changes that are the evolution.  Theory models all have the AML effect related close to power laws on the four fundamental binary properties; $P$ (measured in units of days), $M_{\rm prim}$, $M_{\rm comp}$ (both stellar masses in units of $M_{\odot}$) and $\dot{M}_{-8}$ (measured in units of $10^{-8}$ $M_{\odot}$ yr$^{-1}$).  The three period ranges $<$0.13, 0.13--1.0, and $>$1.0 days likely have separate dominant AML mechanisms, so we need three equations to represent all the XRBs and CVs.  To be specific, I collect the AML laws as:
\begin{eqnarray}
\dot{P}_{\rm AML} = -110\times10^{-12} P^{0.50} M_{\rm prim}^{-0.3}  M_{\rm comp}^{-0.5}  \dot{M}^{0.87}_{-8} \nonumber \\
~~~~~~{\rm for}~P<0.13~{\rm d}, \nonumber \\
\dot{P}_{\rm AML} = -1500\times10^{-12} P^{1.29} M_{\rm prim}^{2.75}  M_{\rm comp}^{-1.00}  \dot{M}^{0.43}_{-8} \nonumber \\
~~~~~~{\rm for}~0.13<P<1.0~{\rm d}, \nonumber \\
\dot{P}_{\rm AML} = -10500\times10^{-12} P^{1.43} M_{\rm prim}^{-0.9}  M_{\rm comp}^{-0.4}  \dot{M}^{0.1}_{-8} \nonumber \\
~~~~~~{\rm for}~P>1.0~{\rm d}.
\end{eqnarray}
These equations now represent the best knowledge of the real AML law for contact binaries.  This AML law applies to all 77 XRBs and CVs of all classes, and so this law can be called `universal' for the contact binaries.

{\bf (6.)}~As an overview, my collection of $\dot{P}$ measures prove that the AML situation is greatly more complex and unknown than any in our community would have expected or hoped.  Three-to-five different AML mechanisms are at work, variously with each dominating in some systems:  {\bf (A)}~The majority of XRBs and CVs show parabolic $O-C$ curves, implying steady AML  Our best knowledge of the  behavior is represented in Equation 18, where each of three period ranges apparently have separate behaviors and separate mechanisms.  For the $P$$<$1 day systems, the strong positive correlation with $\dot{M}$ demonstrates that the mechanism must be driven by the accretion.  The physical mechanism is unknown.  {\bf (B)}~Superposed on the parabolic $O-C$ curves are occasional fast period-change episodes, with periods both increasing and decreasing on timescales of under a few months.  These fast changes are not associated with brightness changes or nova events, so are likely not being driven by accretion.  This separate AML mechanism is mysterious.  {\bf (C)}~Superposed on the parabolic $O-C$ curves are a different type of fast period change that happens at the time of each nova eruption.  For most novae, these fast period changes, once every eruption, dominate the evolution.  All published models for nova period-change have failed, so the real physical mechanism remains unknown.

{\bf (7.)}~How can the empirical AML law be improved?  It will be difficult to get any substantial number of additional XRBs or CVs with well-measured $\dot{P}$, because I have already used all the readily available accreting binaries.  A promising path is to measure and collect $\dot{P}$ values for 77 {\it non}-accreting binaries of all types, as this should reveal the reality of AML independent of $\dot{M}$.  To get small error bars, I judge that the only way is to recognize and prove some mechanism(s) and derive the theory relation.  But the litany of a dozen imaginative conjectures (see the start of Section 8) warns us that merely having an idea and a quick calculation is not adequate.  Success for a theoretical AML model will require that it match the direct $\dot{P}_{\rm AML}$ data, plus other indirect data (like accretion rates and Period Gap edges).  Success for a theoretical AML law requires that it be broadly applicable, and that it makes testable predictions that come true.

\begin{acknowledgments}
Juhan Frank (Louisiana State University), Joe Patterson (Columbia University), Saul Rappaport (Massachusetts Institute of Technology), Mike Shara (AMNH),  Tom Maccarone (Texas Tech), and Christian Knigge (University of Southampton) all provided their expert knowledge in discussions on broad issues and on technical details. 
\end{acknowledgments}

%

\vspace{5mm}
\facilities{AAVSO, DASCH, TESS, ZTF, MAST}

{}

\end{document}